\documentclass[12pt]{article}
\usepackage[utf8]{inputenc}
\usepackage{amsmath, amsthm, amssymb, amsfonts}
\usepackage[font=small, labelfont=bf]{caption}
\usepackage{jheppub}
\usepackage{amsmath, amsthm, amssymb, amsfonts}
\usepackage{comment}
\usepackage{graphicx}
\usepackage{float}
\usepackage[numbers,sort&compress]{natbib}
\usepackage{framed}
\usepackage{mathrsfs}

\usepackage{tikz}
\newcommand{\boxalign}[2][0.97\textwidth]{
  \par\noindent\tikzstyle{mybox} = [draw=black,inner sep=6pt]
  \begin{center}\begin{tikzpicture}
   \node [mybox] (box){%
    \begin{minipage}{#1}{\vspace{-5mm}#2}\end{minipage}
   };
  \end{tikzpicture}\end{center}
}

\usepackage{xcolor}

\newtheorem{Def}{Definition}

\newcommand{\roughly}[1]{\mathrel{\raise.3ex\hbox{$#1$\kern-0.85em\lower1ex\hbox{$\sim$}}}}
\newcommand{\lsim}{\roughly<}

\def\cA{{\cal A}}
\def\cB{{\cal B}}
\def\cD{{\cal D}}
\def\cC{{\cal C}}

\def\cG{{\cal G}}
\def\cH{{\cal H}}

\def\cJ{{\cal J}}

\def\cK{{\cal K}}
\def\cL{{\cal L}}
\def\cM{{\cal M}}
\def\cN{{\cal N}}
\def\cO{{\cal O}}

\def\cR{{\cal R}}

\def\cT{{\cal T}}

\def\cV{{\cal V}}
\def\cW{{\cal W}}

\def\mG{\mathcal{G}}

\newbox\charbox
\newbox\slabox
\def\slsh#1{{      
        \setbox\charbox=\hbox{$#1$}
        \setbox\slabox=\hbox{$/$}
        \dimen\charbox=\ht\slabox
        \advance\dimen\charbox by -\dp\slabox
        \advance\dimen\charbox by -\ht\charbox
        \advance\dimen\charbox by \dp\charbox
        \divide\dimen\charbox by 2
        \raise-\dimen\charbox\hbox to \wd\charbox{\hss/\hss}
        \llap{$#1$}
}}
\def\dsl{\slsh{\partial}}

\def\exd{{\hbox{d}}}
\def\d{\exd}

\def\mfy{\mathfrak{y}}

\def\bea{\begin{eqnarray}}
\def\eea{\end{eqnarray}}
\def\be{\begin{equation}}
\def\ee{\end{equation}}

\def\ol#1{\overline{#1}}

\def\ssA{{\scriptscriptstyle A}}
\def\ssB{{\scriptscriptstyle B}}
\def\ssC{{\scriptscriptstyle C}}
\def\ssD{{\scriptscriptstyle D}}
\def\ssE{{\scriptscriptstyle E}}
\def\ssF{{\scriptscriptstyle F}}
\def\ssG{{\scriptscriptstyle G}}

\def\ssI{{\scriptscriptstyle I}}
\def\ssJ{{\scriptscriptstyle J}}
\def\ssK{{\scriptscriptstyle K}}
\def\ssL{{\scriptscriptstyle L}}
\def\ssM{{\scriptscriptstyle M}}
\def\ssN{{\scriptscriptstyle N}}
\def\ssP{{\scriptscriptstyle P}}
\def\ssQ{{\scriptscriptstyle Q}}
\def\ssR{{\scriptscriptstyle R}}
\def\ssS{{\scriptscriptstyle S}}
\def\ssT{{\scriptscriptstyle T}}
\def\ssU{{\scriptscriptstyle U}}
\def\ssV{{\scriptscriptstyle V}}
\def\ssW{{\scriptscriptstyle W}}
\def\ssX{{\scriptscriptstyle X}}
\def\ssY{{\scriptscriptstyle Y}}

\def\KK{{\scriptscriptstyle KK}}

\def\DBI{{\scriptscriptstyle D \hspace{-0.3mm}B \hspace{-0.3mm}I}}
\def\CS{{\scriptscriptstyle CS}}

\def\WZ{{\scriptscriptstyle W \hspace{-0.6mm}Z}}
\def\IIA{{\scriptscriptstyle IIA}}

\def\dsl{\slsh{\partial}}

\def\vo{\mathcal{V}}

\def\nn{\nonumber}
\def\ov{\overline}

\def\d{\mathrm{d}}

\def\({\left(}
\def\){\right)}

\def\pref#1{(\ref{#1})}

\def\d{\mathrm{d}}

\title{UV Shadows in EFTs: Accidental Symmetries, Robustness and No-Scale Supergravity}

\author[a,b]{C.P.~Burgess,}
\author[c,d]{M. Cicoli,}
\author[c,d]{D. Ciupke,}
\author[e]{S. Krippendorf}
\author[f]{and F. Quevedo}
\affiliation[a]{Department of Physics \& Astronomy, McMaster University\\ $\phantom{wd}$ 1280 Main Street West, Hamilton Ontario, Canada L8S 4M1}
\vspace{2mm}
\affiliation[b]{Perimeter Institute for Theoretical Physics\\  $\phantom{wd}$ 31 Caroline Street North, Waterloo Ontario, Canada  N2L 2Y5 }
\vspace{2mm}
\affiliation[c]{Dipartimento di Fisica e Astronomia, Universit\`a di Bologna,\\  $\phantom{wd}$ via Irnerio 46, 40126 Bologna, Italy.}
\vspace{2mm}
\affiliation[d]{INFN, Sezione di Bologna, Italy.}
\vspace{2mm}
\affiliation[e]{Arnold Sommerfeld Center for Theoretical Physics, LMU,\\ $\phantom{wd}$ Theresienstr.~37, 80333
M\"unchen, Germany.}
\vspace{2mm}
\affiliation[f]{DAMTP, University of Cambridge, Wilberforce Road,  Cambridge, CB3 0WA, UK.}
\vspace{2mm}

\preprint{LMU-ASC 23/20}

\abstract{We argue that accidental approximate scaling symmetries are robust predictions of weakly coupled string vacua, and show that their interplay with supersymmetry and other (generalised) internal symmetries underlies the ubiquitous appearance of no-scale supergravities in low-energy 4D EFTs.  
 We identify 4 nested types of no-scale supergravities, and show how leading quantum corrections can break scale invariance while preserving some no-scale properties (including non-supersymmetric flat directions). We use these ideas to classify corrections to the low-energy 4D supergravity action in perturbative 10D string vacua, including both bulk and brane contributions. Our prediction for the K\"ahler potential at any fixed order in $\alpha'$ and string loops agrees with all extant calculations.  $p$-form fields play two important roles: they spawn many (generalised) shift symmetries; and space-filling 4-forms teach 4D physics about higher-dimensional phenomena like flux quantisation. We argue that these robust symmetry arguments suffice to understand obstructions to finding classical de Sitter vacua, and suggest how to get around them in UV complete models.
}

\begin{document}

\maketitle

\section{Introduction}

Effective field theories (EFTs) are particularly well-suited to the present situation in fundamental physics.  In essence, EFTs streamline the process of making predictions at energies well below a system's characteristic scale, $M$ \cite{TheBook,PetrovBlechman}. Being able to do so simplifies most problems because it allows one to ignore the myriad of irrelevant higher energy scales and concentrate on the degrees of freedom relevant to phenomena at a particular scale. Indeed, a microscopic understanding of quarks is not important when describing atomic or condensed matter physics, and general principles of symmetry, locality and unitarity bring us very far in understanding these phenomena without need for an underlying theory.

Gravitational physics in general -- and string theory\footnote{Although nobody knows for sure what the right theory of quantum gravity is, we here take string theory as our guide since it is the only proposal so far within which our questions can be asked with sufficient precision.} in particular -- seems not to be an exception. On one hand the observational successes of General Relativity (GR) are understood to be robust consequences of almost {\it any} theory of quantum gravity, regardless of UV details  \cite{Weinberg:1978kz, Donoghue:1994dn, Burgess:2003jk, Donoghue:2017ovt}. On the other hand, much of what we know about string theory was obtained by properly identifying the low-energy degrees of freedom and their EFT description, including some of string theory's deepest properties like duality symmetries.

What is unusual about gravity is the enormous hierarchy between currently accessible energies, $E \lsim E_w \sim 10$ TeV, and the much higher energies indicated by the gravitational scale $M_p = (8\pi G_\ssN)^{-1/2} \sim 10^{15}$ TeV. The enormity of this hierarchy has spawned two opposite perspectives about how to understand the world around us. A conservative extreme asserts the hierarchy is so large that gravitational physics is irrelevant. In this viewpoint the focus is on EFTs in their own right, without bothering with possible UV completions. What makes this view difficult is the ubiquity of gravity amongst the clues -- the evidence for dark matter, dark energy and the observed pattern of primordial fluctuations -- we have for how our current theories must be modified.

The other extreme --- the `swampland' -- hypothesis  \cite{Vafa:2005ui} -- instead asserts that for most EFTs sensible UV completions do not exist, so one should instead concentrate on the (more predictive) subset of EFTs for which they do. The focus then becomes an ever-evolving list of conjectures about the properties an EFT must have to allow such a completion \cite{Palti:2019pca}. This point of view is driven by the continued success of the Standard Model at LHC energies, the apparent evidence that de Sitter space is relevant to understanding both primordial fluctuations and the present-day dark energy, and the apparent difficulty in finding either of these within convincing UV completions. 

We here pursue an intermediate point of view, that builds on a traditional EFT strength. EFTs are powerful, but only if you build into them all of the symmetries of the underlying physics. Once this is done EFTs efficiently separate `universal' low-energy predictions from `model-dependent' ones. That is, some low-energy predictions ({\it e.g.}~the Meissner effect for a superconductor \cite{Weinberg:1986cq}; the weak-coupling of statistically degenerate fermions \cite{Polchinski:1992ed, Shankar:1993pf}; or soft-pion theorems for Quantum Chromodynamics (QCD) \cite{Weinberg:1978kz}) are robust consequences of essentially any microscopic description that shares the same low-energy (quasi-)particle and symmetry content. Other predictions ({\it e.g.}~the value of the superconducting transition temperature $T_c$; or the band structure of conducting electrons) are much more model-dependent, and are therefore more informative about what is going on over shorter scales.

Using perturbative string theory as a guide, we argue here (for instance) that any scarcity of de Sitter vacua in string theory is not evidence that many EFTs lie in a swampland. We instead show why de Sitter vacua are always scarce in completely arbitrary EFTs that share the symmetries intrinsic to string theory (and the symmetry reasons for this show how such vacua might be constructed). Furthermore, these symmetries provide a robust foundation for the hierarchies of masses and interactions often found in explicit string constructions. Furthermore, the interplay of these symmetries seem to provide interesting new ways to think about naturalness problems, and why small masses and scalar potentials are sometimes surprisingly robust to UV details. Because our arguments rest on symmetry grounds, and because these symmetries have striking echoes in many of the low-energy puzzles we seek to understand, they can also be useful for UV agnostics who do not care about short-distance completions.

Approximate scale invariance plays a key role in our arguments. On the phenomenological side, there are several reasons why approximate scale invariance (more precisely defined below) seems relevant to fundamental physics. One of these is the nearly scale-invariant pattern of primordial fluctuations.  Another is the long-standing electroweak-hierarchy and vacuum-energy naturalness problems associated with the Standard Model. Scale invariance can be relevant to naturalness problems like these, which hinge on the small size of a scalar mass or vacuum energy, both of which are controlled by dimensionful contributions to a theory's scalar potential.  

This paper does not start from a phenomenological perspective, however. Instead, we argue that several specific types of approximate scale invariance are generic predictions for perturbative string vacua\footnote{Indeed, the existence of these approximate scale invariances is not itself a new observation~\cite{Witten:1985xb, Burgess:1985zz, Nilles:1986cy}.} (and higher-dimensional supersymmetric models in general) and that it is the interplay between these and 4D supersymmetry that give the resulting EFTs unusual naturalness properties. They in particular provide new mechanisms for suppressing corrections to scalar potentials, which lie at the root of the special properties satisfied by `no-scale' supergravity models~\cite{Cremmer:1983bf, Barbieri:1982ac, Chang:1983hk, Ellis:1983sf, Barbieri:1985wq}. We show how these mechanisms arise in the low-energy limit of explicit higher-dimensional supergravity and string compactifications, illustrating their generic nature by using examples taken from type IIA, IIB and heterotic supergravity.

The ubiquity of scale invariance in perturbative string vacua is easy to understand. The key observation is that because string theory has no dimensionless parameters, all perturbative expansions ultimately involve powers of fields, with the action given in the regime $\phi, \psi \ll 1$ as a sum 
\be
   S_{\rm eff} = \sum_{n,m} \cA_{nm} \phi^n \psi^m \,,
\ee
for some fields $\phi$ and $\psi$. Any particular term in this expansion automatically scales in a particular way $S_{mn} \to \lambda^{np_\phi + mp_\psi} S_{mn}$ under a rescaling of the form $\phi \to\lambda^{p_\phi}\phi$ and $\psi \to \lambda^{p_\psi} \psi$. The double field series of this type that we use below arises in practice because of the generic expansion in string loops and in the generic low-energy $\alpha'$ expansion, that are always present for weak-coupling string compactifications.\footnote{These generic string scaling symmetries (with supersymmetry) explain why scale invariances like \pref{metricscaling} are generic in supergravity in six or more spacetime dimensions~\cite{Salam:1989fm, Burgess:2011rv, BMvNNQ, GJZ}. }

We extend on older ideas~\cite{Witten:1985xb, Burgess:1985zz, Nilles:1986cy} that show how scaling arguments efficiently organize how scale-breaking arises order-by-order within perturbative corrections. Once combined with other accidental symmetries (supersymmetry and shift and `shift-like' -- defined below -- symmetries), scaling arguments can account for many of the hierarchies of scale seen in string compactifications, and underlie non-renormalisation theorems for all three of the primary functions that define 4D $\cN=1$ supergravities. While this has long been known~\cite{Witten:1985bz, Burgess:2005jx} for the superpotential, $W$, and gauge kinetic function, $\mathfrak{f}_{ab}$, we show how it can also be true  -- in a sense more precisely explained below -- for the K\"ahler potential, $K$.

\subsection{Scaling, supersymmetry and naturalness}

For the present purposes scale invariance is taken to mean any rigid symmetry that rescales the metric 
\be 
g_{\mu\nu} \to \lambda^2 g_{\mu\nu} \,,
\label{metricscaling}
\ee
and possibly transforms other fields similarly, $\chi^i \to \lambda^{w_i} \chi^i$ (no sum on `$i$') for some weights $w_i$, where $\lambda$ is a constant positive real scale parameter. We call such a transformation a classical symmetry\footnote{Although not strictly speaking a symmetry (since the action is not invariant, and is usually anomalous to boot), this behaviour suffices to ensure invariance of the classical equations of motion.} if under it the Lagrangian density transforms as $\mathcal{L} \to \lambda^{w_\ssL} \mathcal{L}$, for some weight $w_\ssL$. If the action contains the Einstein-Hilbert action, $\mathcal{L} \supset \sqrt{-g} \; \cR$, then $w_\ssL = D-2$ in $D$ spacetime dimensions. We note for future use that if two such transformations are symmetries --- distinguished from one another by acting differently on the non-metric fields --- then they can always be combined in such a way as to write one symmetry as not acting on the metric. 

In later sections special roles are played by fields whose non-zero vacuum values break the scaling symmetry. The metric need not be one of these, despite the appearances of \pref{metricscaling}, because a background metric can preserve scale invariance if there exists a diffeomorphism that, when combined with \pref{metricscaling}, leaves it invariant. The infinitesimal version of the required diffeomorphism, $\delta x^\mu = V^\mu(x)$, defines a homothetic vector field,\footnote{Homothetic vector fields are special cases of conformal Killing vector fields -- {\it i.e.}~vector fields for which $\nabla_\mu V_\nu + \nabla_\nu V_\mu$ is proportional to $g_{\mu\nu}$ -- with $\nabla^\mu V_\mu$ also required to be a constant.} for which $\nabla_\mu V_\nu + \nabla_\nu V_\mu = c \, g_{\mu\nu}$ for constant $c$. Homothetic fields need not exist for generic background metrics, and it is only when they do not that the metric becomes a scale-breaking field.

\subsubsection{Scaling and scalar potentials}

Scale invariance is perhaps the only known symmetry that can enforce the vanishing of a vacuum energy {\em even if it is spontaneously broken}. This is one of the things that makes studies of scale invariance so compelling. Physically, this occurs because --- like for any spontaneously broken global symmetry --- scale transformations continuously relate different scale-breaking field configurations. That is, if $\langle \chi^i \rangle \neq 0$ solves the classical field equations then the scale-invariance of these equations implies $\langle \tilde\chi^i \rangle = \lambda^{w_i} \langle  \chi^i \rangle$ must also be a solution, giving rise to one-parameter families of scale-breaking classical vacua. However --- unlike for internal, rephasing symmetries --- the scale-invariant vacuum (for which $\langle \chi^i \rangle = 0$) also lies in this one-parameter family (corresponding to the $\lambda \to 0$ limit). But the absence of scales forces $V$ to vanish when evaluated at a scale-invariant configuration, and the fact that all the non-zero $\langle \chi^i \rangle$ are related to this point by a symmetry forces $V$ to vanish for all of them as well. 

A more formal way to see this proceeds as follows. Scale invariance of a potential typically means
\be \label{ScInvV}
   V(\lambda^{w_i} \chi^i) = \lambda^{w_\ssV} V(\chi^i) \,,
\ee
for some non-zero\footnote{The weight $w_\ssV$ is typically non-zero to ensure the combination $\sqrt{-g} \; V$ transforms properly, given that the measure, $\sqrt{-g}$, also transforms under the scaling \pref{metricscaling}.} weight $w_\ssV$. Differentiation of this expression with respect to $\lambda$ then implies 
\be \label{ScInvV2}
   \lambda \frac{\exd}{\exd \lambda} V(\lambda^{w_i} \chi) = \sum_i w_i \lambda^{w_i} \chi^i \, \left( \frac{\partial V}{\partial \chi^i} \right)_{\lambda^w \chi} = w_\ssV \lambda^{w_\ssV}  V(\chi) \,.
\ee
Evaluating this at $\lambda = 1$ then shows why $V(\chi_c^i)$ necessarily vanishes (provided $w_\ssV \neq 0$) for any configuration $\chi_c^i$ that is a stationary point. That is, if $(\partial V/\partial \chi^i)_c = 0$ for all $\chi^i$ for which $w_i \neq 0$, then $V(\chi_c^i) = 0$. Clearly this in particular implies the absence of any stationary point with $V (\chi_c^i) \neq 0$, such as would be required for an anti-de Sitter or de Sitter minimum. 

The conclusion that $V(\chi_c^i)$ vanishes holds regardless of whether or not the extremum occurs at the scale-invariant point since it does not assume $\chi^i_c = 0$. Furthermore, any scale invariant point is necessarily an extremum for any field whose weight satisfies $w_i \neq w_\ssV$. To see this it suffices to differentiate \pref{ScInvV2} with respect to $\chi^j$ (and again evaluate the result at $\lambda = 1$), since this implies
\be
  w_j  \frac{\partial V}{\partial \chi^j} + \sum_i w_i \chi^i  \frac{\partial^2 V}{\partial \chi^i \partial \chi^j}  = w_\ssV  \frac{\partial V}{\partial \chi^j}  \,.
\ee
If $\chi^i = 0$ then $(w_\ssV - w_j) (\partial V/\partial \chi^j) = 0$, from which the result follows.

Unfortunately, despite early exploration~\cite{EarlyScale} (see also \cite{Salvio:2014soa}) these observations have not yet proven useful for solving naturalness problems, for several reasons. First, Weinberg's no-go argument~\cite{Weinberg:1988cp} states that although scale invariance can ensure $V$ vanishes along a family of scale-breaking minima, it cannot guarantee the {\em existence} of the scale-breaking minima: small radiative corrections consistent with scale invariance can lift the flat direction along which $\lambda$ varies, leaving only the scale-invariant solution $\chi^i_c = 0$. (See {\em e.g.}~\cite{Burgess:2013ara} for a more recent review of this argument.)

But it is usually even worse than this since quantum corrections typically do not respect scale invariance at all. Although scale invariance is easily arranged to be a symmetry of the classical field equations, it rarely survives quantisation. As mentioned above, most often \pref{metricscaling} does not leave the classical action invariant. Instead one usually finds 
\be 
\label{Sscale}
  S(\lambda^{w_i} \chi^i) = \lambda^{w_\ssL} S(\chi^i) 
\ee
with $w_\ssL \ne 0$. Although \pref{Sscale} is sufficient to ensure invariance of the classical equations of motion, $\delta S/\delta \chi = 0$, it is not a quantum symmetry, so quantum corrections to $S$ need not satisfy \pref{Sscale}. This is typically true even if the classical action were invariant -- {\it i.e.}~if $w_\ssL = 0$ in \pref{Sscale} -- since scale transformations are usually anomalous~\cite{ScaleAnomaly}.

\subsubsection{Supersymmetry and no-scale}

Although very generic, these counter-arguments in themselves do not specify how big any quantum corrections to a would-be flat scaling direction must be. Since supersymmetry famously can keep flat directions flat, even including quantum corrections, one might hope the lifting of flat scaling directions might be suppressed if scale invariance were combined with supersymmetry. Indeed this certainly happens if scale breaking occurs without also breaking supersymmetry, since then supersymmetric non-renormalisation theorems~\cite{SUSYNR} ensure that the scalar potential's flat directions remain flat. But the real challenge is when scale invariance and supersymmetry both break (since both must in any description of the real world).

Intriguingly, there is a broad class of supergravity models for which the classical potential is precisely flat even though supersymmetry breaks along this flat direction. These are models of  the `no-scale' form~\cite{Cremmer:1983bf, Barbieri:1982ac, Chang:1983hk, Ellis:1983sf, Groh:2012tf, Ferrara:1994kg, Covi:2008ea}, whose supersymmetric K\"ahler potential, $K$, by definition satisfies
\be
\label{KKKnoscale}
   K^{i\bar\jmath} \, K_i K_{\bar\jmath} = 3 \,.
\ee
Here subscripts denote partial derivatives --- as in $K_i := \partial K/\partial T^i$ and $K_{\bar\jmath} := \partial K/\partial \ol T^j$ --- and $K^{\bar \jmath i}$ is the inverse matrix to $K_{i\bar\jmath} := \partial^2 K/\partial T^i \partial \ol T^j$. 

No-scale supergravities are known to have special properties, such as having a non-negative $F$-term potential, $V_\ssF \ge 0$, whenever the superpotential $W$ is independent of $T^i$: $W_i = 0$ (something also enforceable with axionic symmetries). To see why recall that\footnote{We follow standard supergravity practice and use units for which $M_p = 1$.} 
\be 
\label{VFdef}
  V_\ssF := e^K \; \left( K^{\ssA\bar\ssB} D_\ssA W D_{\overline{\ssB}} \overline{W} - 3 |W|^2 \right) \,,
\ee
where $\{ z^\ssA \} := \{ T^i, Z^a \}$ denote a collection of fields, for which \pref{KKKnoscale} is satisfied only for the subset of fields $T^i$. Here $D_\ssA W$ denotes the K\"ahler derivative of the superpotential defined by
\be
D_\ssA W := W_\ssA + K_\ssA W \,. 
\ee
Importantly, supersymmetric minima for the `other' fields, $Z^a$, satisfy $D_a W = 0$. Using this, $W_i = 0$ and \pref{KKKnoscale} in \pref{VFdef} implies $V$ vanishes for all $T^i$ at these minima, even if $W$ itself does not. Furthermore, non-zero $W$ implies that supersymmetry is generically broken along these flat directions because the supersymmetry-breaking diagnostic, $D_i W$, typically does not vanish. 

No-scale models turn out to arise very naturally whenever scale invariance and supersymmetry are both present. As shown in more detail below, if the scaling fields $\chi^i$ are the real parts of the chiral multiplets $T^i$, then it can happen that scale invariance requires $e^{-K/3}$ to be a homogeneous degree-one function of the scaling fields
\be
e^{-K/3} \to \lambda \, e^{-K/3} \quad \hbox{when} \quad T^i \to \lambda\, T^i \,.
\ee
As is shown below --- see also Appendix  A of ref.~\cite{Burgess:2008ir} --- when $e^{-K/3}$ is a homogeneous degree-one function that depends only on the real part $\chi^i = T^i + \overline{T}^i$ (as often happens due to axionic shift symmetries), the K\"ahler potential necessarily satisfies \pref{KKKnoscale}.

At face value the no-scale condition \pref{KKKnoscale} seems not so useful for naturalness questions because it is usually not preserved by quantum corrections. This mirrors the statement that scale invariance is itself only approximate, partly because the transformation \pref{metricscaling} transforms the action according to \pref{Sscale}. Furthermore, in the higher-dimensional examples discussed below the scale invariance often acts only on a subset of the fields $\{ \chi^i \}$ and does not extend to act on the $Z^a$. Part of the story to follow therefore is to track how such sources of scale-invariance breaking control the form found for the low-energy 4D effective theory. Of particular interest is the size of loop corrections to the effective potentials, which are not protected by non-renormalisation theorems when supersymmetry breaks along a flat direction.\footnote{In detail this happens because non-renormalisation theorems do not protect the K\"ahler potential from quantum corrections, and these can ruin the no-scale condition \pref{KKKnoscale}.}

\subsubsection{Subleading suppression: beyond minimal no-scale}
\label{ssec:ExtraSupp}

Although scale invariance and the no-scale condition, \pref{KKKnoscale}, are not in general preserved by loop corrections, we now argue below that quantum corrections to the scalar potential's flat directions in these models can nevertheless be smaller than a generic one-loop size. This argument relies on the loop-counting parameter itself being one of the scaling fields.

Additional suppression turns out to arise for two reasons. First, it sometimes happens that quantum corrections that break scale invariance sometimes nonetheless continue to respect the no-scale identity \pref{KKKnoscale}. This happens because although scale invariance can be sufficient for no-scale supersymmetry, it is actually not necessary. 

Second, it also happens that the scalar potential can remain flat even if \pref{KKKnoscale} is violated, so traditional no-scale models form only a subset of supersymmetric models with supersymmetry-breaking but flat potentials. It turns out that the broadest criterion for flat potentials in 4D $\cN=1$ supergravity require
\be 
\label{ExtendedNoScale}
\det M = 0 \quad \hbox{with} \quad M_{i \bar\jmath} := \partial_i \partial_{\bar\jmath}\; e^{-\cG/3} \,,
\ee
where 
\be
\cG(z, \bar z) := K(z,\bar z) + \ln|W(z)|^2\,,
\label{G}
\ee
is the usual K\"ahler-invariant function built from $K$ and $W$. We call \pref{ExtendedNoScale} the `generalised no-scale' condition, and show below (following \cite{Barbieri:1985wq}) that it is the necessary and sufficient condition for the vanishing of the $F$-term potential, $V_\ssF = 0$. Eq.~\pref{ExtendedNoScale} is a `generalised' condition because although \pref{KKKnoscale} can imply \pref{ExtendedNoScale} (such as when $W$ is independent of $T^i$) the converse need not be true. 

Whenever the leading correction to $K(T,\ov T)$ satisfies \pref{KKKnoscale} or \pref{ExtendedNoScale} it does not lift the scalar potential's flat direction, which therefore survives to one higher order than would naively have been expected. We now sketch a cartoon of how this actually happens in practical examples (with concrete realisations from specific low-energy string vacua given in later sections). In known examples the suppression comes when the loop-counting parameter is itself one of the scaling fields,\footnote{As elaborated below, it is precisely when a field is an expansion parameter that scaling symmetries arise at lowest order~\cite{Witten:1985xb, Burgess:1985zz, Nilles:1986cy}, and this is what makes approximate scale invariances ubiquitous for compactifications of perturbative string vacua.} {\it e.g.} $\rho := \chi^{i_0} = T^{i_0} + \ov T^{i_0}$ for some $i_0$. In this case the Lagrangian --- and so also the function $e^{-K/3}$ --- arises as a series of schematic form 
\be 
\label{FieldLoopL}
e^{-K/3} = \cA_0\, \rho + \cA_1 + \frac{\cA_2}{\rho} + \cdots \,,
\ee
where the functions $\cA_k$ do not depend on $\rho$ (but can depend on scale-invariant combinations of the other fields). Because the leading term in the action, $S_0/\hbar$, depends on $\hbar$ only through the combination $\rho/\hbar$, the same arguments that establish that each loop order is associated with an additional factor of $\hbar$ ensure that an $\ell$-loop graph computed using the first term of \pref{FieldLoopL} must be proportional to $(\hbar/\rho)^{\ell-1}$. Consequently tree-contributions from a term like $\cA_k/\rho^{k-1}$ compete with $k$-loop contributions computed using the $\cA_0$ term, and so on. For such theories it is the large-$\rho$ limit that is well-described by semi-classical methods. 

However, the $\rho$-dependence given by \pref{FieldLoopL} also guarantees classical scale invariance for the leading term, $\cA_0\,\rho$, under the transformation $\rho \to \lambda\, \rho$ with all other fields held fixed (including\footnote{The metric acquires a non-trivial transformation property like \pref{metricscaling} once one transforms to Einstein frame by rescaling $\hat g_{\mu\nu} = \psi^p g_{\mu\nu}$ to remove all powers of $\rho$ from the action's Einstein-Hilbert term.} the metric $\hat g_{\mu\nu}$). Higher-loop contributions also scale homogeneously, though differently than does the leading $\cA_0$ term, with an $\ell$-loop contribution scaling like $\cA_\ell/\rho^{\ell-1} \to \lambda^{(1-\ell)} \cA_\ell/\rho^{\ell-1}$. 

Now comes the main argument: for supersymmetric theories, the leading term, $e^{-K_0/3} = \cA_0 \,\rho$, is linear in $\rho$ and so the leading-order K\"ahler potential is 
\be  
K_0 = - 3 \ln  \rho  - 3 \ln  \cA_0
\ee
where $\cA_0$ is $\rho$-independent. As is easily checked, this form for $K_0$ satisfies $3K_{i_0 \bar\imath_0}  = K_{\rho\rho} = 9/\rho^2 = (K_\rho)^2 = K_{i_0} K_{\bar\imath_0}$ as an identity, and so satisfies the no-scale condition \pref{KKKnoscale} for the field $T^{i_0}$. If $W$ is also independent of  $T^{i_0}$ then the $F$-term potential \pref{VFdef} has a flat direction parameterised by $\rho$, with all other fields, $Z^a$, fixed by their field equations $D_a W = 0$. 

Loop corrections might normally be expected to lift this flat direction,\footnote{At least when $W \neq 0$ since then supersymmetry is broken along the flat direction.} but in this case keeping both tree-level and one-loop terms in $K$ gives
\be 
\label{expK1loo}
e^{-K_1/3} =  \cA_0\, \rho + \cA_1  \,.
\ee
But because $\cA_1$ does not depend on $\rho$ this means the corrected result \pref{expK1loo} continues to satisfy the no-scale condition \pref{KKKnoscale}, even though it is no longer homogeneous degree-one in $\rho$. The no-scale condition is easiest to see by noting that $W$ is (by a tree-level assumption, protected by non-renormalisation theorems) independent of  $T^{i_0}$ and
\be
\partial_{i_0} \partial_{\bar\imath_0} e^{-K_1/3} = 0 \,.
\ee
Consequently lifting of the $\rho$ flat direction first arises at {\it second} order in the semi-classical expansion in powers of $1/\rho$, rather than at first order. 

The remaining sections flesh these arguments out  in detail by extracting the all-orders implications of the generic accidental scale invariances associated with the string loop and low-energy expansions. We argue that these scaling symmetries (with supersymmetry) explain why it is fairly generic for scale invariances like \pref{metricscaling} to arise in supergravity in six or more spacetime dimensions~\cite{Salam:1989fm, Burgess:2011rv, BMvNNQ, GJZ}. It also explains why no-scale structure arises so often when these are compactified to 4 dimensions. The implications of scale invariance in 4D then accounts nicely for some of the generic mass hierarchies that arise within these models, particularly for Large Volume Scenarios (LVS) compactifications~\cite{LV, LVcorr} for Type IIB string vacua. 

Along the way we provide  multiple examples of the above loop-suppression mechanism at work in these low-energy string-vacuum EFTs (and possibly also in  Supersymmetric Large Extra Dimensional (SLED) models~\cite{SLED}). For both of these types of models radiative corrections to the scalar potential are non-zero, but are known to be smaller than normally would be expected \cite{vonGersdorff:2005bf,LVLoops, LVLoops1,SLED, SLEDLoops, SLEDLoops1, SLEDLoops2, SLEDLoops3, Burgess:2015gba, Burgess:2015kda, Burgess:2015lda}. By relating the size of corrections to approximate symmetries we hope to allow more systematic searches for circumstances where the suppressions they provide might be more dramatic.

The fact that there are multiple field expansions (and so multiple scale invariances) is also conceptually important for evading the Dine-Seiberg problem~\cite{Dine:1985he}. This asks how the expansion field itself can ever be stabilised in a regime for which the expansion is valid. (That is, in the absence of a hierarchy among the coefficients $v_k$, a potential of the form $V = v_0 \rho + v_1 + v_2/\rho + \cdots$ is generically stabilised for $\rho \sim \cO(1)$, which is too small to trust the $1/\rho$ expansion.) A loophole to this argument arises once there are multiple expansions involving both string coupling and other small quantities, and in this case different types of fields compete with one other in the potential without the multiple expansions needing to break down. This kind of mechanism is realised explicitly in LVS models, with the extra-dimensional volume stabilising at large values without destroying the weak-coupling or low-energy (large-volume) expansions. In detail the consistency of these solutions relies on the no-scale structure of the type IIB field equations, which in turn follows from the multiple scale invariances.  

\subsection{Other uses for scaling and no-scale supergravity}

Besides being useful in their own right, we now summarize how these scaling symmetries lie at the root of some of the phenomenologically attractive features of the compactifications of string vacua, many of which build on the hierarchies of masses that such models inherit from their underlying scaling structure. 

\subsubsection*{Scale-invariant inflationary potentials} 

Scale invariant models very often produce exponential potentials~\cite{EarlyScale, Goncharov:1985yu, Burgess:2016owb} whose shallowness is natural inasmuch as it is protected by effective non-compact rescaling shift symmetries~\cite{Burgess:2014tja} (in the same way that compact shift symmetries protect axionic inflationary models~\cite{Freese:1990rb}). Interestingly, these exponential potentials are also known to provide more successful descriptions of cosmological observations~\cite{Martin:2013tda, Martin:2013nzq, Kallosh:2014rga} than do axion-based models. 

The ubiquity of scale invariance in higher-dimensional supergravity helps ensure these kinds of potentials also have UV completions in string theory~\cite{Burgess:2001vr, Conlon:2005jm, Cicoli:2008gp, Cicoli:2011ct, Burgess:2013sla, Broy:2015zba, Cicoli:2016chb}. Indeed the scaling arguments for type IIB compactifications given below identify a broad class of moduli --- namely fibre moduli --- that first receive masses at high enough order in the string-coupling and $\alpha'$ expansions to that they are naturally light enough to be lighter than the inflationary Hubble scale (and so are natural inflaton candidates)\footnote{ For other recent discussions of no-scale inspired phenomenology and cosmology see for instance~\cite{Ellis:2020xmk} and references therein.} \cite{Cicoli:2008gp}.

Scale invariance of supergravity is also crucial for the more detailed consistency of the extra-dimensional versions of these models. It is crucial because for such models the desired result (the accelerated expansion of the 4D gravitational field) is the same size as (or smaller than) the physics that stabilises the size and shape of the extra dimensions~\cite{Kachru:2003sx} (making it inconsistent to neglect the physics of this stabilisation when model building). But scale invariance ensures that this stabilisation physics (typically involving the gravitational backreaction of any extra-dimensional sources, such as branes or fluxes, that consistent solutions very often require) is much simpler than it could have been in that scale invariance ensures the cancellation of most effects in the 4D curvature, leaving a result largely controlled by local physics near any source branes~\cite{BMvNNQ, GJZ}.

\subsubsection*{Mass hierarchies from no-scale}

Scale invariance and the potential's no-scale structure also plays a central role in the viability of the phenomenology of models based on supergravity compactifications, some examples of which are listed here.

The first requirement to trust the low-energy limit of any 4D string compactification is that the energy scale associated to the scalar potential has to be lower than the string scale $M_s \sim 1/\sqrt{\alpha'}$ and the Kaluza-Klein scale $M_\KK \sim M_s/\vo^{1/6}$, where $\vo$ denotes the dimensionless Calabi-Yau volume in string units and direct dimensional reduction relates the 4D Planck mass $M_p$ to the string scale as $M_p \sim M_s\,\sqrt{\vo}$. However, at tree-level the K\"ahler potential turns out to be $K = -2\ln \vo$ (focusing on the type IIB case) and the vast majority of string vacua in the landscape feature a flux-generated superpotential of order $|W_0| \sim \mathcal{O}(10-100)$~\cite{Denef:2004ze, MartinezPedrera:2012rs, Cicoli:2013cha}. Therefore 
the scale of the tree-level scalar potential (\ref{VFdef}) would in general be larger than the string scale since (restoring Planck units):
\be
V_\ssF \sim e^K |W_0|^2\, M_p^4 \sim |W_0|^2 \,M_s^4 \gtrsim M_s^4\,.
\label{Vscaling}
\ee
However the no-scale relation (\ref{KKKnoscale}), where the sum is over the K\"ahler moduli $T^i = \chi_i+{\rm i}\,\theta_i$, guarantees that the EFT is still under controls since it forces the scalar potential to vanish (and so the coefficient in (\ref{Vscaling}) is zero) after the dilaton $S$ and the complex structure moduli $U^\alpha$ have been stabilised supersymmetrically at $D_S W_0 = D_U W_0 = 0$. In fact, the moduli space for $S$, $T$ and $U$-moduli factorises at tree-level and $W_0$ does not depend on the $T$-moduli since the axions $\theta_i$ enjoy a shift symmetry which is exact in perturbation theory and $W$ has to be a holomorphic function of the moduli. Hence:
\be
V_\ssF  = e^K \left[|D_S W_0|^2 + |D_U W_0|^2 + \left(K^{i\bar\jmath} \, K_i K_{\bar\jmath}-3\right)|W_0|^2\right] = 0\,.
\label{Vcancel}
\ee

The no-scale structure is also crucial to induce a generic hierarchy in the moduli mass spectrum which can have several important phenomenological applications. This can be seen from noticing that the supersymmetric stabilisation at tree-level generates masses for the $S$ and $U$-moduli of order the gravitino mass $m_{3/2}=e^{K/2} |W|$, while the K\"ahler moduli tend to be generically lighter. In fact, as can be seen from (\ref{Vcancel}), the $T$-moduli are flat at tree-level, and so they can be lifted by either perturbative corrections to the K\"ahler potential $K = -2\ln\vo + K_{\rm p}$, or non-perturbative corrections to the superpotential $W = W_0 + W_{\rm np}$. Unless some tuning of the underlying parameters is performed, the $\chi_i$'s are fixed by $K_{\rm p}$, involving an interplay between $\alpha'$ and $g_s$ corrections to $K$, since at weak coupling perturbative physics dominates over non-perturbative terms.\footnote{On the other hand, the $\theta_i$'s are stabilised by $W_{\rm np}$ given that they cannot appear in $K_{\rm p}$ due to the axionic shift symmetry. Hence these axions in general turn out to be very light~\cite{Cicoli:2012sz}.} Taking into account the need to go to canonically normalised states and the fact that the inverse K\"ahler metric in general scales as $K^{i\bar\imath} \sim \chi_i^2$~\cite{Cicoli:2018tcq}, the mass of the $\chi$-moduli is expected to scale as:
\be 
\label{VFnoscale}
m^2_{\chi_i} \sim K^{i\bar\imath}\, \frac{\partial^2 V_\ssF}{\partial \chi_i^2} \sim K^{i\bar\imath}\, \frac{V_\ssF}{\chi_i^2} \sim e^K |W_0|^2 \left( K^{i\bar\jmath} \, K_i K_{\bar\jmath} - 3  \right) = m_{3/2}^2 \left( K^{i\bar\jmath} \, K_i K_{\bar\jmath} - 3  \right).
\ee
This shows that the moduli tend to get a mass of order $m_{3/2}$, unless (\ref{KKKnoscale}) is satisfied by the leading order K\"ahler potential. When this is so, moduli masses become additionally suppressed by no-scale breaking effects generated by $K_{\rm p}$ which are small since $K_{\rm p}$ is an expansion in inverse powers of the K\"ahler moduli~\cite{vonGersdorff:2005bf,LVLoops, LVLoops1, BHK, BBHL,CLW} which have to be all larger than unity to be able to neglect stringy effects, i.e. $\chi_i\gg 1$ $\forall\,i$. Hence we obtain (see also~\cite{Burgess:2010sy} for a discussion of how the no-scale relation protects moduli masses below $m_{3/2}$ from quantum effects):
\be
K^{i\bar\jmath} \, K_i K_{\bar\jmath} - 3 \simeq  K_{\rm p} \sim \epsilon \ll 1 \qquad \Rightarrow \qquad m_{\chi_i}^2 \simeq \epsilon\, m_{3/2}^2 \ll m_{3/2}^2\,,
\label{no-scaleBreak}
\ee
where for example $\epsilon \sim \vo^{-1}$ for the dominant $\alpha'$ correction~\cite{BBHL}, $\epsilon \sim \vo^{-4/3}$ for the leading string loop effects~\cite{vonGersdorff:2005bf,LVLoops, LVLoops1, BHK} and $\epsilon \sim \vo^{-5/3}$ for higher order $\alpha'$ terms~\cite{CLW} (where for simplicity we consider an isotropic limit where all $\chi$-fields are of the same order of magnitude, i.e. $\chi_i \sim \vo^{2/3}$ $\forall\,i$). Explicit examples where the mass spectrum of the K\"ahler moduli has been shown to take this generic behaviour are~\cite{LV, LVLoops1, Berg:2005yu, CLW,AbdusSalam:2020ywo} for a small number of $T$-fields, and~\cite{Cicoli:2016chb, Cicoli:2014sva} for an arbitrarily large number of K\"ahler moduli.

Two non-generic counter-examples where the mass of the K\"ahler moduli is instead of order $m_{3/2}$, are blow-up modes in LVS models and KKLT moduli. In fact, as pointed out in~\cite{Cicoli:2018tcq, Cicoli:2011it}, blow-up modes $\chi_{\rm bu}$ are exceptional since they correspond to resolutions of point-like singularities, and so their inverse K\"ahler metric scales as $K^{i\bar\imath}\sim \vo \sqrt{\chi_{\rm bu}}\sim 1/\epsilon$. On the other hand, in KKLT models, $K_{\rm p}$ is negligible since $W_0$ is tuned exponentially small, $W_0 \sim W_{\rm np}$. In this scenario, the scalar potential therefore scales as $V_\ssF \sim e^K\,W_0 \,W_{\rm np} \sim e^K\,W_0^2 \sim m_{3/2}^2$, inducing moduli masses of order $m_{3/2}$.

\subsubsection*{Absence of phenomenological problems}

If some moduli in the low-energy spectrum of realistic compactifications have masses near the weak scale, it is known that these potentially introduce a number of cosmological problems~\cite{Coughlan:1983ci, Banks:1993en,deCarlos:1993wie} that generally require the moduli $\chi^i$ to be rather heavy: $m_{\chi^i} \gtrsim \mathcal{O}(50)$ TeV. Having moduli this heavy can also cause other problems, however, such as by reintroducing the cosmological gravitino problem~\cite{Endo:2006zj, Nakamura:2006uc}, that arises if the decay of $\chi^i$ into gravitinos is kinematically allowed. However, as we have seen above, the no-scale structure of the low-energy EFT is crucial to avoid the gravitino problem since it makes the moduli naturally lighter than the gravitino, and so forbids their decay into gravitinos.

A mass hierarchy $m_{3/2} > m_\chi > \mathcal{O}(50)$ TeV raises still other problems because in supergravity models soft-breaking terms tend to be of order the gravitino mass, and so $M_{\rm soft}\sim m_{3/2} \gg \mathcal{O}(50)$ TeV makes supersymmetry largely irrelevant to TeV-scale hierarchy issues\footnote{Which at present perhaps should be regarded as a successful prediction.} (and tends to give rise to neutralino dark matter overproduction). This can also be ameliorated by near no-scale structure~\cite{BCKMQ, Aparicio:2014wxa, Cicoli:2012sz, Reece:2015qbf}, with soft scalar masses often given by expressions like
\be \label{m0vsm32}
  m_0^2 = m_{3/2}^2\left(1-\frac13 \underbrace{K^{\ssT\bar{\ssT}}K_{\bar{\ssT}}K_\ssT}_{=3+\epsilon}\right)\simeq \epsilon \,m_{3/2}^2\,,
\ee
which, as in (\ref{no-scaleBreak}), is dominated by subleading no-scale-breaking effects.

\subsubsection*{Sequestering} 

Extra-dimensional models hold out the hope of reducing unwanted consequences of supersymmetry breaking by allowing this to be `sequestered' from Standard Model fields by displacing it at a distance within the extra dimensions~\cite{Randall:1998uk}. The way sequestering appears in the low-energy 4D EFT is by having sequestered sectors `$A$' and `$B$' appear additively in supergravity K\"ahler function: $e^{-K/3} = e^{-K_\ssA/3} + e^{-K_\ssB/3}$. But this type of sequestering has proven to be hard to realise, partly due to the effects of bulk gravitational auxiliary fields~\cite{Anisimov:2001zz, Anisimov:2002az}. 

Those sequestering effects that are found in extra-dimensional models partly rely for their existence on no-scale properties of the low-energy supergravity~\cite{Jockers:2004yj, Jockers:2005zy, Grimm:2004uq, Grimm:2005fa, Kachru:2003sx}. We note in passing that this is partly because the no-scale condition directly imposes conditions on the quantity $e^{-K/3}$, but also partly because no-scale models constrain whether some supersymmetry-breaking auxiliary fields in the gravitational sector -- like that of the compensator for instance -- can acquire {\it v.e.v.}s. These sequestering properties have been exploited in attempts to suppress scalar soft-breaking masses (leading {\it e.g.}~to \pref{m0vsm32}) and attempts to realise the QCD axion as part of a closed-string modulus $T_i$~\cite{Cicoli:2012sz}, where they can help allow much lower axion decay constants, $f_a \ll M_{\KK}, M_s$,  than are usually found~\cite{Cicoli:2013cha}.

\subsubsection*{This paper: a road map}

We marshal our arguments as follows. The next section, \S\ref{sec:noscaleprops}, starts by summarising the definitions of no-scale and generalised no-scale models that are commonly used. The main focus is to extend the above arguments to include more than a single scaling field and to generalise the concept of no-scale beyond scale invariance. This section closes by describing explicit examples of concrete generalised no-scale models that do not fall into the standard no-scale category. We provide several levels of generalisation of no-scale which are summarised in Fig.~\ref{fig:Summary} to which the impatient reader is encouraged to go. 

This is followed in \S\ref{sec:descent2}, \S\ref{sec:descent} and \S\ref{sec:IIA} by a variety of examples from the heterotic, type IIB and type IIA string respectively where the relevant scale invariances arise through dimensional reduction of a higher-dimensional supergravity. In each case we identify the different scaling symmetries in 10D and compactified ${\mathcal{N}}=1$ supersymmetric 4D EFT, including both bulk and localised sources such as DBI and WZ D-brane actions. We then obtain the full information of the tree-level 4D action by fixing the K\"ahler potential, superpotential and gauge kinetic function extending the previous treatments~\cite{Witten:1985xb, Burgess:1985zz, Nilles:1986cy, Witten:1985bz} and comparing with explicit calculations. These sections also quantify the breaking of scale invariance arising from loop corrections, to extract how loops and $\alpha'$ corrections depend on the scaling fields in the low-energy 4D effective theory. We test our general expressions for these perturbative corrections by contrasting them with the perturbative corrections that have been computed so far and with the known non-renormalisation theorems. We close in \S\ref{sec:openquestions} with a summary of results and description of open directions.

We complement our presentation with a series of self-contained appendices that cover details of some of the material not fully covered in the main text. Appendix ~\ref{4D_no_scale_app} exploits linear multiplets to write in a simple way a generalisation of the standard no-scale which, which expressed in terms of chiral multiplets, is given by Eq.~\pref{noscalechiralcomplicated3_app}. Appendix ~\ref{app:accidentalshift} describes in detail the generalised accidental shift symmetries of the underlying heterotic and type IIB string theories that are behind the fact that tree-level K\"ahler potentials depend on particular combinations of the moduli and matter fields. Appendix ~\ref{app:10Dscaling} expands the discussion of the scaling behaviour of different contributions to the type IIB tree-level action including Wess-Zumino terms and non-Abelian D-brane actions, and examines a pitfall that can complicate identifying the scaling behaviour of a magnetically coupled brane.

\section{Categories of no-scale supergravity} 
\label{sec:noscaleprops}

\begin{figure}[t]
\centering
\includegraphics[width=0.9\textwidth]{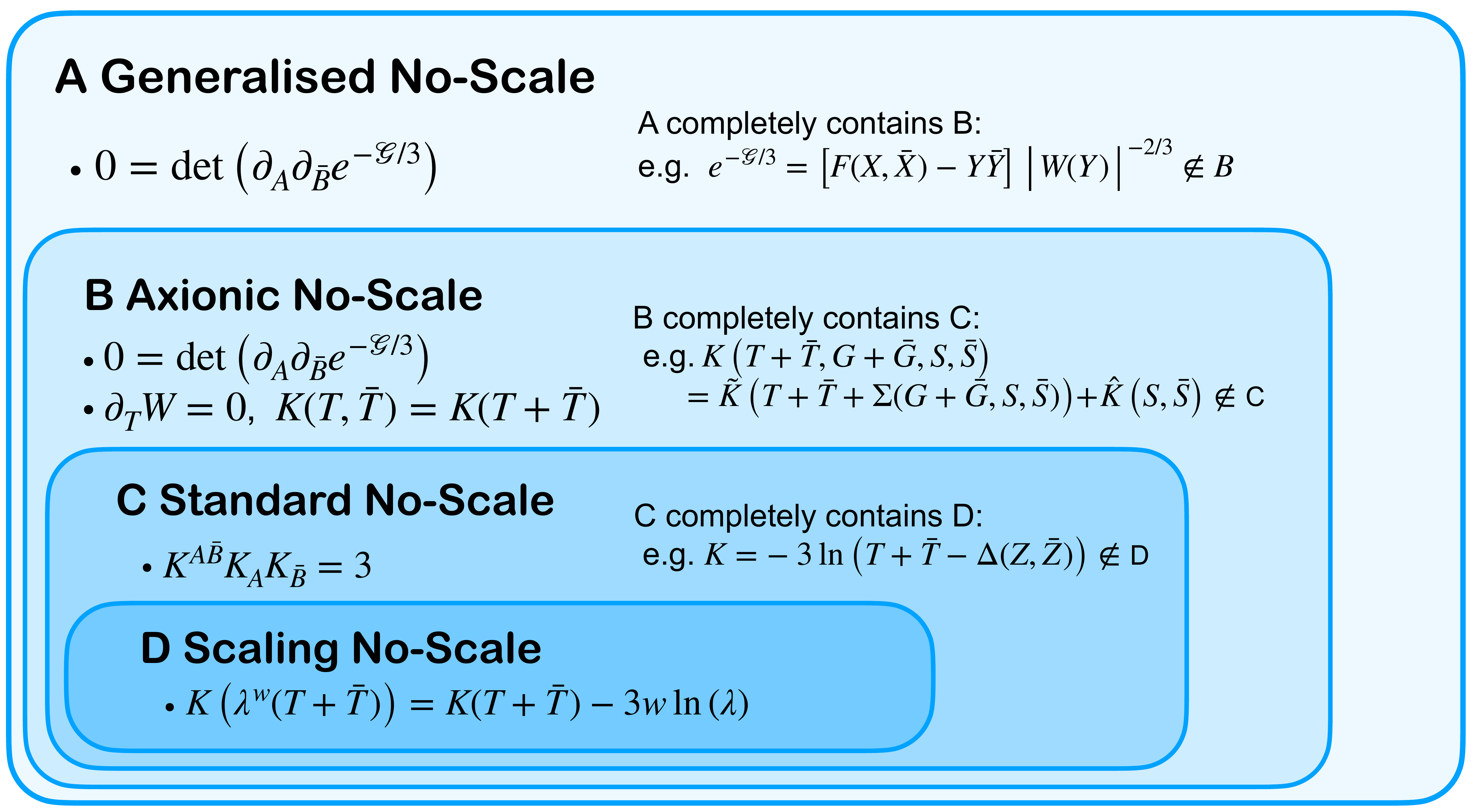}
\caption{\label{fig:Summary} Summary of the nested classes of no-scale models described in the text. The precise definitions are given in \S~\ref{sec:definitions}. The sample models highlighting the relationship between these categories are discussed in \S~\ref{sec:Relations}.}
\end{figure}

Since we argue that no-scale models play a major role in expressing how high-energy scaling symmetries get expressed at low energies, we start by clarifying the relationship between several different notions of `no-scale' that exist in the literature. We place these into four nested categories:
\be
  \hbox{(Scaling no-scale)} \subset \hbox{(Standard no-scale)} \subset \hbox{(Axionic no-scale)} \subset \hbox{(Generalised no-scale)} \,.
\ee
We here define these categories and show how they are related, presenting along the way characteristic examples for each. A pictorial representation of this categorisation is given in Fig.~\ref{fig:Summary}. All of these categories share an attractive defining feature: their classical scalar potential is non-negative, $V \ge 0$, often with $V$ vanishing (despite the spontaneous breaking of supersymmetry) over a many-parameter set of field configurations. 

\subsection{Definitions}
\label{sec:definitions}

This section defines four categories of no-scale models that naturally arise,\footnote{We concentrate on $F$-term scalar potentials. Another generalisation of no-scale models includes the possibility of cancellations between $F$- and $D$-terms~\cite{DallAgata:2013jtw} that we do not consider here.} starting from the most restrictive models and working our way out to the most general case.

\subsubsection*{Scaling no-scale models}

We define `scaling no-scale' models as those that satisfy:

\begin{Def}
A supergravity theory is a {\rm `scaling no-scale' model} if there is a subset, $\{ T^i \}$, of its $n$ chiral multiplets, $\{ z^\ssA \} = \{ T^i , Z^a \}$, for which the following four conditions hold simultaneously for a non-trivial range of $T^i$: 

\medskip
$(i)$ $D_a W = 0$ for all $Z^a$; 

$(ii)$ $K$ depends only on the real part of the $T^i$; 

$(iii)$ the $T^i$ do not appear in the (non-zero) superpotential $W$; 

$(iv)$ the quantity $e^{-K/3}$ is homogeneous degree-one in the $T^i$, that is
\be
\label{ScalNoScaleKcond0} 
e^{-K/3} \to \lambda^{w_\ssT} \, e^{-K/3}  \quad \hbox{whenever} \quad T^i \to \lambda^{w_\ssT} \, T^i \,.
\ee
\end{Def}
\noindent
Notice that conditions ($ii$) and ($iii$) can be arranged if shifts of the imaginary parts of $T^i$ are symmetries. This definition is motivated by the fact that (\ref{ScalNoScaleKcond0}) can be ensured by a system's classical scale invariances, and because \pref{ScalNoScaleKcond0} implies the standard no-scale condition, \pref{KKKnoscale} (see \S\ref{sec:scalesufficient} below).

\subsubsection*{Standard no-scale models}

Standard (or traditional) no-scale models~\cite{Cremmer:1983bf, Barbieri:1982ac, Chang:1983hk, Ellis:1983sf} are defined by:\footnote{Recently models have been called `no-scale' when they satisfy the weaker condition that $K^{i \bar\jmath} K_i K_{\bar\jmath}$ is a constant, not necessarily equal to 3~\cite{Ciupke:2015ora}. We do not use this nomenclature here.} 

\begin{Def}
\label{def2}
A supergravity theory is a {\rm `standard no-scale' model} if there is a subset, $\{ T^i \}$, of its $n$ chiral multiplets, $\{ z^\ssA \} = \{ T^i , Z^a \}$, for which the following three conditions hold simultaneously for a non-trivial range of $T^i$: 

\medskip
$(i)$ $D_a W = 0$ for all $Z^a$; 

$(ii)$ the $T^i$ do not appear in the (non-zero) superpotential $W$;

$(iii)$ the K\"ahler potential satisfies the identity \pref{KKKnoscale}: $K^{i \bar\jmath} K_i K_{\bar\jmath} = 3$.
\end{Def}

\noindent
This definition is motivated by the form of the $\cN=1$ supergravity $F$-term scalar potential (\ref{VFdef}). It can be shown that $D_a W = 0$ always solves the $Z^a$ field equations, and so using this to eliminate the $Z^a$ from the scalar potential leads to the remaining terms
\be
\left. V_\ssF \right|_{D_aW = 0} := e^K |W|^2 \left( K^{i\bar\jmath} K_i K_{\bar\jmath} - 3 \right)  \,,
\ee
which vanishes whenever \pref{KKKnoscale} is satisfied.

\subsubsection*{Generalised and axionic no-scale models}

The broadest category on our list is the most general one that has a non-negative classical $F$-term potential minimised along a flat direction at $V_\ssF = 0$~\cite{Barbieri:1985wq}. We therefore define `generalised no-scale' models as those that satisfy:\footnote{Notice that the terminology `generalised no-scale' has also been used with a different meaning in~\cite{Blaback:2015zra} where the authors considered a superpotential which in general depends on the $T$-fields but effectively enjoys a shift symmetry in the $T^i$ when the $Z$-moduli are fixed supersymmetrically.} 

\begin{Def}
A 4D $\cN=1$ supergravity is called {\rm `generalised no-scale'}  when the matrix
\be 
\label{MmatDef}
M_{\ssA \bar \ssB} := \partial_\ssA \partial_{\bar \ssB} \Bigl( e^{-\cG/3} \Bigr) 
\ee
has a zero eigenvalue, where $\cG$ is the K\"ahler-invariant potential defined in (\ref{G}). A vanishing eigenvalue equivalently implies $\det M = 0$. 
\end{Def}
\noindent
{\it Axionic no-scale models} are the special case of these generalised no-scale models for which the imaginary parts of the moduli also enjoy a shift symmetry that keeps the superpotential from depending on $T^i$ and keeps the K\"ahler potential from depending on ${\rm Im}(T^i)$. 

Condition \pref{MmatDef} is shown in~\cite{Barbieri:1985wq} to be both necessary and sufficient for having a classical $F$-term potential that satisfies $V_\ssF \ge 0$, with non-trivial (possibly supersymmetry breaking) configurations along which it is minimised at $V_\ssF = 0$. We pause briefly to review their arguments, whose main point is to express the scalar potential in terms of the function
\be
\Omega:=e^{-{\mathcal G}/3}
\ee
and its derivatives $\Omega_\ssA := \partial_\ssA\Omega$ and $M_{\ssA{\bar \ssB}} := \Omega_{\ssA{\bar \ssB}} := \partial_{\ssA}\partial_{\bar \ssB}\Omega$. These definitions imply
\be \label{GmatvsM}
{\mathcal{G}}_{\ssA{\bar \ssB}}=-\frac{3}{\Omega^2}\left(\Omega M_{\ssA{\bar \ssB}}-\Omega_\ssA \Omega_{\bar \ssB}\right)
\ee
whose inverse matrix, $\cG^{-1}$, has components
\be \label{GinvComp}
{\mathcal{G}}^{{\bar \ssB}\ssA}= -\frac{\Omega}{3}\left( M^{{\bar \ssB}\ssA} + \frac{ M^{{\bar \ssC}\ssA}\, \Omega_{\bar \ssC}\,  \Omega_\ssD \,  M^{{\bar \ssB}\ssD}}{\Omega -\Gamma }\right)
\ee
where $M^{{\bar \ssB}\ssA} M_{\ssC{\bar \ssB}} = \delta^\ssA_\ssC$ and $\Gamma := M^{{\bar \ssA}\ssB} \Omega_{{\bar \ssA}} \Omega_\ssB$. Since $\Omega > 0$ and since positivity of the kinetic terms implies $\mG_{\ssA\bar\ssB}$ is positive definite, (\ref{GmatvsM}) implies $M$ can have precisely one non-negative eigenvalue while all others must be negative.

Assembling these expressions, the $F$-term scalar potential is then written as
\be \label{VFBarb}
V_\ssF =e^{\mathcal G}  \left(\mG^{\bar \ssB \ssA } \,\mG_{\bar \ssB} \mG_\ssA -3\right)= \frac{3}{\Omega^2(\Gamma-\Omega)} = -\frac{3}{\Omega^3}\det\left(-\frac{3}{\Omega}\, \mG^{-1}\, M\right)
\ee
where the second equality uses \pref{GinvComp} and the last equality evaluates the determinant using $\det(\delta_{a}^b+k_ak^b)= 1+k_ak^a$. Since $M$ can have at most one non-negative eigenvalue, it is the sign (or vanishing) of this eigenvalue that controls the sign of $V_\ssF$.

This last expression reveals the scalar $F$-term potential to be non-negative if $\Gamma>\Omega$ --- or if $\det(-M)\leq 0$ (given that positive kinetic terms require $\mG_{\ssA \bar \ssB}$ and $\mG^{{\bar \ssB}\ssA}$ to be positive definite). But positivity of the kinetic terms also ensures 
\be \label{GGGpos}
   \mG^{\bar\ssA \ssB} \mG_{\bar\ssA} \mG_\ssB =  \frac{3\Gamma}{\Gamma-\Omega}  
\ee
is strictly positive and so it must always be true that either $\Gamma < 0$ (in which case $V_\ssF$ is negative) or $\Gamma > \Omega > 0$ (for which $V_\ssF$ is positive).\footnote{Eq.~\pref{VFBarb} with positive $V_\ssF$ is particularly interesting in view of the observational evidence for positive vacuum energies (both at present and in the distant past). Such potentials may help understand the dissonance between this and difficulties obtaining de Sitter space in standard supergravity theories.} Eq.~\pref{VFBarb} can only vanish without \pref{GGGpos} also vanishing if $\Gamma \to \pm\infty$, which indeed is the limit approached when an eigenvalue of $M_{\ssA\bar\ssB}$ tends to zero. 

\subsection{No-scale without no-scale}
\label{sec:Relations}

This section explores the motivations for the above definitions and identifies how they are related to one another. We illustrate how each category is a proper subset of its parent by providing concrete examples -- those listed in Fig.~\ref{fig:Summary} -- for each category that is not also an element of the next-smaller category.

\subsubsection{Scaling no-scale $\subset$ Standard no-scale}
\label{sssec:standardnoscale}

The fact that scaling no-scale models must also be standard no-scale (as in Definition 2), follows by direct calculation~\cite{Burgess:2008ir}. The starting point is the observation that Definition 1 requires $K$ to depend only on the real parts, $\chi^i := T^i + \ov T^i$, and to satisfy the property \pref{ScalNoScaleKcond0}, rewritten here as
\be 
\label{homoKcondold0}
K(\lambda^{w_\ssT} \chi) = K(\chi) - 3 w_\ssT \ln \lambda \,.
\ee
This holds as an identity for all $\chi^i$ and $\lambda$. As mentioned earlier, (\ref{homoKcondold0}) is suggestive of a scale invariance, as is explored in more detail below.

The sufficiency of \pref{homoKcondold0} for \pref{KKKnoscale} is shown by direct differentiation. In detail, taking the first and second derivatives of \pref{homoKcondold0} with respect to $\lambda$ gives (after taking $\lambda \to 1$) the following two conditions:
\be 
\label{valone0}
w_\ssT \chi^i K_i = -3 w_\ssT \qquad \hbox{and} \qquad w_\ssT^2 \chi^i \chi^j K_{ij} + w_\ssT (w_\ssT -1) \chi^i K_i = 3w_\ssT  \,.
\ee
Using the first of these in the second then implies (when $w_\ssT \neq 0$)
\be 
\label{interim0}
\chi^i \chi^j K_{ij} =  3 \,.
\ee 
Finally, differentiating the first of \pref{valone0} with respect to $\ov T^j$ gives $K_{\bar\jmath} + \chi^i K_{i\bar\jmath} = 0$, which when used to eliminate $\chi^i$ converts \pref{interim0} into \pref{KKKnoscale}.

To prove that not all standard no-scale models are of the scaling no-scale type consider the example
\be 
K = - 3 \ln(T + \ov{T} - \Delta) \,,
\label{KTDelta}
\ee
with chiral fields $\{ z^\ssA \} = \{ T, Z^a \}$, where $\Delta = \Delta(Z,\ov Z)$ is arbitrary (but does not depend on $T$) and the superpotential is a constant: $W = W_0$. This is not of the scaling no-scale type because the presence of $\Delta$ means 
\be
e^{-K/3} = T + \ov{T} - \Delta
\ee
need not be homogeneous degree-one under rescalings of $z^\ssA$ ({\it e.g.}~when $\Delta$ is constant). 

For this model $\Omega = e^{-\mG/3} = e^{-K/3} |W|^{-2/3}$ and so because $W$ is field-independent the eigenvalues of $M_{\ssA\bar \ssB}$ are proportional to those of the matrix 
\be \label{NmatDef}
N_{\ssA \bar \ssB} := \partial_\ssA \partial_{\bar \ssB} \Bigl( e^{-K/3} \Bigr) 
= \left( \begin{array}{ccc}
  0  && 0  \\  0 &&  -\Delta_{a \bar b}
\end{array}
\right) \,,
\ee
which has an obvious zero eigenvalue for any value of the $z^\ssA$. Consequently the potential $V_\ssF$ must vanish identically. The same conclusion also follows from direct differentiation. The first derivatives are
\be
K_\ssT = K_{\bar \ssT} =  - 3\,e^{K/3} \,, \quad  K_a =  3\Delta_a\,e^{K/3} \quad \hbox{and} \quad K_{\bar a} =   3\Delta_{\bar a}\,e^{K/3} \,,
\ee
and so the matrix of second derivatives and its inverse are
\be
   K_{\ssA \bar \ssB} = 3\,e^{2K/3}  \left( \begin{array}{ccc}
  1  && - \Delta_{\bar b} \\ - \Delta_a &&  \Delta_{a \bar b}e^{-K/3} + \Delta_a \Delta_{\bar b}
\end{array}
\right) \quad \hbox{and} \quad 
   K^{\bar \ssB \ssC} = \frac{e^{-K/3}}{3}  \left( \begin{array}{ccc}
  e^{-K/3} + \mathfrak{n}  &&  \Delta^{c} \\   \Delta^{\bar b} &&  \Delta^{\bar b c}  
  \end{array}
\right) \,,
\ee
where $\mathfrak{n} := \Delta^{\bar a b} \Delta_{\bar a} \Delta_b$, $\Delta^a := \Delta^{\bar b a} \Delta_{\bar b}$ and $\Delta^{\bar a} := \Delta^{\bar a b} \Delta_b$ with $\Delta^{\bar a b}$ defined as the inverse matrix: $\Delta_{a \bar b} \Delta^{\bar b c} = \delta^c_a$. These expressions imply
\be
  K^{\bar \ssA \ssB} K_{\bar \ssA} K_\ssB = 3\,e^{K/3}  \left( \begin{array}{c}
  -1  \\  \Delta_{\bar b}  \end{array} \right)^\ssT  \left( \begin{array}{ccc}
  e^{-K/3} + \mathfrak{n}  &  \Delta^{c} \\   \Delta^{\bar b} &  \Delta^{\bar b c}  
  \end{array}
\right)  \left( \begin{array}{c}
  -1 \\ \Delta_{c}  \end{array} \right) = 3 \,,
\ee
showing that the model is of standard no-scale type. 

Having established that $V_\ssF$ vanishes for {\it all} fields whenever $W$ is independent of both $T$ and the $Z^a$, one can also ask how these flat directions are lifted if $W=W(Z)$ with $W_\ssT$ remaining zero. In this case the $F$-term potential evaluates to
\be
   V_\ssF = e^K K^{\bar{a} b}  \ol{W}_{\bar a} W_b = \frac13 \, e^{2K/3} \Delta^{\bar{a} b}  \ol{W}_{\bar a} W_b  \,,
\ee
which is strictly non-negative whenever $\Delta^{\bar a b}$ is positive definite. This potential vanishes for all $T$ once the $Z^a$ are minimised by solving the $T$-independent conditions $W_a = 0$. Notice that $D_a W$ in general need {\it not} vanish at this solution,\footnote{If solutions to $D_\ssA W = 0$ exist, they are always stationary points of the potential.} indicating that the auxiliary fields for both $T$ and $Z^a$ break supersymmetry along this flat direction. The existence of this flat direction is also consistent with the K\"ahler potential \pref{KTDelta}, now regarded as being a function only of $T$ with $Z^a = Z^a_0$ fixed to the solution $W_a(Z_0)=0$, since this is also of the standard no-scale type.

Variations of this example include some cases of practical later interest, such as
\be 
e^{-\cG/3} = \left[ \cA (T+ \ov T) - \Delta(Z^a, \ov Z^a) \right]\,| h(Z^a) |\, \Omega(U^\alpha,\ov U^\alpha)
\label{sequester2}
\ee
with $\cA(T+ \ov T)$ a homogeneous degree-one function of its argument, $h$ a holomorphic function of $Z^a$ and $\Delta(Z, \ov Z)$ and $\Omega(U,\ov U)$ real functions of a different collection of scalar fields. When $h$ and $\Omega$ are constants, this reduces to the example given in (\ref{KTDelta}) and so is no-scale with flat directions parameterised by both $T$ and $Z^a$. More generally, even in the presence of non-trivial functions $h$ and $\Omega$, this is a case for which the scalar potential is guaranteed to be non-negative --- because $\det \partial_\ssA \partial_{\bar\ssB}(-e^{-\cG/3} )\leq 0$~\cite{Barbieri:1985wq}  --- and is minimised for configurations along which $V_\ssF = 0$ corresponding to a flat direction in the $T$-field direction. 

As shown in more detail below, this situation is typically realised in string compactifications, with the fields $Z^a$ corresponding to matter or D-brane position moduli while the $U^\alpha$ represent complex-structure moduli and $T$ is the K\"ahler modulus corresponding to the overall breathing mode. As we shall see later, when the fields $Z^a$ are D3-brane moduli, the function $\Delta(Z,\ov Z)$ is given by the K\"ahler potential for the extra-dimensional geometry through which the D3-branes move.

\subsubsection{Standard no-scale $\subset$ Axionic no-scale}

We next show that standard no-scale models form a proper subset of the class of both axionic and generalised no-scale models. There is no mystery that standard no-scale models are a subset of axionic no-scale models because standard no-scale models assume an axionic symmetry and because generalised no-scale models are the most general ones with flat directions in $V_\ssF$; showing them to be a proper subset is the interesting thing. We do so here by demonstrating that standard no-scale models satisfying \pref{KKKnoscale} do not exhaust the class of axionic no-scale models. 

In detail, standard no-scale is sufficient for generalised no-scale because the constancy and non-vanishing of $W = W_0$ ensures the matrices $M$ and $N$ defined in \pref{MmatDef} and \pref{NmatDef} are related by $M_{i \bar \jmath} = N_{i \bar \jmath}\, |W_0|^{-2/3}$, and so $M$ and $N$ share any zero eigenvectors. But direct differentiation shows that the matrix $N_{i \bar \jmath}$ has a zero eigenvalue if and only if the no-scale condition \pref{KKKnoscale} is satisfied~\cite{Barbieri:1985wq}, and so the standard no-scale Definition 2 always implies the generalised no-scale property of Definition 3.

We now provide a class of examples that show that axionic no-scale models need not also be standard no-scale (in the sense that they do not satisfy \pref{KKKnoscale}). To this end we follow~\cite{Ciupke:2015ora} and focus on examples where the flat direction fields, $T^i$, and the other fields, $Z^a$, do not have a simple product-manifold structure ({\it i.e.}~when the kinetic terms mix these two kinds of fields). As argued in Appendix ~\ref{4D_no_scale_app}, this broader class of examples is more easily built starting from a dualised representation for which each axion, $a$, is traded for a 2-form gauge potential, $B_{\mu\nu}$, with field strength $H = \exd B$, according to $H_{\mu\nu\lambda} \propto \epsilon_{\mu\nu\lambda\rho} \partial^\rho a$. For supersymmetric theories this involves trading ordinary chiral supermultiplets (containing $a$) for linear supermultiplets (containing $B$)~\cite{LinearMult, NewMinimal, Cecotti:1987nw, Ovrut:1988fd, Burgess:1995kp}. The point of doing so is to use this construction to find examples of flat supersymmetry breaking potentials for which property \pref{KKKnoscale} does {\it not} hold. 

To this end consider an $\mathcal{N}=1, D=4$ supergravity with three different types of chiral multiplets: $\{ T^i , G^{\hat{\imath}} , S^a \}$, where $i=1,\dots,n \ , \hat{\imath}=1,\dots,m$ and $a=1, \dots, p$. Of these, the fields $\{T^i, G^{\hat{\imath}}\}$ are moduli with axionic shift symmetries 
\be
T^i \rightarrow  T^i + i \lambda^i \ , \qquad G^{\hat{\imath}} \rightarrow G^{\hat{\imath}} + i \lambda^{\hat{\imath}}
\ee
that ensure $K = K(T+\ol T, G+\ol G)$ and that these fields do not appear in the superpotential: $W = W(S^a)$. We further ask the K\"ahler potential for these fields to have the `coordinate-degenerate' form
\be \label{coordinatedegeneracy}
 K(T + \ol{T}, G + \ol{G}, S,\ol{S}) = \cK\Bigl[ T^i + \ol{T}^{\,i} + \Sigma^i(G + \ol{G},S^a,\ol{S}^{\,a}) \Bigl] + \hat{K}(S^a,\ol{S}^{\,a}) \,,
\ee
where $\Sigma^i$ are real-valued functions of $G^{\hat{\imath}} + \ol{G}^{\,\hat{\imath}}$, $S^a$ and $\ol{S}^{\,a}$.  

In general, these assumptions allow kinetic mixing between the fields $S^a$, $T^i$ and $G^{\hat{\imath}}$, and it is this mixing that complicates identifying the correct no-scale description. This is simpler to see within the dual formulation that trades the chiral multiplets $T^i$ and $G^{\hat{\imath}}$ for linear multiplets because the dualisation disentangles the chiral multiplets $S^a$ from the shift-symmetric directions $T^i$ and $G^{\hat{\imath}}$, greatly simplifying the scalar potential (see Appendix ~\ref{4D_no_scale_app} for details).  

In particular, a sufficient condition for the scalar potential of this model to be positive semi-definite is
\be 
K^{\ssI \bar\ssJ} K_{\bar\ssJ} K_{\ssK} {\mathcal{M}^{\ssK}}_{\ssI} = 3\,,  
\label{noscalechiralcomplicated3}
\ee
where $\Psi^\ssI = \{T^i, G^{\hat{\imath}}\}$ collectively denotes all of the axionic multiplets, and the matrix $\cM$ is defined by
\be  
{(\mathcal{M}^{-1})^{\ssI}}_{\ssJ} = \delta_{\ssJ}^{\ssI} - K^{\ssI {\bar{a}}} K_{\ssJ {\bar{a}}}- K^{a \bar\ssI} K_{a \bar\ssJ} \ .
\label{noscalechiralcomplicated2}
\ee
The potential has flat directions with $V_\ssF = 0$ once $S^a$ is minimised at a supersymmetric configuration using $D_a W = 0$. Eq.~\pref{noscalechiralcomplicated2} generalises the standard no-scale condition \pref{KKKnoscale}, reducing to it in the special case where $\Sigma^i$ is independent of $S$, so that the $S^a$ do not mix with the fields $\Psi^\ssI$ (in which case $K_{a\bar\ssJ} = K_{\ssJ \bar a} = 0$ and so ${\cM^\ssI}_\ssJ = \delta^\ssI_\ssJ$).

\subsubsection{Axionic no-scale $\subset$ Generalised no-scale}
\label{sssec:NoScaleToy}

Finally, we provide examples that show that generalised no-scale models need not require axionic shift symmetries, and that flat directions can exist for which \pref{KKKnoscale} does not hold when restricted just to the flat direction fields in themselves. Among other things this shows that standard no-scale and axionic no-scale models form a proper subset of generalised no-scale models. 

To this end consider the following concrete class of models involving two chiral-scalar supermultiplets, $X$ and $Y$. Define
\be \label{OmegaToy}
\Omega= e^{-{\mathcal G}/3}=\Bigl[ F(X,\ol{X})-Y \ol{Y}\Bigr] \, \left |W(Y)\right|^{-2/3} \,,
\ee
with the function $F$ chosen to satisfy $F > Y \ol Y$ so $\Omega > 0$, and $|F_\ssX|^2 > F F_{\ssX \bar\ssX}$ in order to guarantee positive kinetic terms, since 
\be
   \mG_{\ssA \bar \ssB} = \frac{3}{\cD^2}  \left( \begin{array}{ccc}
  -\cD F_{\ssX\bar\ssX}+ | F_\ssX |^2&& - F_\ssX Y \\ - F_{\bar\ssX} \ol{Y} &&  \cD + Y \ol{Y}
\end{array}
\right)  \,,
\ee
where $\cD := F - Y\ol{Y}$ and
\be
\det\left({\mathcal G_{\ssA{\bar \ssB}}}\right)= \frac{9}{(F-|Y|^2)^3}\, \left( |F_\ssX|^2-FF_{\ssX\bar \ssX}\right) \,.
\ee
The standard no-scale diagnostic for this theory evaluates to
\be \label{ToyNSCrit}
   K^{\bar \ssA \ssB} K_{\bar \ssA} K_\ssB = \frac{3(|F_\ssX|^2 - F_{\ssX\bar\ssX} |Y|^2)}{|F_\ssX|^2 - F F_{\ssX \bar\ssX}} \,.
\ee

The $F$-term scalar potential for this model then is
\be
V_\ssF  =  \frac{1}{3(F-|Y|^2)^2}\, \left[|W_\ssY|^2\, +\, \frac{F_{\ssX\bar\ssX}}{ |F_\ssX|^2-FF_{\ssX\bar\ssX}}\, |3W-YW_\ssY|^2 \right] \,.
\label{positive}
\ee
This is manifestly non-negative as long as the kinetic terms are positive and $F_{\ssX\bar\ssX}\geq 0$. When these conditions are satisfied the global minimum of the potential is found by setting $V_\ssF =0$, which fixes the values of $X$ and $Y$ so that each of the two terms in the square bracket of \pref{positive} vanishes. A variety of possibilities arise, depending on the choices of the functions $F$ and $W$:

\medskip\noindent
{\bf Case 1:} If $F=h(X)+\ol{ h( X)} $ and $W = W_0$ is constant, the potential vanishes for all $X$ and $Y$ (a special case of the standard no-scale cancellation, as is seen by using $F_{\ssX\bar\ssX} = 0$ in \pref{ToyNSCrit}). Supersymmetry is generically broken along these flat directions if $W_0 \neq 0$. Although for generic $h(X)$ it might appear that no symmetries survive in the $X$ sector (and in particular no axionic symmetries) this is mistaken because one may always perform a field redefinition from $X$ to $Z := h(X)$, after which a shift symmetry for $Z$ does exist.

\medskip\noindent
{\bf Case 2:} If $F=h(X)+\ol{ h(X)}$ and $W=W(Y)$ is non-trivial, then 
\be
V_\ssF  =  \frac{|W_\ssY|^2}{3(F-|Y|^2)^2} \,,
\label{positive2}
\ee
and so $Y = Y_0$ is fixed by solving $W_\ssY(Y_0) =0$ while the $X$ direction remains flat (again corresponding to a standard no-scale model). Supersymmetry is broken along this flat direction if $W(Y_0)$ does not vanish.

\medskip\noindent
{\bf Case 3:} For general $F$ but with $W=W_0$ a non-zero constant, the potential becomes
\be
V_\ssF  =  \frac{3|W_0|^2 F_{\ssX\bar\ssX}}{(F-|Y|^2)^2( |F_\ssX|^2-FF_{\ssX\bar\ssX})} \,,
\label{positive3}
\ee
and the $Y$ dependence comes only from the overall prefactor $(F-|Y|^2)^{-2}$. Classical solutions $X = X_c$ and $Y = Y_c$ are found by solving  $V_\ssX(X_c,Y_c)=V_\ssY(X_c,Y_c)=0$. In the $Y$ direction (for fixed $X$) this gives a local de Sitter minimum at $Y_c =0$ and a runaway to $V_\ssF = 0$ as $Y_c \to \pm\infty$, provided $F$ is chosen so that $V \geq 0$. 

In the special case that solutions to $F_{\ssX\bar\ssX}(X_c) =0$ exist (with $F_\ssX(X_c) \neq 0$, so the kinetic terms remain non-degenerate), the potential vanishes (and so is minimised) at $(X_c,Y)$ and the $Y$ direction is flat (and can be trusted for $|Y|^2 < F(X_c)$). This again gives a non-supersymmetric standard no-scale flat direction, but this time parameterised by $Y$ rather than $X$.  Should the condition $ F_{\ssX\bar\ssX}(X_c) =0$ only fix one of the two real directions in $X$, then the other combination is also a flat direction. Although this flat direction is standard no-scale in the sense that \pref{ToyNSCrit} is satisfied, notice that $K(Y, \ol{Y})$ obtained by truncating $K(X,Y,\ol{X},\ol{Y})$ at $X = X_c$ does {\it not} satisfy the no-scale criterion for $Y$ alone, since $K^{\bar\ssY \ssY} K_{\bar\ssY} K_\ssY \neq 3$.

\medskip\noindent
{\bf Case 4:} Suppose next that $F$ is general but the superpotential admits simultaneous solutions to the two conditions $W(Y_c) =W_\ssY(Y_c) =0$. In this case $Y_c$ represents a global minimum at zero vacuum energy with unbroken supersymmetry. The $X$ direction remains arbitrary along a supersymmetric flat direction. Since $F$ is arbitrary there need not be any internal (phase rotation or shift) symmetry along the $X$ direction, in which case neither of the two real fields in $X$ can be dualised to a linear multiplet.

\medskip\noindent
{\bf Case 5:} Finally suppose there is no value of $Y$ that satisfies both $W=0$ and $W_\ssY=0$ but there is a value $X = X_c$ that satisfies $F_{\ssX\bar\ssX} (X_c) = 0$. Then both $X$ and $Y$ are generically fixed (as joint solutions to $F_{\ssX\bar\ssX} =0$ and $W_\ssY=0$). The potential vanishes at this point, which is therefore a global minimum (for $F$ chosen so that $V_\ssF \geq 0$), and supersymmetry is broken because $W(Y_c)$ is non-zero.

This example can still be contrived to have a flat direction, however, if the condition $F_{\ssX\bar\ssX} = 0$ does not fix both real components of $X$. For example, if $F$ has the form 
\be
   F=(X+\ol{X}-\xi)^{3+n}|X|^2 + h(X)+\ol{h(X) }\,,
\ee
with $\xi$ a real number, $n$ a positive number and $h$ a holomorphic function chosen so that ${\rm Re}\left(h(\xi)\right)$ and $|h_\ssX(\xi)|$ are large and positive. Then $F_{\ssX\bar\ssX} = 0$ whenever $X+\ol{X} = \xi$, showing that $V_\ssF$ has a minimum (with positive kinetic terms) at $X+\ol{X}=\xi$ and $Y=Y_c$ determined by $W_\ssY(Y_c)=0$. With these values $V_\ssF = 0$ for any value of $X- \ol{X}$, which labels therefore a flat direction even if there is no need to have a shift symmetry for the Lagrangian describing fluctuations about this vacuum (and so no representation in terms of linear multiplets). Furthermore, although (\ref{ToyNSCrit}) shows $K^{\bar a b}K_{\bar a} K_b = 3$ when both $X$ and $Y$ are included, there is no sense in which the potential's flat direction defines a low-energy supersymmetric theory (there is only one real flat direction) for which \pref{KKKnoscale} holds separately once the heavier fields are integrated out.
 
\medskip
The above examples illustrate several things. The five highlighted cases provide several examples of both standard and non-standard no-scale models (with flat directions for both $X$ and $Y$, just $X$, just $Y$ or just a real flat direction -- parameterised by $X-\ol{X}$). Some of these have no axionic shift symmetries (though the examples also show that the freedom to perform field redefinitions can complicate deciding whether such symmetries exist). Other choices for $W$ and $F$ give isolated minima with both $X$ and $Y$ stabilised at a de Sitter or Minkowski minimum, with or without supersymmetry.
 
These examples can be generalised to include many fields $Y$ and $X$ whose kinetic terms and scalar potential are both positive provided the $X$ fields do not appear in $W$ and the invertible matrix $M$ has only one positive eigenvalue.\footnote{We note in passing that the $(1,N-1)$ signature of the matrix $M$ is reminiscent of the same property for the moduli space metric for K\"ahler moduli.} Multiple fields are interesting inasmuch as this is what typically arises in UV completions (which at present are limited to string theory realisations), and generalisations of the above models often do provide their low-energy descriptions, as is described in more detail starting in \S\ref{sec:descent2}.

\subsection{Specifics of scaling sufficiency}
\label{sec:scalesufficient}

This section fleshes out more precisely when approximate scale invariances can suffice for no-scale behaviour in a subsector of 4D $\cN=1$ supergravity, and when they do not. Although we also find more general connections between scaling and no-scale models in later sections, the tools developed in this section prove useful in later sections when identifying how scaling properties of extra-dimensional supergravities constrain the low-energy effective 4D supergravity obtained by dimensional reduction.

The implications of scaling symmetries for 4D supergravity are most directly made using the off-shell formalism of the superconformal tensor calculus~\cite{{Freedman:1976xh, Deser:1976eh, Stelle:1978wj, Ferrara:1978jt, Cremmer:1978hn}}. Within this framework the two-derivative component action can be written in superspace~\cite{Salam:1974jj} in the form
\be
\label{superconf} 
\frac{ \cL}{\sqrt{-g}} = -3 \int \exd^4 \theta \; \mathfrak{D} + \int \exd^2 \theta \; \Bigl( \mathfrak{F}_\ssW + \mathfrak{F}_g \Bigr) + \hbox{c.c.}  \,,
\ee
where, when specialised to zero-derivative terms\footnote{For future reference we remark that the scaling arguments used here and in later sections apply equally well if $\mathfrak{D}$ and $\mathfrak{F}$ also involve higher superspace derivatives. All that fails in this case is the ability to recast the conclusions in terms of its implications for $K$, $W$ and $\mathfrak{f}_{ab}$ of \pref{FDdefsKWf}.}
\be \label{FDdefsKWf}
   \mathfrak{D} = e^{-K/3}  \, \ol \Phi \Phi \,, \quad \mathfrak{F}_\ssW = W \, \Phi^3 \quad \hbox{and} \quad \mathfrak{F}_g = \mathfrak{f}_{ab} \cW^a \cW^b \,,
\ee
where $\cW^a = \lambda^a + F^a_{\mu\nu} \gamma^{\mu\nu} \theta + \cdots$ is the gauge field-strength chiral-spinor superfield (where $\lambda^a$ is the gaugino) and $\Phi =\{\varphi, \psi_\varphi, F_\varphi\}$ is the conformal compensator chiral multiplet. $K(z,\bar z)$, $W(z)$ and $\mathfrak{f}_{ab}(z)$ are functions of the chiral-scalar multiplets, collectively denoted $z^\ssA$. The compensator multiplet enters because the supergravity action is most easily derived for superconformal theories whose bosonic symmetries also include local scale transformations (or Weyl invariance), and $\Phi$ is the `spurion' superfield that breaks superconformal symmetry down to ordinary 4D Poincar\'e supergravity.

It must be emphasised that the component form for the Lagrangian density is obtained from \pref{superconf} using the rules of the conformal tensor calculus and {\it not} simply using the rules of Grassmann integration for global supersymmetry. These differ for several reasons: most notably the appearance of non-minimal couplings to spacetime curvature (with the resulting need to Weyl rescale\footnote{The need to Weyl rescale is the reason the $D$-term is written as $e^{-K/3}$, since this choice ensures the target-space metric appearing in the $z^\ssA$ kinetic terms is $\cG_{\ssA\bar\ssB} = \partial_\ssA\partial_{\bar\ssB}K$.} the metric to go to 4D EF) and the appearance of supergravity-multiplet auxiliary fields (whose elimination contributes important parts to the low-energy theory).\footnote{It can happen that leading terms as $M_p \to \infty$ with $m_{3/2}$ fixed can be obtained applying the rules of global supersymmetry to \pref{superconf}, provided one chooses the compensator $\Phi$ judiciously~\cite{Cheung:2011jp, Kugo:1982mr}.}

These supergravity complications in extracting the component form from \pref{superconf} are largely irrelevant for the present purposes, however, which only ask what scaling properties of the Lagrangian density imply for the scaling of the functions $W$, $\mathfrak{f}_{ab}$ and $e^{-K/3}$. These can be inferred as if one used \pref{superconf} in global superspace partly because the additional terms all scale consistently due to the scaling properties of the supergravity-multiplet auxiliary fields, and partly because the change in the metric scaling weight due to any Weyl rescaling can be captured by a change in the scaling weight assigned to the compensator $\Phi$.

It is also true that the scaling symmetries of interest are at best only accidental symmetries of the classical field equations, and so themselves typically do not hold to all orders in $1/M_p$. We address this below using the old strategy~\cite{Witten:1985xb, Burgess:1985zz, Nilles:1986cy, Witten:1985bz} of imagining the Lagrangian \pref{superconf} to arise order-by-order in small parameters (like $g_s$ or $\alpha'$ in string examples), and following separately how each term scales. 

\subsubsection{No-scale from scale invariance}
\label{ssec:scalingandnoscale}

Returning now to the question of the implications of scale invariance for supersymmetric models, we start by assuming\footnote{As we see below, descent from a higher-dimensional supergravity provides a good motivation for this assumption because scale invariance is a very generic property of most higher-dimensional supergravities at the two-derivative level.} the 4D supergravity action has couplings that --- to leading order in $1/M_p$ --- enjoy a classical scale invariance in the precise sense that 
\be  
\cL \to \lambda^{w_\ssL} \cL \quad \hbox{when} \quad g_{\mu\nu} \to \lambda^2 g_{\mu\nu} \quad \hbox{while} \quad 
T^i \to \lambda^{w_\ssT} T^i \quad \hbox{and} \quad \theta \to \lambda^{1/2} \theta \,,
\label{sym}
\ee
for some constant $\lambda$ and powers ${w_\ssL}$ and $w_\ssT$. Here $g_{\mu\nu}$ is the 4D Einstein frame metric and $T^i$ denotes a collection of 4D chiral superfields. The fermionic coordinate $\theta$ transforms in \pref{sym} in a way that is correlated with the metric transformation because the fermion kinetic terms must be automatically scale invariant whenever the kinetic terms of their scalar partners are. If scale invariance of the scalar field kinetic terms requires a scalar $\cT^i$ to transform as $\cT^i \to \lambda^{w_\ssT} \cT^i$, then its fermionic partner's kinetic term scales correctly when $\psi^i \to \lambda^{w_\ssT-1/2} \psi^i$. This difference in scaling properties precisely compensates for the replacement of $g^{\mu\nu}$ in $(\partial \cT)^2$ by ${e_a}^\mu$ in $\bar\psi \dsl \psi$. This is consistent with the superfield representation $T^i = \cT^i + \theta \psi^i + \cdots$ when $\theta$ scales as in \pref{sym}. There may also be other fields, $Z^a$, present that do not scale but we imagine these to be minimised at supersymmetric vacua for which $D_aW = 0$.

One could change the value of $w_\ssT$ by performing a field redefinition amongst the $T^i$. We do not do so because we are interested in cases where the imaginary part of $T^i$ enjoys a shift symmetry, which is not invariant under such redefinitions. The scale invariance of the Einstein action fixes the scaling of the Lagrangian $w_\ssL.$ In particular, in $d$ spacetime dimensions $\cL_\ssE  \propto \sqrt{-g} \;g^{\ssM\ssN} R_{\ssM\ssN}$ satisfies $\cL_\ssE \to \lambda^{d-2} \cL_\ssE$ and so ${w_\ssL} = d-2$. For $d =4$ we then have ${w_\ssL}=2$. 

Our interest is in what this same transformation rule for the terms in \pref{superconf} requires for the otherwise unknown functions $K$ and $W$. We know (using $d=4$ and ${w_\ssL} = 2$) that when applied to \pref{superconf} the symmetry \pref{sym} requires 
\be 
\label{Phisym}
\bar\Phi \Phi \, e^{-K/3} \to  \bar\Phi \Phi \, e^{-K/3} \qquad \hbox{and} \qquad
\Phi^3 \, W \to \lambda^{-1} \Phi^3 \, W \,,
\ee
since \pref{superconf} relates these quantities to the $D$- and $F$-term Lagrangian densities by $\bar\Phi \Phi \, e^{-K/3} = \theta^2 \bar\theta^2 L_\ssD + \cdots$ and $\Phi^3 \, W = \theta^2 L_\ssF$, where in both cases $L := \cL/\sqrt{-g}$ and so $\cL \to \lambda^2 \cL$ implies $L  \to \lambda^{-2} L$.  

Suppose we know -- perhaps because it involves a specific combination of fields with known scaling properties -- that $W$ transforms as $W \to \lambda^{w_\ssW} W$. Then \pref{Phisym} implies 
\be 
\Phi \to \lambda^{-(1+w_\ssW)/3} \Phi \qquad \hbox{and so} \qquad
e^{-K/3} \to \lambda^{2(1+w_\ssW)/3} e^{-K/3} \,.
\label{TreeExpKSS}
\ee
This states that $e^{-K/3}$ is a homogeneous function of the $T^i$ with degree $q$, where
\be \label{pvswandwT}
q = \frac{2 \left( 1 + w_\ssW \right)}{3 w_\ssT}\,.
\ee
If $K$ is a function only of $\chi^i = T^i+\ov T^i$, then \pref{TreeExpKSS} and \pref{pvswandwT} imply
\be
K(\lambda^{w_\ssT} \chi^i) = K(\chi^i) -3 q w_\ssT \ln \lambda  \,,
\label{New}
\ee
for all $\chi^i$ and $\lambda$, which would reproduce (\ref{homoKcondold0}) if $q=1$. In fact, differentiating (\ref{New}) and repeating the arguments used below \pref{homoKcondold0} implies  
\be
K^{i\bar\jmath} K_i K_{\bar\jmath} =  3 q\,.
\ee
We see that scale invariance implies the no-scale condition only if $e^{-K/3}$ is a homogeneous function of degree $q=1$, which is the same sufficient condition as found in \S\ref{sec:Relations}. From \pref{pvswandwT} this requires
\be 
w_\ssT = \frac23 (1+w_\ssW) \,,
\label{wTvsw}
\ee
and so no-scale behaviour in this example\footnote{A similar argument also holds in the dualised theory using linear multiplets (see Appendix ~\ref{4D_no_scale_app}).} requires the $T$-field scaling weight to be related to the scaling weight of $W$.

This above arguments establish that under some circumstances scale invariance can imply the scaling no-scale condition. The next three sections explore the several types of scale invariance that are inherited from (and are generic to) higher-dimensional supergravity~\cite{Salam:1989fm, Burgess:2011rv}, and argue that these (with low-energy $\cN=1$ supersymmetry) lie at the root of the ubiquity of (and the size of the breaking of) no-scale models in the low-energy limit of known UV completions.

\section{Descent from 10D heterotic supergravity}
\label{sec:descent2}

We start by asking how scale invariance constrains the $\cN=1$ supergravities obtained as 4D EFTs from 10D heterotic supergravity~\cite{Gross:1984dd, Gross:1985fr, Gross:1985rr}, slightly improving old arguments~\cite{Witten:1985xb, Burgess:1985zz, Nilles:1986cy}. Heterotic vacua are in many ways simpler than type IIA or IIB vacua since they do not involve localised sources like D-branes or orientifold planes, and much is known about their low-energy EFT (and their corrections)~\cite{Witten:1985bz, Green:2016tfs, Chamseddine:1980cp, Bergshoeff:1981um, Chapline:1982ww, Cai:1986sa, Gross:1986mw, Russo:1997mk, Bergshoeff:1989de, Chemissany:2007he}.

\subsection{Scaling in heterotic supergravity}
\label{ssec:hetroscale}

The heterotic case starts with the 10D String Frame (10D SF) heterotic Lagrangian, whose bosonic part for the purposes of scaling arguments takes the schematic form\footnote{We drop numerical coefficients here because our only interest is in how each term scales.}
\be \label{HeteroticSF}
\cL = \frac{1}{(\alpha')^4} \sqrt{-\hat g} \, e^{-2\phi} \left[  \hat g^{\ssM\ssN} \Big( \hat R_{\ssM\ssN} + \partial_\ssM \phi \, \partial_\ssN \phi \Bigr) + \alpha' \, \hat F^2  + (\alpha' )^2\, \hat H^2 \right]  \,,
\ee
plus fermionic terms. Both $F$ and $H$ are field strengths given schematically by
\be \label{FieldStrHS}
 F = \exd A + f A^2 \qquad \hbox{and} \qquad
 H = \exd B_2 +  A\, (\exd A + f A^2) \,,
\ee
where $A_\ssM$ is a 10D Yang-Mills potential and $B_{\ssM\ssN}$ is a Kalb-Ramond gauge 2-form. Here $f$ is a dimensionless parameter that depends on how $A_\ssM$ is normalised -- useful to keep track of in what follows -- and so systematically accompanies the structure constants of the non-Abelian gauge group. In \pref{HeteroticSF} hats on squared field strengths indicate contractions are performed using the SF metric, so $\hat F^2 = \hat g^{\ssM\ssN} \hat g^{\ssP\ssQ} F_{\ssM\ssP} F_{\ssN\ssQ}$ and so on. 

Transforming to 10D Einstein Frame (10D EF) using 
\be
\hat g_{\ssM\ssN} = \left( \frac{\alpha'}{\kappa^{1/2}} \right) \, e^{\phi/2}\, \tilde g_{\ssM\ssN} = e^{(\phi-\langle\phi\rangle)/2}\, \tilde g_{\ssM\ssN}
\ee
where $\kappa$ is the 10D Planck parameter, $\kappa = (\alpha')^2 g_s$, and $g_s=e^{\langle\phi\rangle}$ is the string coupling set by the {\it v.e.v.}~of the 10D dilaton, then gives
\be
  \cL =  \sqrt{-\tilde g}  \left[ \frac{1}{\kappa^2} \, \tilde g^{\ssM\ssN} \Big( \tilde R_{\ssM\ssN} + \partial_\ssM \phi \, \partial_\ssN \phi \Bigr) + \frac{e^{-\phi/2}}{\kappa^{3/2}} \; \tilde F^2  + \frac{e^{-\phi}}{\kappa} \; \tilde H^2 \right]  \,,
\ee
plus fermionic terms (where now $\tilde F^2 = \tilde g^{\ssM\ssN} \tilde g^{\ssP\ssQ} F_{\ssM\ssP} F_{\ssN\ssQ}$ and so forth). 

Following~\cite{Burgess:1985zz} we note that this action (including its unwritten fermionic terms) enjoys the following two scaling properties:\footnote{As mentioned in the introduction 10D EFTs for perturbative string vacua have two scaling symmetries, corresponding to the two perturbative expansions of their UV completion: the $\alpha'$ and string loop expansions. By contrast, an EFT like 11D supergravity for a strongly coupled vacuum has only one scaling symmetry: transforming $g_{\ssM\ssN}\to \lambda^2 g_{\ssM\ssN},$ and $A_{\ssM\ssN\ssP}\to \lambda^3 A_{\ssM\ssN\ssP}$ scales the 11D Lagrangian as $\cL_{11} \to \lambda^9 \cL_{11}$. This has its roots in the $\alpha'$ expansion, and is sometimes used to redefine the 11D scale $\kappa_{11}$ (see for instance~\cite{Russo:1997mk} and references therein).}
\begin{itemize}
\item {\it Dilaton scale transformations,} defined by
\be \label{Scale1HS}
  \tilde g_{\ssM\ssN} \to \lambda\, \tilde g_{\ssM\ssN} \,, \quad e^{- \phi/2} \to \lambda\, e^{- \phi/2} \,,
\ee
with all other fields fixed, under which $\cL \to \lambda^4 \cL$. Notice that this transformation is fairly obvious in 10D string frame \pref{HeteroticSF}, given that \pref{Scale1HS} implies the string-frame metric, $\hat g_{\ssM\ssN} \simeq e^{\phi/2} \tilde g_{\ssM\ssN}$, is invariant.

\item {\it Rescaling property,} under which
\be
  F_{\ssM\ssN} \to \xi\,F_{\ssM\ssN} \,, \quad H_{\ssM\ssN\ssP} \to \xi^2\, H_{\ssM\ssN\ssP} \,, \quad e^{\phi/2} \to \xi^2\,e^{\phi/2} \,,
\ee
with all other fields fixed, under which $\cL \to \cL$.  We call this a `property' rather than a symmetry in the sense used in~\cite{Burgess:1985zz}. The point is that the transformation rules are defined acting on the field strengths, $F$ and $H$, rather than as transformations of the field potentials, $A$ and $B_2$. The only obstruction to promoting the transformation to act on the gauge potentials is the presence of the non-Abelian contributions to $F$ and $H$ in \pref{FieldStrHS}. Alternatively, we can take the definition to act as $B_{\ssM\ssN} \to \xi^2 B_{\ssM\ssN}$ and $A_\ssM \to \xi A_\ssM$ if we imagine the dimensionless parameter, $f$, to be a spurion, transforming as 
\be \label{nonabelianspurion}
   f \to \xi^{-1} f\,.
\ee 

\end{itemize}

\subsubsection{Compactified fields}

The 4D EF metric, $g_{\mu\nu}$ is related to the 10D EF metric by
\be
\tilde g_{\mu\nu} = \left( \frac{\kappa^2}{V_\ssE\, \kappa_4^2} \right) \, g_{\mu\nu} \,,
\ee
where $V_\ssE$ denotes the dimensionful extra-dimensional volume as measured with the 10D EF metric. (We denote by $\hat V$ the volume measured using the 10D SF metric. Also useful is the dimensionless quantity $\cV_\ssE := V_\ssE/(\alpha')^3$ obtained by normalising by the string scale.) Since $\tilde g_{\ssM\ssN} \to \lambda \tilde g_{\ssM\ssN}$ we have $V_\ssE \to \lambda^3 V_\ssE$ and so 
\be
g_{\mu\nu} \to \lambda^4 g_{\mu\nu} \,.
\ee
The two universal 4D moduli for heterotic compactifications are given by~\cite{Witten:1985xb, Burgess:1985zz}
\bea 
  S &=& \frac{\hat V}{(\alpha')^3} \; e^{-2\phi} + i \, a = \frac{V_\ssE}{(\alpha')^3} \; e^{-\phi/2}  + i \, a = \cV_\ssE\, e^{-\phi/2}  + i \, a \nn\\
  T &=& \frac{\hat r^2}{\alpha'} + i \, b = \frac{r_\ssE^2}{\alpha'} \; e^{\phi/2} + i \, b = \cV_\ssE^{1/3} \, e^{\phi/2} + i \, b \,, \label{eq:defts}
\eea
where $a$ and $b$ are the model-independent axion fields (coming respectively from $B_{\mu\nu}$ and $B_{mn}$) and we use the notation $\hat V = \hat r^6$ for the SF volume and similarly $V_\ssE = r_\ssE^6$ (also often written $e^\sigma := r_\ssE^2/\alpha'$ so that $\cV_\ssE = e^{3\sigma}$). Because $S+\ol{S} \propto \cV_\ssE \, e^{-\phi/2}$ and $T+\ol{T} \propto \cV_\ssE^{1/3} e^{\phi/2}$, these moduli inherit the following transformation properties under the two scaling symmetries\footnote{The axions scale in the same way, as is clear for $T$ since $b$ is proportional to $B_{mn}$, and as~\cite{Burgess:1985zz} demonstrates for $a$ using the duality transformation that relates $B_{\mu\nu}$ to $a$.}
\be 
\label{SscaleNew}
  S \to \frac{\lambda^4}{\xi^2} \; S \qquad \hbox{and} \qquad 
   T \to \xi^2 \; T  \,.
\ee

\subsubsection{Compactified gauge kinetic terms}

The scaling properties of the $\cN=1$ supersymmetric action in 4D are obtained in leading approximation by truncating the 10D EF action given above. The Maxwell term in particular gives
\be \label{ClassMaxHet}
 \cL_g \propto \sqrt{-\tilde g_{10}} \; e^{-\phi/2} \, \tilde g^{\mu\nu} \tilde g^{\lambda \rho} F_{\mu\lambda} F_{\nu\rho} \propto  \sqrt{-g_4}\; V_\ssE \, e^{-\phi/2} \, g^{\mu\nu}  g^{\lambda \rho} F_{\mu\lambda} F_{\nu\rho} \,.
\ee
The gauge coupling function $\mathfrak{f}$ can be read off from this by comparing to 
\be
 \cL_g \propto   \sqrt{-g_4}\,\Bigl[ \hbox{Re} ( \mathfrak{f})\, g^{\mu\nu}  g^{\lambda \rho} F_{\mu\lambda} F_{\nu\rho} +  \hbox{Im} (\mathfrak{f}) \, \epsilon^{\mu\nu\lambda\rho} F_{\mu\nu} F_{\lambda \rho} \Bigr]\,.
\ee
Since $\cL_g \to \lambda^4 \cL_g$ inherits the scaling property of the 10D action, and since $F_{\mu\nu} \to \xi F_{\mu\nu}$ while $ \sqrt{-g_4}\;  g^{\mu\nu}  g^{\lambda \rho}$ is scale invariant, this shows that 
\be 
\label{ftrans}
\mathfrak{f} \to \frac{\lambda^4}{\xi^2} \; \mathfrak{f} \,,
\ee
and so $\mathfrak{f}$ scales the same way as does $S$ in \pref{SscaleNew}.

\subsubsection{Superspace transformation properties}
\label{sssec:SStransProp}

For the rest of the Lagrangian it is worth identifying how quantities in 4D superspace must scale. To this end we repeat the exercise given in \S\ref{ssec:scalingandnoscale}, but now applied to the two heterotic scaling transformations given above.  Only the main results are quoted here, to indicate what changes.

One change is that the 4D EF metric scales as $ g_{\mu\nu} \to \lambda^4 g_{\mu\nu}$ (rather than as in \pref{sym}) and so, defining $\cL = \sqrt{-g_4} \; L$, the transformations $\cL \to \lambda^4 \cL$ and $\sqrt{-g_4} \to \lambda^8 \sqrt{-g_4}$ of the 4D EFT imply $L \to {L}/{\lambda^4}$. Furthermore, comparing bosonic and fermionic kinetic terms shows that the superspace coordinate must scale as
\be 
\label{thetatrans}
  \theta \to \lambda\, \theta \,.
\ee
Then the superspace relations $\mathfrak{F} = L_\ssF\, \theta^2 + \cdots$ and $\mathfrak{D} = L_\ssD\, \theta^4 + \cdots$ imply
\be 
\label{Ftrans}
\mathfrak{F} \to \frac{\mathfrak{F}}{\lambda^2} \quad \hbox{and} \quad \mathfrak{D} \to \mathfrak{D} \,.
\ee
Since both $g_{\mu\nu}$ and $\cL$ are invariant under the $\xi$ transformations, so must also be $\theta$ and both $\mathfrak{F}$ and $\mathfrak{D}$. Notice that the scaling properties \pref{Ftrans} apply not just for the derivative-independent contributions described by $\mathfrak{f}_{ab}$, $K$ and $W$, but also when $\mathfrak{F}$ and $\mathfrak{D}$ involve higher superspace derivatives.

Since the field strength $F_{\mu\nu} \to \xi F_{\mu\nu}$ and is invariant under $\lambda$-scalings, we learn the 4D gauge field-strength supermultiplet, $\cW = \zeta + F_{\mu\nu} \gamma^{\mu\nu} \theta + \cdots$ (where $\zeta$ is the gaugino field) scales as
\be 
\label{gaugeSFtrans}
 \cW \to \frac{\xi}{\lambda^3} \; \cW \,.
\ee
This gives a second way to compute how $\mathfrak{f}$ scales, since the supersymmetric gauge kinetic term is $\mathfrak{F}_g = \mathfrak{f}_{ab} \, \cW^a \cW^b$, so \pref{Ftrans} and \pref{gaugeSFtrans} imply \pref{ftrans}.

The transformation of the superpotential, $W$, and K\"ahler potential, $K$, both require knowing how the compensator, $\Phi$, scales. To pin this down we use the result that direct truncation shows that the leading contribution to the low-energy superpotential is given by $W = f\, C^3$ where $C$ are matter fields that arise from the extra-dimensional gauge potentials, $C \sim A_m$, and $f$ is the dimensionless spurion appearing in the normalisation of the gauge-field structure constants defined in \pref{FieldStrHS}: $F = \exd A + f\, A^2$. We have seen that the 10D scaling symmetries act with $A \to \xi A$ (with $A$ invariant under $\lambda$-scalings) provided we transform $f$ as in \pref{nonabelianspurion}. Since $C$ arises as a mode in $A$ it transforms as
\be
C \to \xi\, C \,,
\ee
and consistency with $W = f \,C^3$ requires $W$ to scale as
\be
W \to  \xi^2\, W \,,
\ee
and to be invariant under $\lambda$-scalings. Consistency of $\mathfrak{F} = \Phi^3 \, W$ with \pref{Ftrans} then dictates the compensator transform as 
\be 
\label{comptrans}
\Phi \to (\lambda \xi)^{-2/3} \; \Phi \,.
\ee

Finally, the transformations \pref{comptrans} and \pref{Ftrans} for $\mathfrak{D} = \ov\Phi \Phi \, e^{-K/3}$ imply the K\"ahler potential satisfies
\be \label{expKtrans}
  e^{-K/3} \to   (\lambda \xi )^{4/3} \; e^{-K/3} \,.
\ee
This scaling dictates the dependence of $K$ on the fields $S$ and $T$ (once all other fields are combined into scale invariant combinations), giving
\be
  e^{- K/3}= (S+\ol S)^{1/3} (T + \ol T)\; e^{-K_{\rm inv}/3}  \,,
\ee
where $K_{\rm inv}$ can only depend on scale invariant ratios of any other fields (more about which below). The above expression agrees with the result obtained by direct dimensional truncation of the 10D action~\cite{Witten:1985xb, Burgess:1985zz}:
\be 
     K_{\rm tr} = - \ln(S+\ol S) - 3 \ln(T+\ol T) + K_{\rm inv} \,.
\ee  

These two scaling properties do not in themselves determine how $K_{\rm inv}$ depends on other fields, such as the field $C$ discussed above, beyond stating that it must be a function only of invariant combinations like $C \ol C/(T +\ol T)$ or $f \, (C\ol C)^2$ (where invariance under 4D gauge rotations of the $C$'s dictates they enter $K$ only through the combination $C\ol C$, and dependence\footnote{Notice that because $f$ always arises in 10D together with powers of the gauge potential, in 4D $f$ only ever appears together with powers of $C$ and $\ol C$, allowing it to appear within $K$ only through invariant combinations like $f(C\ol C)^2$ [as opposed to $f(T+\ol T)^2$, say].} on the spurion $f$ tracks the ways in which the second scaling symmetry is broken by non-Abelian gauge interactions). This dependence also agrees with direct truncation of the 10D action, which is consistent with (but do not require) the form
\be
\label{STCSequester}
  K = - \ln(S+\ol S) - 3 \ln(T+\ol T - \ol C C) \,.
\ee

Because direct truncation calculations are done at leading order in the $\alpha'$ and string-loop expansions, they cannot in practice distinguish between \pref{STCSequester} and
\be \label{KtrHS}
   K_{\rm tr} = - \ln(S+\ol S) - 3 \ln(T+\ol T) + 3 \ol C C/(T+\ol T)  + \cdots \,.
\ee
These are indistinguishable because (see below) the $\alpha'$ expansion is an expansion in powers of $(T+\ol T)^{-1}$ and direct calculations in practice test only the leading terms. Extra information is required to believe the `sequestered' form \pref{STCSequester} to all orders in $\ol C C/(T+\ol T)$. 

\subsubsection*{Accidental shift symmetries}

A physical motivation for the sequestered form of the K\"ahler potential~\pref{STCSequester} starts from the observation that there should be, at leading order, no energetic preference for any specific value for the volume modulus $T$ (which is, after all, what it means to be a modulus of the leading order solutions). The sequestered form achieves this by having the matter field appear in no-scale form, as is easily seen by comparing \pref{KtrHS} with the models of \S\ref{sssec:standardnoscale}.

Appendix ~\ref{app:accidentalshift} provides a separate symmetry argument in favour of \pref{STCSequester}. This symmetry relies for its existence on the observation that the axion field arises in the 10D theory as a mode of the extra-dimensional 2-form gauge potential $b(x) \in B_{mn}(x,z)$. But $B_{mn}$ itself only appears in the 10D action through the gauge-invariant field strength $H = \exd B_2 - \Xi_\CS$, where $\Xi_\CS$ is the Chern-Simons 3-form built from the 10D Yang-Mills potential $A_\ssM$. Because $C(x) \in A_\ssM(x,z)$ descends from the gauge potential, the systematic appearance of $B_2$ with $A$ within $H$ implies the 4D theory inherits an approximate global shift symmetry that relates the complex scalars $T$ and $C$.

More concretely, adopting complex extra-dimensional coordinates $z^i, \bar z^{\bar\jmath}$ (with $i,\bar\jmath =1,2,3$) the 4D axion $b$ and matter field $C$ arise during compactification as 
\be
  B_2(x,z) = b_\ssI(x) \, \omega^\ssI_{i\bar\jmath} \, \exd z^i \, \exd \bar z^{\bar\jmath} \qquad \hbox{and} \qquad
  A^{i\,r}(x,z) = C^r_\ssI (x) \, {(\omega^\ssI)^{i}}_{\bar\jmath} \, \exd \bar z^{\bar\jmath} \,,
	\label{Decomp}
 \ee
where $I$ labels a basis of harmonic extra-dimensional (1,1)-forms $\omega^\ssI$ and the index pair $(ir)$ is an adjoint gauge label for $E_8$ written as a product of an $SU(3)$ index $i$ and an $E_6$ index $r$.\footnote{Eq.~(\ref{Decomp}) focusses only on the $(\textbf{27},\textbf{3})$ states counted by $h^{1,1}$, though similar arguments apply, for example,  for $(\ov{\textbf{27}},\ov{\textbf{3}})$ states counted by $h^{1,2}$.} The specific axion appearing in $T$ corresponds to a specific harmonic form $\omega$, the compactification's universal K\"ahler form. The label $i$ can appear either as a coordinate or gauge index because the background gauge potentials in the $SU(3)$ subgroup is non-zero and identified with the extra-dimensional spin connection.

The arguments of Appendix~\ref{app:accidentalshift} then show that $H$ is invariant under an accidental approximate shift symmetry of the form $\delta A^{ir} = \eta^r \, {\omega^i}_{\bar\jmath} \, \exd \bar z^{\bar\jmath}$ together with $\delta B_2 \propto \delta A_{ir} \wedge A^{ir} +$ c.c. (for constant complex shift parameters $\eta^r$). This symmetry appears to be only approximate due to the presence of the background $SU(3)$ gauge field, and as a result it does not preserve terms in the 10D action involving the non-Abelian part of the gauge interactions, such as the commutator term in $F = \exd A + f\,[A , A]$.

$\cN=1$ supersymmetry requires the counterpart to this shift symmetry in the low-energy 4D EFT to act holomorphically on the chiral multiplets, with $C^r$ shifting by a complex constant and $T$ (which contains the axion $b$) shifting in a way that is linear in the $C$'s. This dictates a symmetry of the form:
\be\label{4DaccShiftHet}
   \delta C^r_\ssI = \eta^r_\ssI \qquad \hbox{and} \qquad \delta T = {\cT^\ssI}_{\ssJ}\, \bar\eta_r^\ssJ \, C^r_\ssI \,,
\ee
for a constant complex parameter $\eta^r_\ssI$ and an appropriate set of coefficients ${\cT^\ssI}_\ssJ$ (expressible, in explicit compactifications, as integrals over the harmonic forms). Invariance of $K$ under these transformations implies $K$ can depend on $T$ only through the invariant combination $T+\ol T - {\cT^\ssI}_\ssJ \ol C_r^\ssI C^r_\ssJ$, which together with the scaling symmetry implies the sequestered form \pref{STCSequester}.

Notice that the superpotential, $W = \frac16 \, f\,g^{\ssI\ssJ\ssK} d_{rst} C^r_\ssI C^s_\ssJ C^t_\ssK$, obtained by dimensional reduction, is not invariant under \pref{4DaccShiftHet}, which instead transforms as
\be
 \delta W = \frac12\,f \, g^{\ssI\ssJ\ssK} d_{rst} C^r_\ssI C^s_\ssJ\, \eta^t_\ssK \,.
\ee
We understand this to reflect the non-invariance in 10D of the non-Abelian part of the field strength under the 10D shift $\delta A^{ir}$, since in dimensional reductions $W$ descends partly from the non-Abelian part of the gauge kinetic term Tr$\,F_{\ssM\ssN} F^{\ssM\ssN}$ in 10D (as well as the non-Abelian part inside $H^2$). This failure of the superpotential to preserve the shift symmetry is an artefact of the breaking of gauge symmetries by the background gauge field in heterotic compactifications, and is not a property of the shift symmetries we encounter for type IIB theories in later sections.

\subsection{Higher orders in $g_s$ and $\alpha'$}
\label{ssec:HeteroHigherPowers}

To this point the discussion of heterotic scaling symmetries has been restricted to lowest order in the string-coupling ($g_s$) and $\alpha'$ expansions. But scaling symmetries are most informative in their constraints on contributions at higher order in $g_s$ and $\alpha'$. In 10D SF a broad class of these schematically look like\footnote{Circles are used as indices here since for scaling purposes the exact nature of the index contractions is less useful than is the total number of contravariant and covariant indices.}
\be \label{SFradcorHS}
 \cL_{nmr} = \frac{1}{(\alpha')^5} \sqrt{-\hat g_{10}} \; e^{2(n-1)\phi} \Bigl( \alpha' \hat g^{\circ \circ} {{\hat R}^\circ}_{\circ \circ\circ} \Bigr)^{m+1} \Bigl[( \alpha')^2 \hat g^{\circ \circ} \hat g^{\circ \circ} F_{\circ\circ} F_{\circ\circ} \Bigr]^{r} 
\ee
where $n$ counts string loops while $m$ and $r$ count powers of metric curvature and gauge field strength (as proxies for more general terms in the derivative expansion).  

In 10D Einstein frame these become
\bea \label{EFradcorHS}
 \cL_{nmr} &=& \frac{1}{\kappa^{5/2}}\Bigl( e^{5\phi/2} \sqrt{-\tilde g_{10}} \Bigr)\; e^{2(n-1)\phi} \Bigl( \kappa^{1/2} e^{-\phi/2} \tilde g^{\circ \circ} {{\tilde R}^\circ}_{\circ \circ\circ} \Bigr)^{m+1} \Bigl[ \kappa\,e^{-\phi} \tilde g^{\circ \circ} \tilde g^{\circ \circ} F_{\circ\circ} F_{\circ\circ} \Bigr]^{r}  \nn\\
 &=& \kappa^{(m+2r-4)/2} \sqrt{-\tilde g_{10}} \; e^{(4n-m-2r)\phi/2} \Bigl( \tilde g^{\circ \circ} {{\tilde R}^\circ}_{\circ \circ\circ} \Bigr)^{m+1} \Bigl[ \tilde g^{\circ \circ} \tilde g^{\circ \circ} F_{\circ\circ} F_{\circ\circ} \Bigr]^{r} \,.
\eea
Two reality checks for (\ref{EFradcorHS}) are: the case $(n,m,r) = (0,0,0)$ which gives the 10D classical Einstein-Hilbert term (from which $\phi$ drops out in EF), and the case $(n,m,r) = (0,-1,1)$ which corresponds to the classical Maxwell term discussed in \pref{ClassMaxHet} above.

As is easily verified $\cL_{nmr}$ obeys the following scaling transformation property
\be
 \cL_{nmr} \to \lambda^4 \left( \frac{\xi^2}{\lambda} \right)^{4n} \left( \frac{1}{\xi^2} \right)^{m+r} \cL_{nmr} \,,
\ee
which agrees with the lowest-order scaling result discussed above when restricted to string tree level ($n=0$) and terms linear in $\tilde R$ or $F^2$. Writing $\cL_{nmr} = \sqrt{-g} \; L_{nmr}$ and repeating the arguments of the previous section then implies 
\be
 L_{nmr} \to \lambda^{-4} \left( \frac{\xi^2}{\lambda} \right)^{4n} \left( \frac{1}{\xi^2} \right)^{m+r} L_{nmr} \,,
\ee
and so once the factors of $\theta^2$ and $\theta^4$ are extracted to go to superspace,  the quantities $K$ and $W$ must scale as
\be
 \mathfrak{F}_{nmr} \to \lambda^{-2} \left( \frac{\xi^2}{\lambda} \right)^{4n} \left( \frac{1}{\xi^2} \right)^{m+r} \mathfrak{F}_{nmr} 
 \quad \hbox{and} \quad
 \mathfrak{D}_{nmr} \to  \left( \frac{\xi^2}{\lambda} \right)^{4n} \left( \frac{1}{\xi^2} \right)^{m+r} \mathfrak{D}_{nmr} \,.
\ee

Extracting the factors of $\cW^2$ and $\Phi$ as done earlier for the leading terms then gives the scaling properties of $K$, $W$ and $\mathfrak{f}$ for any $n$, $m$ and $r$. For the gauge kinetic function this gives 
\be 
\label{GaugeKloop}
\mathfrak{f}_{nmr} \to \frac{\lambda^4}{\xi^2} \left( \frac{\xi^2}{\lambda} \right)^{4n} \left( \frac{1}{\xi^2} \right)^{m+r} \mathfrak{f}_{nmr} \,,
\ee
(where $r$ is kept open since some of the $F$'s could be background gauge fields, say, in the extra dimensions). Similarly, for the superpotential 
\be
W_{nmr} \to \xi^2 \left( \frac{\xi^2}{\lambda} \right)^{4n} \left( \frac{1}{\xi^2} \right)^{m+r} W_{nmr} \,.
\ee
For both $\mathfrak{f}$ and $W$ the contributions for generic ($n,m,r$) typically vanish because of the non-renormalisation theorems. The K\"ahler potential similarly satisfies
\be
\label{heteroticnmr}
 \left( e^{-K/3} \right)_{nmr} \to (\lambda \xi)^{4/3} \left( \frac{\xi^2}{\lambda} \right)^{4n} \left( \frac{1}{\xi^2} \right)^{m+r} \left( e^{-K/3} \right)_{nmr} \,.
\ee

For comparison purposes, notice that the transformation properties \pref{SscaleNew} imply
\be
\frac{T^3}{S} \to \left(\frac{\xi^2}{\lambda^4} \right) \Bigl( \xi^2 \Bigr)^3  \left( \frac{T^3}{S} \right) = \left( \frac{\xi^2}{\lambda} \right)^4  \frac{T^3}{S} \,.
\ee
Keeping in mind that axion shift symmetries forbid $K$ from depending on the imaginary parts of $S$ and $T$ we find the following double expansion for $e^{-K/3}$ in terms of $S+\ol S$ and $T+\ol T$:
\be
\label{heteroticnmr2}
\setlength\fboxsep{0.25cm}
\setlength\fboxrule{0.8pt}
\boxed{\left( e^{-K/3} \right)_{nmr} =(T+\ol T)(S+\ol S)^{1/3}\sum_{nmr} \cB_{nmr}\left[\frac
{(T+\ol T)^3}{S+\ol S}\right]^n\, \left(\frac{1}{T+\ol T}\right)^{m+r} \,,}
\ee
with scale-invariant coefficients $\cB_{nmr}$. 

In all these expressions the first term ($n=m+r=0$) gives the dependence on $S$, $T$ found earlier, and the correction terms show how $S$ and $T$ must contribute at any specific order. In principle the dependence of the $\cB_{nmr}$ on other fields like $\ov C C$ requires more information, such as by demanding that $T$ and $\ov C C$ enter only through the combination $T+\ol T - \ov C C$ (as is only appropriate if the approximate shift symmetry of Appendix ~\ref{app:accidentalshift} survives at the order of interest). 

These expressions also easily generalise to include more $T$ moduli since every appearance of $T+\ol T$ can be replaced by any homogeneous degree-one function of multiple $T+ \ol T$'s. For example, any appearance of $T+\ol T$ could equally well have been written as a power of $(S+\ol S)$ and the Einstein-frame extra-dimensional volume ${\mathcal V}_E= [(S+\ol S) (T+\ol T)]^{4/3}$ (as is done for the type IIB vacua discussed below).

Non-renormalisation theorems arise when these scaling relations conflict with the requirements of shift symmetry and holomorphy~\cite{Witten:1985bz}. For instance, holomorphy and the scaling of $ \mathfrak{f}_{nmr}$ given by \pref{GaugeKloop} only allows a dependence
\be  \label{fnmrHetero}
\mathfrak{f}_{nmr} \propto S \left( \frac{T^3}{S} \right)^n \frac{1}{T^{m+r}} \,.
\ee
But proper transformation under the axion shift symmetries -- {\it c.f.}~\pref{eq:defts} -- only allows $S$ and $T$ to appear linearly in $\mathfrak{f}$. Linearity is only consistent with \pref{fnmrHetero} for two special cases: ($i$) the tree-level term arising when\footnote{Notice $m+r=0$ can be consistent with $r=1$, as required for the Maxwell term, because $m = -1$ is allowed -- corresponding to there being no curvature terms in \pref{SFradcorHS} or \pref{EFradcorHS}.} $n = m+r=0$, with $\mathfrak{f} = S$; and ($ii$) the case $n=1$ with $m+r = 2$, corresponding to $\mathfrak{f} = T$. 

These inferences are consistent with direct truncation~\cite{Ibanez:1986xy} which suggests that the $n=1$ term $\mathfrak{f} = T$ comes from the anomaly cancelling piece of the 10D action~\cite{Green:1984sg}. However the anomaly cancelling action is not quite in the class of Lagrangians considered above, since it is of the form
\be
\cL_{\rm anom} \sim B \wedge F \wedge F \wedge \hat R \wedge \hat R  \,,
\ee
in the string frame (with no dilaton as appropriate for 1 loop). This gives the anomaly coupling for the axion part of $T$ when $B$ and $\hat R \wedge \hat R$ have components in the extra dimensions. This naively has $m = r = 1$ and so seems to fit into the above discussion, but the factor of $B$ was not included when deriving the above rules. 

This is easily fixed, by keeping in mind the transformation $B_{\ssM\ssN} \to \xi^2 B_{\ssM\ssN}$ and recognizing that the scaling of the factor $\epsilon^{\ssM_1\cdots \ssM_{10}}$ in the wedge product makes this new piece scale like $B_{\ssM\ssN} \hat g^{\ssM\ssN}$. This corresponds to $B_{\ssM\ssN} \tilde g^{\ssM\ssN} e^{-\phi/2}$ in the 10D Einstein frame and so requires \pref{GaugeKloop} to acquire an extra factor of $\xi^2 \lambda^{-1} (\lambda/\xi^2) = 1$ (showing the scaling does not change compared to the above, so the counting $n=m=r=1$ is correct).\footnote{Notice that this discussion differs somewhat from that of~\cite{Nilles:1986cy}, where a similar analysis was done to identify quantum corrections in heterotic vacua using scaling symmetries. There the string-loop expansion parameter was assumed to scale as $T/S$ in order to reproduce the corrections to the gauge kinetic function coming from the anomaly cancelling term. Here we instead find that tracking both loop and $\alpha'$ expansions separately gives expansion parameters $T^3/S$ and $1/T$. Although~\cite{Nilles:1986cy} agrees (by construction) with our power-counting for the one-loop correction to the gauge kinetic function, $\mathfrak{f}=S+T$, the powers of $T+\ol T$ and $S + \ol S$ predicted in \cite{Nilles:1986cy} order-by-order for the K\"ahler potential contain only the subset of terms for which $m+r = 2n$.}

Notice that a leading correction in $\alpha'$ -- such as computed in~\cite{Chang:1986ac} -- to the K\"ahler potential correspond to $n = r = 0$ and $m=1$ and so gives a $1/(T+\ol T)$ correction of the form  
\be  \label{heteroalphap0}
e^{-K/3} \simeq (S+\ol S)^{1/3} (T+\ol T) \left[ 1 + \frac{\mathfrak{K}}{T+\ol T} \right], 
\ee
where $\mathfrak{K}$ is a scale invariant function of any other fields of interest. Since this preserves the no-scale structure for $T$ (once $S$ is minimised at a minimum for which $D_\ssS W = 0$), it does not lift the flatness of the $T$-direction. This kind of flatness preservation at subdominant order in $1/(T+\ol T)$, when encountered for type IIB vacua, has come to be known as `extended no-scale structure'~\cite{vonGersdorff:2005bf,LVLoops, LVLoops1} (see below).

Eq.~\pref{heteroalphap0} also agrees with the usually quoted form for the $\alpha'$-corrected K\"ahler potential obtained by explicit dimensional reduction:
\be\label{heteroalphap}
K = - \ln (S+\ol S) -3 \ln (T+\ol T + \mathfrak{K}_{\rm tr}) \,,
\ee
where $\mathfrak{K}_{\rm tr}$ is now a constant. From the point of view presented here the correction $\mathfrak{K}_{\rm tr}$ naturally appears inside the logarithm because eqs.~\pref{superconf} and \pref{FDdefsKWf} together show that when the Lagrangian comes as an expansion in powers of $\alpha'$ and $g_s$ this implies the same for the quantity $e^{-K/3}$ rather than for $K$ itself.

\section{Descent from 10D type IIB supergravity}
\label{sec:descent}

The connection between scale invariance, supersymmetry and no-scale effective theories also works for perturbative type IIB string vacua, since (like all perturbative weakly curved vacua) the 10D type IIB supergravity also enjoys the accidental classical scale symmetries whose roots lie in the underlying $g_s$ and $\alpha'$ expansions~\cite{Burgess:2011rv, Burgess:2005jx}.  An important new complication that arises here (but not for the heterotic models of the previous section) is the important role played by localised D-brane and orientifold plane sources, and the different dependence that these sources have on the string coupling $g_s \propto e^{\phi}$. 

\subsection{Scaling in type IIB supergravity}
\label{ssec:IIBSGScaling}

Type IIB supergravity in 10 dimensions famously involves a self-dual 5-form field, and this complicates formulating its dynamics in terms of an action. The scale invariance highlighted here can be formulated as a symmetry of the classical field equations, but we instead describe it in terms of an action through the common artifice of temporarily ignoring the 5-form self-duality condition, which is sufficient for the purposes of identifying the scale symmetries. 

The 10D type IIB action has the form
\be
 S_{\rm IIB} =   S_{\rm bulk} +  S_{\rm loc}   \ ,
\ee
where $S_{\rm bulk}$ describes the supergravity degrees of freedom while $S_{\rm loc}$ contains degrees of freedom localised on specific spacetime surfaces, such as those associated with D-branes and O-planes. We  examine the scaling properties of each of these in turn.

\subsubsection{Scale invariance of the bulk}

The bosonic part of the type IIB bulk action in 10D EF is (schematically)\footnote{As before numerical factors are suppressed since they are irrelevant when establishing scaling properties. We focus also on the bosonic sector though we checked that the symmetries we find do extend to the complete action involving the fermions. }
\be
\label{TypeIIBBulk}
 S_{\rm bulk} = \int \sqrt{-\tilde g} \; \left\{\tilde R - \frac{\lvert \partial \tau \rvert^2}{(\mathrm{Im}\,\tau)^2} - \frac{\lvert G_3 \rvert^2}{\mathrm{Im}\, \tau} - \tilde{F}_5^2 \right\}  + \int \frac{1}{\mathrm{Im} \,\tau}\, C_4 \wedge G_3 \wedge \bar{G}_3   \ ,
\ee
where we use units for which $\kappa = 1$ and (as before) $\tilde g_{\ssM\ssN}$ denotes the 10D EF metric (with $g_{\mu\nu}$ reserved for the 4D EF metric $g_{\mu\nu}$ and the 10D SF metric being $\hat g_{\ssM\ssN}$). As usual $\tau = C_0 + i e^{-\phi}$ is the complex scalar 10D axio-dilaton field and $G_3 = F_3 + \tau H_3$ is the complex 3-form field strength, where $F_{p+1} = \exd C_p$, $H_3 = \exd B_2$ and $\tilde F_5 = F_5 - \frac12\, C_2 \wedge H_3 + \frac12 \, B_2 \wedge F_3$. 

This classical bulk action enjoys an accidental $SL(2,\mathbb{R})$ symmetry under which 
\be\label{SL2Rdef}
  \tau \to \frac{a \tau + b}{c \tau + d} \quad \hbox{and} \quad 
  G_3 \to \frac{G_3}{c\tau + d}  \,,
\ee
where $ad-bc =1$. Note that \pref{SL2Rdef} already includes a scale transformation (that does not act on the metric) as the special case $b=c=0$ and $a = 1/d$. As we see in \S\ref{ssec:IIBHiOrd} below, this $SL(2,\mathbb{R})$ symmetry is broken by generic $\alpha'$ and $g_s$ corrections, with the subgroup of axionic shifts of $C_0 = \hbox{Re}\,\tau$ thought to survive to all orders in perturbation theory. A discrete duality subgroup, $SL(2,\mathbb{Z})$ (including the S-duality transformation $\tau \leftrightarrow -1/\tau$ that takes  $\phi \leftrightarrow - \phi$ and so $g_s \leftrightarrow g_s^{-1}$), is also believed to survive non-perturbative corrections, as an exact symmetry.

To identify other classical scale invariances of this system consider scaling the bulk fields with weights assigned as follows
\be
\label{IIBscalings}
 \tilde g_{\ssM\ssN} \rightarrow \lambda^{w_g} \tilde g_{\ssM\ssN} \ , \qquad \tau \rightarrow \lambda^{w_{\tau}} \tau \ , \qquad B_2 \rightarrow \lambda^{w_{B_2}} B_2 \,, \qquad C_p \rightarrow \lambda^{w_{C_p}} C_p \,.
\ee
The weights of the various form fields and dilaton must be related to one another in such a way as to allow all terms in $G_3$ and $\tilde F_5$ to scale the same way which requires
\be \label{F5B2reln}
 w_{C_4} = w_{C_2} + w_{B_2}   \quad \hbox{and} \quad
  w_{C_2}   = w_{\tau} + w_{B_2} \ .
\ee
Under the transformations \pref{IIBscalings} the bulk action \pref{TypeIIBBulk} scales homogeneously
\be
S_{\rm bulk} \rightarrow \lambda^{4w_g} S_{\rm bulk}
\label{10DLscale}
\ee
provided that the various weights are related by the conditions
\be
  w_{C_4} = 2w_g \ , \qquad   w_{B_2} = 2w_g - w_{C_2} \quad \hbox{and} \quad  w_\tau = 2(w_{C_2} - w_g) \ .
\label{weights_F5_etc}
\ee

The above conditions leave two of the weights, $w_g$ and $w := w_{C_2}$ say, arbitrary (although one of these can be set to an arbitrary conventional value as part of the definition of $\lambda$). The particular combination with $w_g = 0$ corresponds to the scale transformation included within the $SL(2,\mathbb{R})$ invariance mentioned above. The other is a scale transformation that in particular scales the metric and so changes the classical action by the non-zero factor given in \pref{10DLscale}. 

\subsubsection{Scaling of localised sources}

The action for a D$p$-brane in 10D EF is $S_{\rm loc} = S_\DBI  + S_\WZ$ where $S_\WZ$ denotes the Wess-Zumino-type contribution (to which we return below) and
\be\label{DBIaction}
S_\DBI  = - \mu_p \int_{\ssS_p} \mathrm{d}^{p+1} \zeta \; \mathrm{Im}(\tau) \left\{-\mathrm{det}\left[ \frac{\tilde g_{ab}}{(\mathrm{Im}\,\tau)^{1/2}} + B_{ab} + 2\pi \alpha' F_{ab} \right] \right\}^{1/2} \,.
\ee
Here $\mu_p$ is the brane tension and the integration is over the $(p+1)$-dimensional brane world-volume $S_p$, whose embedding into 10D spacetime is given by $x^\ssM = y^\ssM(\zeta)$, where the $\zeta^a$ parameterize the brane's world-volume. Bulk fields carry indices $a,b$ ({\it e.g.}~the EF metric, $\tilde g_{ab}$, and Neveu-Schwarz 2-form gauge field, $B_{ab}$) to indicate that they are pulled back to the brane world-volume, as in 
\be
  \tilde g_{ab}(\zeta) = \tilde g_{\ssM\ssN} \partial_a y^\ssM \partial_b y^\ssN  \qquad \hbox{and} \qquad B_{ab}(\zeta) = B_{\ssM\ssN} \partial_a y^\ssM \partial_b y^\ssN \,,
\ee
where the left-hand sides are regarded as functions of $\zeta^a$ while the fields $\tilde g_{\ssM\ssN}$ and $B_{\ssM\ssN}$ on the right-hand side are evaluated at $x^\ssM = y^\ssM(\zeta)$. Similar pull-back factors are not required for $F_{ab}(\zeta)$, since this is the field-strength for open-string gauge fields defined on the brane world-volume. 

A necessary condition for \pref{DBIaction} to be scale invariant is to have $B_{\ssM\ssN}$ and the SF metric $\hat g_{\ssM\ssN} = \tilde g_{\ssM\ssN}/(\hbox{Im }\tau)^{1/2}$ share the same weight, so $w_{B_2} = w_g - \frac12 \, w_\tau$, which is an automatic consequence of eqs.~\pref{F5B2reln} and \pref{weights_F5_etc}.  Furthermore, $F_{ab}$ must share the same weight as the other fields in the determinant, and so
\be \label{openstringmaxwellwt}
 F_{ab} \rightarrow  \lambda^{2w_g-w} F_{ab} \,.
\ee

The DBI part of the action then scales as $S_\DBI  \rightarrow \lambda^{w_\DBI} S_\DBI $ with
\be \label{wDBIdef}
 w_\DBI(p)  
 =   (p-1)w_g - \frac12 (p-3)w   \,.
\ee
As shown in Appendix \ref{app:WZscaling} the Wess-Zumino part of the action scales in precisely the same way (for all $p$) with no additional assumptions.

The localised terms of the action therefore scale like $S_{\rm loc} \to \lambda^{w_\DBI(p)} S_{\rm loc}$, which differs\footnote{In principle one might imagine arranging $S_{\rm loc}$ to scale in the same way as the bulk by scaling the brane position fields, $y^\ssM$, appropriately. In the end such attempts get thwarted by the implicit dependence on $y$ that the bulk fields carry because they are evaluated in $S_{\rm loc}$ at the brane positions.} from the bulk scaling \pref{10DLscale} for all $p$. That there should be a difference in scaling is perhaps most obvious in the case $w = 2w_g$ (for which the SF metric does not scale) since then $w_\DBI = w =2w_g$ (for all $p$) and the transformations $S_{\rm loc} \to \lambda^{w} S_{\rm loc}$ and $S_{\rm bulk} \to \lambda^{2w} S_{\rm bulk}$ simply reflect that in string frame $\cL_{\rm bulk} \propto e^{-2\phi}$ and $\cL_{\rm loc} \propto e^{-\phi}$.  In this sense $S_{\rm loc}$ can be regarded as just a particular piece of the action's general expansion in powers of $g_s$ and $\alpha'$ (as described in more detail in \S\ref{ssec:IIBHiOrd}). 

\subsubsection{Dimensional reduction}

We require the implications of the above 10D scaling symmetries for the low-energy 4D effective theory found after dimensional reduction near a supersymmetric vacuum. Because IIB and heterotic supergravities have different amounts of supersymmetry in 10D the give differing amounts of 4D supersymmetry when compactified. In particular, the simplest Calabi-Yau constructions used to obtain 4D $\cN=1$ supergravity for heterotic vacua lead to $\cN=2$ supergravity\footnote{Residual effects of this would-be $\cN=2$ supersymmetry sometimes leave a residue in compactifications, such as by producing sequestered forms in $K$ for the couplings between various sectors.} when used in IIB supergravity. $\cN=1$ supersymmetry can be obtained once D-branes and orientifold planes are also included.

In what follows we highlight the interplay between background fluxes and the presence of source branes for $\cN=1$ vacua because this is central to the physics of modulus stabilisation, whose properties appear at low energies through the 4D scalar potential.\footnote{Much of the interest in IIB vacua is because this stabilisation physics is so well developed~\cite{Giddings:2001yu, Dasgupta:1999ss}.} 

Similar to heterotic vacua, we consider compactifications with metric
\be
\label{GKP}
 \mathrm{d}\tilde s_{(10)}^2 = \cW^{-1}(z) \tilde g_{\mu \nu}(x)\, \mathrm{d}x^\mu \mathrm{d}x^\nu + \cW(z) \tilde g_{mn} (x,z) \,\mathrm{d} z^m \mathrm{d} z^n \,,
\ee
where $\cW(z)$ is the warp factor (whose detailed form plays no role in what follows because it does not scale). The 4D burden of 10D metric scaling \pref{IIBscalings} is shared by the 4D metric $\tilde g_{\mu \nu} \rightarrow \lambda^{w_g} \tilde g_{\mu \nu}$ and the extra-dimensional dimensionless warped volume moduli
\be
\label{VEweight}
  \mathcal{V}_\ssE^{(n)} := \frac{1}{(\alpha')^3}\int \mathrm{d}^6 y \;\sqrt{-\tilde g_6} \; \cW^n \quad \hbox{for which} \quad
  \cV_\ssE^{(n)} \rightarrow \lambda^{3w_g} \mathcal{V}_\ssE^{(n)} \,,
\ee
for all $n$ [so the superscript `$(n)$' is henceforth dropped]\footnote{Including warping into the volume modulus is convenient once branes are present because it happens that it is the warped 4-cycle volume, $\int_{\cC_4} \exd^4z \, \cW^2$, that appears as the real part of the holomorphic field in expressions like \pref{TvschiaIIB}, once brane back-reaction on the metric is included~\cite{Kachru:2003sx,Giddings:2005ff, Baumann:2006th, Berg:2004ek}.} (which uses $\tilde g_{mn} \rightarrow \lambda^{w_g} \tilde g_{mn}$). The 4D EF metric, $g_{\mu\nu}$, then scales as does $g_{\mu\nu} := \mathcal{V}_\ssE \,\tilde g_{\mu\nu}$, while the 4D effective Lagrangian density inherits the scaling of the 10D action, and so $g_{\mu\nu}  \rightarrow \lambda^{4w_g} g_{\mu\nu}$ and $\cL \to \lambda^{w_\ssL} \cL$, with classical reductions giving
\be \label{wLbulkbrane}
   w_\ssL = 4w_g \;\; \hbox{(bulk)} \quad
    \hbox{or} \quad w_\ssL = (p-1)w_g - \frac12 (p-3)w \;\; \hbox{(brane)}\,. 
\ee

The chiral superfields describing the orientifold-even $\mathcal{N}=1$ K\"ahler moduli in the 4D theory are 
\be \label{TvschiaIIB}
   T_\ssI = \chi_\ssI + i a_\ssI \,,
\ee
where $\chi_\ssI$ denote extra-dimensional 4-cycle volumes and the axions $a_\ssI$ arise as duals of the mixed components, $C_{\mu\nu mn}$ of the Ramond-Ramond 4-form field. These are related to the total (unwarped) volume, $\cV_\ssE^{(0)}$, and the two-cycle volumes, $t^\ssI$, by the standard relations
\be
\mathcal{V}^{(0)}_\ssE = \frac16 \,\kappa_{\ssI\ssJ\ssK} t^\ssI t^\ssJ t^\ssK  \quad \hbox{and} \quad \chi_\ssI = \frac{\partial\cV^{(0)}_\ssE}{\partial t^\ssI} = \frac12 \kappa_{\ssI\ssJ\ssK} t^\ssJ t^\ssK \ ,
\label{2cyclerelations}
\ee
that also follow from the condition that one supersymmetry survives into the 4D world. The calculable coefficients $\kappa_{\ssI\ssJ\ssK}$ are characteristic of the Calabi-Yau space of interest. The scaling weight of $a_i$ therefore is the weight of $C_4$, which eq.~\pref{weights_F5_etc} implies is $w_{C_4} = 2w_g$ while \pref{VEweight} and \pref{2cyclerelations} together show that $\chi_i$ scales consistently, so $T_\ssI \rightarrow \lambda^{2w_g}\, T_\ssI$. 

All told, the scaling of the 10D action implies the 4D Lagrangian obtained by classical reduction transforms as $\cL \to \lambda^{w_\ssL} \cL$ with $w_\ssL$ given by \pref{wLbulkbrane} under the field transformations
\be
g_{\mu\nu} \rightarrow \lambda^{4w_g} g_{\mu\nu} \ , \quad  T_\ssI \rightarrow \lambda^{2w_g}\, T_\ssI \ , \quad
  \tau \rightarrow \lambda^{2(w - w_g)} \tau \ , \quad  Z^\ssA \rightarrow Z^\ssA \ , 
\label{4D_scalingwwg}
\ee
with $Z^\ssA$ collectively denoting all other non-scaling chiral multiplets.

\subsubsection*{No-scale condition}

To make contact with \S\ref{ssec:scalingandnoscale} we temporarily focus on a subset of values for the weights $w$ and $w_g$. We first (in this section only) use the freedom to redefine $\lambda$ to choose $w_g = \frac12$ since this ensures $g_{\mu\nu} \to \lambda^2 g_{\mu\nu}$, as was the convention used in \S\ref{ssec:scalingandnoscale}. 

With this choice the IIB bulk Lagrangian transforms as $\cL_{\rm bulk} \to \lambda^2 \cL_{\rm bulk}$ while \pref{4D_scalingwwg} becomes
\be
g_{\mu\nu} \rightarrow \lambda^{2} g_{\mu\nu} \ , \quad  T_\ssI \rightarrow \lambda\, T_\ssI \ , \quad
  \tau \rightarrow \lambda^{2w - 1} \tau \ , \quad  Z^\ssA \rightarrow Z^\ssA \,.
\label{4D_scaling}
\ee
The imaginary parts of the $T_\ssI$ also inherit a shift symmetry from the freedom to shift $C_4$ by harmonic forms, as was also done in \S\ref{ssec:scalingandnoscale}. To make the comparison with \S\ref{ssec:scalingandnoscale} more explicit choose the special case $w = w_g (=\frac12)$ so that the axio-dilaton is invariant. In this case \pref{4D_scaling} becomes (\ref{sym}) with the prediction $w_\ssT=1$. 

Furthermore, dimensional reduction of the bulk action of IIB supergravity gives the superpotential~\cite{Gukov:1999ya}
\be
W = \int G_3 \wedge \Omega \ ,
\label{WGVW}
\ee
where $\Omega_{ijk}$ is the universal extra-dimensional holomorphic 3-form (whose existence follows on general grounds from the requirements of unbroken $\cN=1$ supersymmetry in 4D~\cite{Candelas:1985en}). This implies the superpotential scales\footnote{We imagine all extra-dimensional harmonic forms are normalised 
so they do not themselves scale.} with weight $w_\ssW = w$:
\be
W \rightarrow \lambda^w W   \,,
\label{Wscaling}
\ee
(and so $w_\ssW = \frac12$ given the present choice $w = \frac12$).

Repeating the arguments of \S\ref{ssec:scalingandnoscale} therefore shows that this particular 10D symmetry implies the 4D compensator and K\"ahler potential must scale as in \pref{TreeExpKSS}, which in turn implies that $e^{-K/3}$ is a homogeneous function of degree $q$, so
\be
K^{\ssI\bar\ssJ} K_\ssI K_{\bar{\ssJ}} =  3 q \qquad \hbox{with} \qquad 
q = \frac{2 \left( 1 + w \right)}{3 w_\ssT}\, = 1\,,
\ee
where the last equality uses $w=w_g = \frac12$ (as well as \pref{4D_scaling}, which tells us $w_\ssT=1$). The no-scale structure for the (orientifold even) K\"ahler moduli of 4D type IIB supergravity (once the other fields are evaluated at their supersymmetric minima, $D_\tau W = D_{Z^\ssA} W = 0$) is thereby seen to be a consequence of the classical scale invariance of the UV theory combined with supersymmetry and axionic shift symmetries. 

\subsubsection{$K$ and $W$ for bulk moduli}

The two 10D scaling symmetries say much more than just that the $T_\ssI$ fields have a no-scale structure. Repeating the arguments of \S\ref{sec:descent2} shows that these two scaling symmetries combine with supersymmetry and axionic shift symmetries to completely fix how $K$ can depend on $\tau$ and one combination of the $T_\ssI$ (and do so order-by-order in the string coupling and $\alpha'$ expansions).  

To show this we return to general $w$ and $w_g$, in which case the metric scaling $g_{\mu\nu} \to \lambda^{4w_g}g_{\mu\nu}$ given in \pref{4D_scalingwwg} implies the superspace fermionic coordinate must scale like $\theta \to \lambda^{w_g} \theta$ to capture the difference between the superpartner scaling properties. Combined with the Lagrangian scaling $\cL \to \lambda^{w_\ssL} \cL$ it follows that the superfields $\mathfrak{F}$ and $\mathfrak{D}$ of \pref{superconf} must scale as
\be \label{FDtrans}
  \mathfrak{F} \to \lambda^{w_\ssL -6 w_g} \mathfrak{F}   \quad \hbox{and} \quad
  \mathfrak{D} \to \lambda^{w_\ssL - 4w_g} \mathfrak{D}   \,.
\ee
To see the utility of these transformation properties consider in turn dimensional reductions of both bulk and brane actions.

\subsubsection*{Orientifold-even moduli}

Consider first the bulk, for which $w_\ssL = 4w_g$ and \pref{FDtrans} becomes
\be \label{FDtransB}
  \mathfrak{F} \to \lambda^{ -2 w_g} \mathfrak{F}   \quad \hbox{and} \quad
  \mathfrak{D} \to   \mathfrak{D}  \qquad \hbox{(bulk)} \,.
\ee
Since \pref{Wscaling} implies $w_\ssW = w$ the K\"ahler potential and compensator scale as
\be \label{BulkCompScale}
  \Phi \to \lambda^{- \frac13(w+2w_g)} \Phi \quad \hbox{and} \quad
   e^{-K/3} \to \lambda^{\frac23(w+2w_g)} e^{-K/3} \quad\hbox{(bulk)} \,.
\ee

To see what this means for the dependence on the moduli, write the 4-cycle volumes, $\chi_\ssI$, as a single scaling field  -- $\cV_\ssE^{2/3}$, say, regarded as the homogeneous degree-one function of the $\chi_\ssI$ found by solving eqs.~\pref{2cyclerelations} -- plus a collection of scale-invariant ratios: $\{ \chi_\ssI \} = \{ \cV_\ssE^{2/3}, \cR_\alpha \}$. Keeping in mind that axionic shift symmetries preclude $K$ from depending on $T_\ssI -\ol T_\ssI$ or $\tau+\bar\tau$, comparing the transformations (\ref{BulkCompScale}) with \pref{4D_scalingwwg} shows $K$ at leading order must be 
\be 
e^{-K/3} =  \mathfrak{K}_{\ssB} \left( \tau-\bar\tau\right)^{1/3} \cV_\ssE^{2/3} \,,
\label{KvstauVEscale}
\ee
where $\mathfrak{K}_{\ssB}$ is an arbitrary function that depends only on the {\em invariant} fields ({\em i.e.}~functions of the $\cR_\alpha$ and of the $Z^a$). Eq.~\pref{KvstauVEscale} is consistent with the expression obtained by dimensional truncation at lowest order in the string and $\alpha'$ expansion
\be \label{KbulkIIB}
 K  = -  \ln \left( \tau-\bar\tau\right) - 2 \ln \cV_\ssE  - \ln\left(\int \Omega \wedge \ol{\Omega}\right)   \,,
\ee
where the holomorphic 3-form $\Omega$ is independent of both $\tau$ and the $T_\ssI$ (which implies $\mathfrak{K}_\ssB$ does not depend on the $\cR_\alpha$).

Notice in particular that the transformation laws (\ref{Wscaling}) and (\ref{BulkCompScale}) together imply that $\cG$, defined in \pref{G}, scales with weight
\be \label{cGscalingIIB}
 e^{\cG} = e^K |W|^2 \rightarrow \lambda^{-4w_g} e^{\cG} \,,
\ee
and so comparison with \pref{4D_scalingwwg} shows that $e^{\cG}$ has the same weight as does $g^{\mu\nu}$. This is a consistency check\footnote{Alternatively, \pref{cGscalingIIB} can be regarded as a way to infer the scaling of $K$ given $W$'s scaling (or vice versa), in a way that avoids use of \pref{superconf}.} since it is necessary for 4D scale invariance, as can be seen from the requirement that terms like $\sqrt{-g}\; e^K |W|^2 \subset V_\ssF$ scale in the same way as does the 4D Einstein action, $\sqrt{-g}\; g^{\mu\nu} R_{\mu\nu}$. 

\subsubsection*{Orientifold-odd moduli: generalised no-scale models}

Type IIB vacua have other low-energy closed-string modes besides the K\"ahler moduli discussed above. These other modes include complex-structure moduli and orientifold-odd moduli (which are distinct from the K\"ahler moduli discussed above by transforming differently under the orientifold involution needed to obtain a chiral $\mathcal{N}=1$ spectrum in type IIB vacua). We briefly pause here to describe these latter moduli because they concretely realise the axionic generalised (as opposed to standard) no-scale form.

We focus specifically on orientifold-odd $(1,1)$-forms can contribute low-energy 4D states in this way, composed of axions dual to 10D 2-forms where one of the axionic shift symmetries is broken by orientifolding. The holomorphic field describing these additional moduli can be written
\be
G^{\hat{\ssI}} = c^{\hat{\ssI}} - \tau\, b^{\hat{\ssI}}   \,,
\ee
where $c^{\hat{\ssI}}$ and $b^{\hat{\ssI}}$ are axions associated with $C_2$ and $B_2$. The presence of these fields in the low-energy EFT changes the relation between the moduli $\chi_\ssI$ and the holomorphic fields $T_\ssI$ to~\cite{Grimm:2004uq, Cicoli:2012vw}  
\be
 \chi_\ssI =  T_\ssI + \ol{T}_\ssI -i\,\kappa_{\ssI \hat{\ssJ}\hat{\ssK}} \, 
\frac{(G^{\hat{\ssJ}} -\ol{G}^{\,\hat{\ssJ}})(G^{\hat{\ssK}}-\ol{G}^{\,\hat{\ssK}})}{2\left(\tau-\bar \tau\right)} \,,
\ee
where all we use about the coefficients (much like those appearing in \pref{2cyclerelations}) is that they do not scale. The K\"ahler potential is now regarded as being a function of the non-axionic fields $T_\ssI + \ol{T}_\ssI$ and $G^{\hat{\ssI}} -\ol{G}^{\,\hat{\ssI}}$. 

The scaling arguments made above go through as before (with $G^{\hat{\ssI}} \to \lambda^w G^{\hat{\ssI}}$), and this is a case where the formulation with linear multiplets is convenient. Constructing the dual theory by dualising $(T_\ssI, G^{\hat{\ssI}})$ to linear multiplets,\footnote{In~\cite{Grimm:2004uq, Grimm:2005fa} the dualisation was performed only for the orientifold-even moduli $T_\ssI$. In this case also a no-scale statement holds, but it is slightly misleading as the moduli $G_{\hat{\ssI}}$ actually correspond to flat directions in the respective Minkowski vacuum.} the dual linear multiplets become (in the notation of Appendix ~\ref{4D_no_scale_app})
\be
L_{(\ssT)}^{\underline \ssI} =  \frac{t^\ssI}{2\mathcal{V}_E} \ , \qquad L^{(\ssG)}_{\underline{\hat{\ssI}}} = - \frac{1}{4 \mathcal{V}_E} \kappa_{\hat{\ssI}\hat{\ssJ}} \, b^{\hat{\ssJ}} \,,
\ee
where $t^\ssI$ are the 2-cycle volumes encountered in \pref{2cyclerelations} and $\kappa_{\hat{\ssJ}\hat{\ssK}} := \kappa_{\ssI \hat{\ssJ}\hat{\ssK}} \, t^\ssI$. The K\"ahler potential (and frame function, $F$, of Appendix ~\ref{4D_no_scale_app}) are
\be
\begin{aligned}
K &= -\ln(\tau-\bar\tau) + \ln\left(\tfrac{4}{3}\kappa_{\ssI\ssJ\ssK} L_{(\ssT)}^{\underline \ssI} L_{(\ssT)}^{\underline \ssJ} L_{(\ssT)}^{\underline \ssK} \right) + K_{\rm cs}(U^\ssA,\ol{U}^\ssA) \\
F &= \frac{16 \mathcal{K}}{3} \; \kappa^{\hat{\ssI}\hat{\ssJ}} L^{(\ssG)}_{\underline{\hat{\ssI}}} L^{(\ssG)}_{\underline{\hat{\ssJ}}} + \frac{64  \mathcal{K}^2}{3 \,\mathrm{Im}\,\tau}\;  \kappa_{\ssI \hat{\ssJ}\hat{\ssK}} \kappa^{\hat{\ssJ}\hat{p}} \kappa^{\hat{\ssK}\hat{q}} L_{(\ssT)}^{\underline \ssI} L^{(\ssG)}_{\underline{\hat{p}}} L^{(\ssG)}_{\underline{\hat{q}}} \,,
\end{aligned}
\ee
where $\mathcal{K} = (\tfrac{4}{3} \kappa_{\ssI\ssJ\ssK} L_{(\ssT)}^{\underline \ssI} L_{(\ssT)}^{\underline \ssJ} L_{(\ssT)}^{\underline \ssK})^{-1/2}$ and $\kappa^{\hat{\ssI}\hat{\ssJ}}$ denotes the inverse of $\kappa_{\hat{\ssI}\hat{\ssJ}}$ (understood as a function of $L_{(\ssT)}^{\underline \ssI}$). 

These expressions have the form of eq.~\eqref{eq:no_scale_lm_explicit_app} and coordinate degeneracy is manifest via the property in eq.~\eqref{eq:coord_degen_app}. Furthermore, $e^K$ is a homogeneous function of degree $3$ in the linear multiplets with condition (\ref{eq:weight_requirement_app}) --- {\it i.e.}~$w_\ssT = \frac23(1+w)$ --- satisfied for $w_\ssT=1$ and $w=1/2$. This in turn implies that the orientifold-even and orientifold-odd moduli satisfy the no-scale relation (\ref{eq:no_scale_lm_app}) in terms of the linear multiplets $L_{(\ssT)}^{\underline \ssI}$ and $L^{(\ssG)}_{\underline{\hat{\ssI}}}$, which is equivalent to the more complicated version (\ref{noscalechiralcomplicated3}) in terms of the original chiral multiples $T_\ssI$ and $G^{\hat{\ssI}}$. We see that type IIB models with orientifold-odd moduli furnish at tree-level an example of generalised (axionic) no-scale structure, as claimed.

\subsection{Open-string moduli}
\label{ssec:IIBopenstringmod}

Localised sources contribute their own moduli to the low-energy EFT, in addition to the contributions they make to the action of bulk moduli discussed above. This section extends our earlier considerations to include the implications of the scaling symmetries for these moduli. These include D3/D7-brane gauge potentials and position moduli as well as D7 Wilson-line moduli.

\subsubsection{Gauge kinetic function}

The simplest light brane-localised degree of freedom to handle are 4D gauge fields, for which scaling constrains how bulk moduli can appear in the kinetic function $\mathfrak{f}_{ab}$. Because these come at leading order from classical reductions of the brane actions the only change from the arguments used above is the need to use a different Lagrangian scaling weight, $w_\ssL   =   (p-1)w_g - \frac12 (p-3)w$, given in \pref{wLbulkbrane}.

We start with the gauge kinetic function, $\mathfrak{f}_{ab}$, which because $\mathfrak{F}_g = \mathfrak{f}_{ab} \cW^a \cW^b$ requires the scaling properties of $\cW^a$. If $F_{\mu\nu} \to \lambda^{w_\ssA} F_{\mu\nu}$ then the superfield $\cW \ni  F_{\mu\nu} \Gamma^{\mu\nu} \theta$ must scale like $\cW \to \lambda^{w_\ssA - 3 w_g} \cW$, and so using this with \pref{FDtrans} shows the gauge coupling function therefore scales with weight $w_{\mathfrak{f}} = (w_\ssL - 6 w_g) - 2(w_\ssA - 3w_g) = w_\ssL - 2w_\ssA$.  Using our earlier result \pref{openstringmaxwellwt} that $w_\ssA = 2w_g - w$ for gauge fields on D$p$-branes as well as the brane Lagrangian transformation \pref{wLbulkbrane} then implies $\mathfrak{f}_{ab}$ scales with weight 
\be \label{gaugekinweight}
  w_{\mathfrak{f}} =   (p-5) w_g - \frac12( p-7 )   w \,,
\ee
which for D3 and D7 branes implies the scalings 
\be \label{fabD3D7scale}
  \mathfrak{f}_{\alpha\beta} \to \lambda^{2(w-w_g)} \mathfrak{f}_{\alpha\beta} \quad \hbox{(D3)} \qquad
  \hbox{or} \qquad  \mathfrak{f}_{\alpha\beta} \to \lambda^{2w_g} \mathfrak{f}_{\alpha\beta} \quad \hbox{(D7)}  \,.
\ee
These respectively scale like $\tau$ (D3) and like $T_i$ (D7), consistent with direct dimensional reduction. 

The usual non-renormalisation theorems follow once the appropriate transformation properties under the axionic shift symmetries are imposed. These state that $\mathfrak{f}_{ab}$ cannot depend on fields whose shift symmetries are exact and can depend at most linearly on fields whose shift symmetries have a gauge anomaly.

\subsubsection{Brane contributions to the K\"ahler potential} 

Position moduli associated with the transverse position field, $y^k$, of space-filling branes are perhaps the next simplest cases of brane-localised states. Low-energy couplings of these moduli with closed-string (bulk) modes are constrained by the scaling properties of brane contributions to the 4D supergravity K\"ahler potential $K$. 

Repeating the arguments given shows how brane contributions in the 4D supergravity K\"ahler function scale. Using \pref{BulkCompScale} for the scaling of $\Phi$ and using $w_\ssL$ from \pref{wLbulkbrane} in \pref{FDtrans} for $\mathfrak{D} = e^{-K/3} \, \ol\Phi \Phi$ shows that $e^{-K/3} \to \lambda^{w_\ssK} e^{-K/3}$ with weight
\be \label{KScaleGenBr}
   w_\ssK = \left( p - \frac{11}3 \right) w_g - \frac12\left(p-\frac{13}3 \right) w \,.
\ee
For D3 and D7 branes this implies the transformation properties
\be \label{D3D7Kscale}
  e^{-K/3} \to \lambda^{2(w - w_g)/3} e^{-K/3} \;\; \hbox{(D3)} \quad \hbox{and} \quad
  e^{-K/3} \to \lambda^{2(5w_g-2w)/3} e^{-K/3} \;\; \hbox{(D7)} \,.
\ee
Comparing with the transformation properties of the bulk moduli \pref{4D_scalingwwg} shows that the brane contribution to the low-energy 4D supergravity action has the general form
\be\label{D3D7KSymForm}
   e^{-K/3} = \mathfrak{K}_3 \; (\tau - \bar\tau)^{1/3} \;\; \hbox{(D3)}\quad \hbox{and} \quad
   e^{-K/3} = \mathfrak{K}_7 \; (\tau-\bar\tau)^{-2/3} \cV_\ssE^{2/3} \;\; \hbox{(D7)} \,,
\ee
where $\mathfrak{K}_p$ again denote arbitrary scale invariant functions. 

\subsubsection*{D3 position moduli}

The simplest comparison of these results is with leading-order calculations of the low-energy action for the transverse position field, $y^k$, for a single space-filling D3-brane situated at a point in the extra dimensions. For D3 branes this can be compared with explicit truncation of higher-dimensional actions, which when expanded in powers of $y^i$ leads to a K\"ahler function of the form~\cite{Grana:2003ek}
\be \label{prob1}
 e^{-K/3} = (\tau - \bar \tau)^{1/3} \Bigl[\cV_\ssE^{2/3} - \omega_{i\bar\jmath} \, y^i \bar y^{j} \Bigr] \,,
\ee
where $y^i$ are the complex brane position moduli and $\omega_{i\bar\jmath}$ is a non-scaling extra-dimensional 2-form evaluated at the D3 brane position. The $\cV_\ssE^{2/3}$ term in the square bracket of \pref{prob1} comes from the bulk action, as in \pref{KvstauVEscale}. The rest comes from the brane action and agrees with \pref{D3D7KSymForm}, with the choice $\mathfrak{K}_3 = - \omega_{i\bar\jmath} \, y^i \bar y^{j}$ for the scale-invariant function. 
 
Although scale invariance in itself does not constrain at all the dependence of $K$ on the brane positions, other symmetries can. Similar to the heterotic case, the sequestered form \pref{prob1} can be regarded as being a 4D residue of an underlying accidental shift symmetry (as in \S\ref{sssec:SStransProp}). Appendix ~\ref{app:accidentalshift} traces how this symmetry arises from the 10D point of view, and in the simplest case where there is only one K\"ahler modulus, $T = \cV_\ssE^{2/3} + i b$, this symmetry has the holomorphic form
\be \label{lowestyIIBshift}
  \delta y^i = a^i   \quad \hbox{and} \quad \delta T = \omega_{i\bar\jmath} \, \bar a^j y^i \,,
\ee
to lowest order in powers of $y^i$. Here $a^i$ is a complex constant shift parameter. As we see below in \S\ref{sssec:BranyW}, for IIB theories these shifts are less badly broken by background fields than in the heterotic case, because (unlike for gauge fields) $y^i$ does not participate in the Calabi-Yau construction's identification of background gauge and spin connections. 

As Appendix ~\ref{app:accidentalshift} also argues, these symmetry arguments plausibly extend beyond leading order in powers of $y$, modifying \pref{lowestyIIBshift} to 
\be \label{betteryIIBshift}
  \delta y^i = \xi^i(y)   \quad \hbox{and} \quad \delta T = r_\xi(y) \,,
\ee
where $\xi^i(y)$ is the universal Reeb holomorphic Killing vector of the supergravity target space corresponding to an $R$-symmetry\footnote{$\xi^i$ is {\it not} a Killing vector of the Calabi-Yau space itself (which does not exist) because the Calabi-Yau space is only the projection of the supergravity target space after removal of the compensator field (see {\it e.g.}~\cite{FvP}).} under which the Calabi-Yau K\"ahler potential transforms as $\delta k(y,\bar y) = r_\xi(y) + \ol{r_\xi(y)}$. (Transformation \pref{lowestyIIBshift} is retrieved from \pref{betteryIIBshift} at leading order in the expansion about the brane's background position, $y^i = y^i_{(c)} + \mfy^i(x)$, with $a^i = \xi^i(y_{(c)})$.) This supports the arguments of~\cite{Kachru:2003sx,Giddings:2005ff, Baumann:2006th, Baumann:2007ah} that including higher powers of $y^i$ modifies \pref{prob1} into the expression
\be \label{Kahlerkahler}
 e^{-K/3} = (\tau - \bar \tau)^{1/3} \Bigl[  \cV_\ssE^{2/3} - k(y,\bar y) \Bigr] \,,
\ee
where $k(y,\bar y)$ is the K\"ahler potential of the Calabi-Yau space itself that satisfies $\omega_{i \bar\jmath} = \partial_i \partial_{\bar\jmath} \, k$. This form is again consistent with the scaling predictions \pref{KvstauVEscale} (bulk) and \pref{D3D7KSymForm} (brane) for the specific choice for $\mathfrak{K}_3 = - k(y,\bar y)$. 

The virtue of having a symmetry derivation with a 10D provenance is that it makes it easier to identify the leading order where symmetry-breaking effects can arise. Although a definitive determination of the robustness of \pref{Kahlerkahler} to $\alpha'$ and $g_s$ expansions is not yet possible, one might hope that because the underlying symmetry leading to \pref{lowestyIIBshift} and \pref{betteryIIBshift} is the axionic shift symmetry, $b \to b +\,$(constant) it might hold to all orders in $\alpha'$. Explicit calculations~\cite{Berg:2005yu} show that symmetries \pref{lowestyIIBshift} and \pref{betteryIIBshift} do not survive string-loop corrections, however, since these generate a $y$-dependence in a way that does not arise purely in the combination $\cV_\ssE^{2/3} - \cC_{i\bar\jmath}\, y^i \bar y^j$.

Since $\cV_\ssE^{2/3}$ is homogeneous degree one in the K\"ahler moduli $T_\ssI$ it provides a standard (non-scaling) no-scale example for which both the $T_\ssI$ and $y^k$ are flat directions. This is most easily seen for the case where only a single K\"ahler modulus exists, since in this case the K\"ahler potential becomes
\be
K=- \ln(\tau - \bar \tau) - 3\ln\left[T+\bar T-k(y,\bar y)\right] 
\label{KTphi}
\ee
and so has the form given in (\ref{KTDelta}). This is true for {\it any} Calabi-Yau space despite its K\"ahler potential being potentially very complicated. In this case the interplay between scaling-, shift- and super-symmetry ensure all of the corrections in powers of $y$ that take one from \pref{prob1} to \pref{Kahlerkahler} -- at fixed order in $\alpha'$ and $g_s$ -- preserves the no-scale structure, despite classical bulk scale invariance not being an invariance of the brane and bulk Lagrangians, providing an example of no-scale persistence in the presence of corrections. As we see below this no-scale robustness also survives the leading $\alpha'$ corrections to \pref{Kahlerkahler}.  

Of course, none of this says that flatness in the $y^k$-directions cannot be lifted in the underlying 10D theory. One way to do so is to include supersymmetry-breaking imaginary anti self dual (IASD) background fluxes -- {\it c.f.}~\cite{ibanezuranga}. Another approach places stacks of branes at coincident points in the extra dimensions, allowing inter-brane open-string exchange to generate a low-energy superpotential~\cite{Grana:2003ek} (whose scaling can also be understood -- see \S\ref{sssec:BranyW} below).

\subsubsection*{D7 position moduli}

A similar comparison exists for the moduli associated with positions of space-filling D7 branes that are wrapped about extra-dimensional 4-cycles. The moduli correspond to expansions of the 2-form $\omega_{mn}(y) := \Omega_{mnp} y^p$ in powers of harmonic 2-forms, $\omega^{\ssA}_{mn}$, with
\be \label{yAdef}
  \omega_{mn}(y) := \Omega_{mnp} y^p = y_\ssA \omega^\ssA_{mn} \,.
\ee
The natural metric for the kinetic terms of the $y_\ssA$ in the 4D theory is 
\be \label{D7modmet}
  \cG^{\ssA\ssB} \propto \int_{\cC_4}  \omega^\ssA \wedge \omega^\ssB 
\ee
where the integral is over the extra-dimensional 4-cycle $\cC_4$ about which the brane is wrapped, pulled back to its world-volume. So far as scaling is concerned what is important is that none of $y^i$, $\Omega_{mnp}$ or $\omega^\ssA_{mn}$ scale. 

Explicit semiclassical dimensional reduction~\cite{Jockers:2004yj, Jockers:2005zy} reveals the contribution of $y_\ssA$ to $K$, which is quoted as
\bea \label{7prob1}
 e^{-K/3} &\propto& \Bigl[\tau - \bar \tau -  \cG^{\ssA\bar\ssB} \, y_\ssA \bar y_\ssB \Bigr]^{1/3} \cV_\ssE^{2/3} \nn\\
 &\simeq&  ( \tau - \bar \tau )^{1/3} \cV_\ssE^{2/3} - \frac{1}3 \, \cG^{\ssA\bar\ssB} \, y_\ssA \bar y_\ssB\; \frac{\cV_\ssE^{2/3}}{(\tau - \bar\tau)^{2/3}} + \cdots\,,
\eea
where we switch to complex coordinates and the second line here keeps only the leading powers of $y_\ssA$, since these are what were actually computed. As before, the first term gives the contribution of the bulk action and the second is consistent with the scaling prediction \pref{D3D7KSymForm} for contributions from the brane action, this time with the invariant function $\mathfrak{K}_7 =  -\frac{1}3 \, \cG^{\ssA\bar\ssB} \, y_\ssA \bar y_\ssB$. 

The first line of \pref{7prob1} is again suggestive of an accidental shift symmetry, and Appendix ~\ref{app:accidentalshift} indeed shows how the same arguments used for D3 branes also produce a similar symmetry for D7 position moduli. While in the D3 case the brane WZ couplings realise non-linear shift symmetries for axions coming from the field $C_4$, the D7 case does so for shifts of the axion, $C_0$, that sits inside $\tau$, implying a similar accidental symmetry
\be
   \delta y_\ssA = a_\ssA  \quad
    \hbox{and} \quad \delta \tau = -i \cG^{\ssA \bar\ssB} \bar a_\ssB y_\ssA \,.
\ee
 
The result \pref{7prob1} is not a no-scale form, showing that D7 position moduli are more easily stabilised by the interplay between brane wrapping and background fluxes. As we see below they in general also appear in the low-energy superpotential \pref{WGVW} generated at tree level by extra-dimensional fluxes. 

\subsubsection*{D7 Wilson-line moduli} 

Wilson-line moduli form a category of D7 moduli that also correspond to flat directions (and so appear in no-scale form in their contributions to the effective 4D supergravity. Wilson-line moduli are massless states that arise in the brane-based gauge sector because of the presence of harmonic 1-forms. Although harmonic 1-forms do not exist for Calabi-Yau spaces, they can exist in a 4D sub-surface of a Calabi Yau that is spanned within the extra dimensions by a stack of space-filling D7 branes. 

Consider then such a D7 stack wrapped around an extra-dimensional 4-cycle $\mathcal{C}_4$. The zero-mode expansion of the on-brane gauge field $A_\ssM (x,z)$ is
\be
A(x,z) =A_\mu (x) \exd x^\mu + a_\ssI (x) \cA^\ssI (z) + \bar{a}_{\bar{\ssJ}}(x) \bar{\cA}^{\bar{\ssJ}}(z)~,
\label{WilsonDef}
\ee
where $x^\mu$ and $z^m$ respectively denote the four large and four extra-dimensional on-brane directions (with brane coordinates $\zeta$ chosen so that $x^\mu = y^\mu(\zeta) = \zeta^\mu$ and $z^m = y^m(\zeta) = \zeta^m$). Here $A_\mu(x)$ appears as a spin-one gauge potential in the low-energy 4D theory, while $a_\ssI(x)$ are the spinless 4D Wilson line moduli. The  $\cA^\ssI$ represent a basis of harmonic 1-forms that live on the surface $\mathcal{C}_4$. 

Explicit dimensional reduction of the DBI action gives the following form for the K\"ahler potential for the D7 Wilson line moduli,
\be
e^{-K/3} = (\tau-\bar{\tau})^{1/3}\left[\cV_\ssE^{2/3} - c_\ssW\,\mathcal{C}^{\ssI\bar{\ssJ}} a_\ssI \bar{a}_{\bar{\ssJ}}\right]
\label{KW}
\ee
where $c_\ssW$ is a constant and $\mathcal{C}^{\ssI\bar\ssJ}$ is given by 
\be
\mathcal{C}^{\ssI\bar{\ssJ}} = \int_{\mathcal{C}_4} \omega_2 \wedge \cA^\ssI \wedge \bar{\cA}^{\bar{\ssJ}}
\ee
where $\omega_2$ is the pull-back onto the D7 world-volume of the 2-form $\omega_2$ that is Poincar\'e dual to the 4-cycle $\mathcal{C}_4$. 

As usual the first term in \pref{KW} is the bulk contribution, and the second term is the brane contribution. This once again is consistent with the prediction \pref{D3D7KSymForm}, though in a slightly less trivial way: in this case the required invariant function is
\be
  \mathfrak{K}_7 = - c_\ssW\, \cC^{\ssI\bar\ssJ} a_\ssI \bar{a}_{\bar\ssJ} \,\left( \frac{ \tau - \bar \tau  }{ \cV_\ssE^{2/3} } \right) \,,
\ee
which is scale-invariant because the gauge-field scaling rule \pref{openstringmaxwellwt} implies $A_\ssM$ (and so also $a_\ssI$) scales with weight $2w_g-w$ (with $\cC^{\ssI\bar\ssJ}$ scale invariant).  Because \pref{KW} implies
\be
K = - \ln (\tau-\bar{\tau}) - 3 \ln \left[\cV_\ssE^{2/3} - c_\ssW\,\mathcal{C}^{\ssI\bar{\ssJ}} a_\ssI \bar{a}_{\bar{\ssJ}}\right] \,,
\label{KWlog}
\ee
and this has the same form as \pref{KTDelta} --- at least once $\tau$ is eliminated using $D_\tau W = 0$ --- Wilson-line moduli provide yet another example of a standard no-scale compactification~\cite{Jockers:2004yj}. The resemblance between \pref{KW} and the result \pref{prob1} for D3-brane moduli also makes sense since these moduli are related to these by T-duality.   

Table~\ref{tab:survey} summarises the no-scale behaviour of both closed and open string moduli of type IIB 4D models with the presence at the tree-level of background fluxes according to the classification presented in \S\ref{sec:noscaleprops}.

\begin{table}[!htb]
 \centering
 \begin{tabular}{ c  c  c }
 Modulus & Weight & No-scale type \\ \hline
 Dilaton & $2(w-w_g)$ & Not no-scale \\
Complex str., D7-deformations & $0$ & Not no-scale \\
 Orientifold-even K\"ahler moduli  & $2 w_g$ & Scaling no-scale \\
 D3-position & $0$ & Standard no-scale \\ 
 D7 Wilson-line & $2w_g-w$ & Standard no-scale \\ 
 Orientifold-odd K\"ahler moduli  & $w$ & Generalised (axionic) no-scale \\ 
 \end{tabular}
  \caption{Summary of scaling and no-scale behaviour of type IIB closed string and D3/D7-brane open string moduli in the presence of background fluxes.}
  \label{tab:survey}
\end{table}

\subsubsection{Brane contributions to the superpotential}
\label{sssec:BranyW}

The bulk transformation law \pref{BulkCompScale} for $\Phi$ and the scaling rule \pref{FDtrans} for $\mathfrak{F}$ found using the scaling weight $w_\ssL$ of \pref{wLbulkbrane} for the brane Lagrangian also fixes the scaling weight of superpotential contributions obtained by dimensionally reducing the brane action:
\be
  w_\ssW = (w_\ssL - 6w_g) - 3\left[-\frac13 ( w+2w_g) \right]  = \frac12 (p-5)(2w_g - w)  \,. 
\ee
Recalling that  \pref{openstringmaxwellwt} assigns the brane gauge fields the scaling weight $w_\ssA = 2w_g - 2$, this implies that brane contributions to $W$ scale as
\be \label{D3D7WTrans}
  W \to \lambda^{-w_\ssA} W \;\; \hbox{(D3)} \qquad \hbox{and} \qquad
  W \to \lambda^{+w_\ssA} W \;\; \hbox{(D7)} \,.
\ee

The first of these agrees with explicit calculations for stacks of $N$ D3 branes, for which the brane positions, $Y^i$, take values in the Lie algebra of $U(N)$, $SO(N)$ or $USp(N)$ -- depending on the orientifold involution -- and have a superpotential of the form \cite{Jockers:2005pn} 
\be \label{WD3stack}
  W \propto f\, \hbox{Tr} \Bigl( Y^i  \, [ Y^j  \,, Y^k ] \Bigr) \Omega_{ijk} \,,
\ee
where $\Omega_{ijk}$ is the Calabi-Yau holomorphic 3-form evaluated at the brane position. $f$ is the dimensionless non-Abelian spurion defined as in the discussion around \pref{nonabelianspurion} in the heterotic case (which arises when one promotes a scaling rule for the non-Abelian field strength $F = \exd A + f\, A^2$ to a scaling rule for the gauge potential $A$). Since $f$ always scales to cancel scalings of the gauge field the spurion has weight $- w_\ssA$. This, together with scale invariance of the brane positions, $Y^k$, ensures \pref{WD3stack} is consistent with \pref{D3D7WTrans}.

The superpotential of \pref{WD3stack} is also consistent with a multi-brane generalisation of the D3 shift symmetry \pref{lowestyIIBshift}:
\be \label{lowestyIIBshiftnonabel}
  \delta Y^i = a^i  \quad \hbox{and} \quad \delta T = \omega_{i\bar\jmath} \,\hbox{Tr} \,(  \bar a^j Y^i ) \,,
\ee
where the constant shift parameter, $a^i$, is proportional to the unit matrix.

A similar story goes through for D7 moduli. Direct dimensional reduction of the D7 action reveals a contribution to the superpotential of the form \cite{Jockers:2005pn}
\be\label{D7Wformula}
 W \sim  \int_{\cC_4} F_2 \wedge \omega(y) 
 \ee
where the integration is over the 4-cycle $\cC_4$ wrapped by the space-filling D7 brane and $F_2$ is the on-brane gauge field-strength. The 2-form $\omega(y)$ is as defined in \pref{yAdef}. Again we find consistency with \pref{D3D7WTrans} given that $\omega(y)$ is scale-invariant and \pref{openstringmaxwellwt} implies $F_2$ scales with weight $w_\ssA$.
 
\subsection{Flux quantisation and space-filling 4-forms}
\label{sssec:SpaceFill}

Before extending the above to open-string moduli and to higher orders in $g_s$ and $\alpha'$ we pause to highlight a part of the low-energy EFT that plays an important conceptual role in allowing it capture the physics of its higher-dimensional UV completion. This part of the EFT is also worth attention because it has unusual features that may prove to be part of any ultimately successful approach to naturalness issues. (Some of these were flagged as potentially useful for natural issues in inflationary models in~\cite{Kaloper:2008fb}.)

The starting point is recognition of the important role played in the UV theory by flux quantisation in the extra dimensions. Competition between flux quantisation and brane wrapping is part of what `stiffens' the extra dimensions against deformations and thereby allows its various moduli to be stabilised~\cite{Giddings:2001yu}. Many low-energy properties depend strongly on the values of these fluxes ({\it e.g.} 4D supersymmetry might be broken for some values of fluxes and not for others). This is seen explicitly from formulae like \pref{WGVW} which give the low-energy superpotential explicitly in terms of extra-dimensional form-fields like $G_3$, whose expectation values are quantised within topologically non-trivial extra dimensions. Flux quantisation is also important for understanding scaling properties since quantised background values can break the scaling of non-trivially scaling form-fields, like $C_p$ and $B_2$.  

From the low-energy perspective the puzzle is: how does the 4D EFT `learn' about flux quantisation? It cannot do so directly through expressions like \pref{WGVW}, which gives $W$ in terms of integrals over extra-dimensional fields like $F_3$, and $H_3$ that do not themselves appear as low-energy fields in the 4D theory. From the 4D point of view the dependence of expressions like \pref{WGVW} on low-energy fields is only given implicitly inasmuch as the background values for $F_3$, and $H_3$ also depend on some of the low-energy moduli. This cannot be the whole story because the 4D effective theory should be self-contained, and not require detailed knowledge of UV fields that are already integrated out. 

It turns out that there is a general mechanism whereby 4D EFTs learn about extra-dimensional flux quantisation, both for electromagnetic flux in phenomenological compactifications~\cite{Burgess:2015lda} and for the more general higher-dimensional string vacua of interest here~\cite{Bousso:2000xa,Bielleman:2015ina, Herraez:2018vae}. They do so through the appearance in the 4D EFT of space-filling 3-form gauge potentials, $\cB_{\mu\nu\lambda}^{(a)}$, whose presence usually is not that interesting in 4D because it does not describe a propagating degree of freedom. But although 3-form potentials do not propagate, their field strengths do capture all of the magic of the higher-dimensional flux quantisation. 

\subsubsection{Space-filling 4-forms}

Space-filling 4-forms are generic whenever extra-dimensional form fields are present (such as those whose background fluxes are quantised). To see why, consider the example relevant to the IIB vacua of interest here, where the background value $\oint_\cC G_3$ is quantised for some topologically non-trivial 3-cycle $\cC$ in the extra dimensions. (There are a number of such cycles, for Calabi-Yau spaces given by the Hodge numbers, $h^{3,0} = 1$ and $h_{2,1} \geq 0$, that respectively count the number of independent harmonic holomorphic $(3,0)$ forms, $\Omega_{ijk}$, and $(2,1)$ forms, $\beta^{(a)}_{jk\bar \imath}$, with $a = 1, \cdots, h_{2,1}$.) In 10D $G_3$ dualises to a 7-form field strength,\footnote{As before, the hat on $\hat *$ indicates duality is performed using the 10D SF metric.} $G_7 = \hat * G_3$, whose components in the large non-compact 4 dimensions are space-filling. In particular, the decomposition
\be \label{G7vsG4harm}
    G_7 = G_4^{(0)} \wedge \Omega + G_4^{(a)} \wedge \ol \beta^{(a)} \,, 
\ee
shows how $G_7$ produces a variety of 4D 4-form field-strengths, $G^{(0)}_4$ and $G^{(a)}_4$ in the low-energy 4D theory. At lowest order these appear quadratically in the action in the dimensional reduction of the $\ol G_3 G_3$ kinetic terms of the higher-dimensional bulk action.  

To see how such terms influence low-energy four-dimensional dynamics, we follow~\cite{Burgess:2015lda} and consider a stripped-down example of the 4D Lagrangian that results if there were only one such a 4-form field: 
\bea \label{EF4D}
  \cL_4 &=& - \sqrt{-  g} \; \left[ \frac{1}{2\kappa_4^2} \,  g^{\mu\nu} \Bigl(  R_{\mu\nu} + \cG_{ab} \,\partial_\mu \chi^a \, \partial_\nu \chi^b \Bigr) + V(\chi) \right. \\
  && \qquad\qquad\qquad \left. + \frac{1}{2\cdot 4!} \, Z(\chi) \, G_{\mu\nu\lambda\rho} G^{\mu\nu\lambda\rho} - \frac{1}{4!} \, X(\chi) \, \epsilon^{\,\mu\nu\lambda\rho} G_{\mu\nu\lambda\rho} \right] + \cL_{st4} \,,\nn
\eea
where $g_{\mu\nu}$ is the 4D EF metric and $\chi^a$ represent 4D scalars (such as moduli or axions) while $G_4 = \exd \cB_3$ is the 4-form field strength. The functions\footnote{These functions become matrices when multiple 4-forms are present, but the manipulations presented here go through unchanged.} $\cG_{ab}(\chi)$, $V(\chi)$, $Z(\chi)$ and $X(\chi)$ are calculable (in principle) by dimensionally reducing the full theory (see~\cite{Burgess:2015lda} for a worked example of how this is done). Notice in particular that the presence of $\chi^a$ in these functions -- and in $X$ in particular -- generically breaks any axionic shift symmetry it might have had, bringing the news to 4D of how flux quantisation effects can remove the protection such shift symmetries would otherwise have given.

The last term in \pref{EF4D} is a `surface' term, $\cL_{st4}$, defined by
\be \label{Lst4def}
  \cL_{st4} := \frac{1}{3!} \, \partial_\mu \Bigl( \sqrt{-g} \; Z\,  \check G^{\mu\nu\lambda\rho} \cB_{\nu\lambda\rho} \Bigr) \,,
\ee
that arises when explicitly performing the duality transformation in the higher-dimensional theory (and can be regarded as a Gibbons-Hawking term for the 3-form potential), with
\be
 \check G_{\mu\nu\lambda\rho} := G_{\mu\nu\lambda\rho} - \frac{X}{Z} \, \epsilon_{\mu\nu\lambda\rho} \,.
\ee
Although \pref{Lst4def} is a total derivative it must be kept when evaluating the action at a solution to the field equations (as is done when integrating out the 4-form) to the extent there are boundaries (including spatial infinity). Spatial infinity is effectively always a non-negligible boundary because the field $G_{\mu\nu\lambda\rho}$ does not fall off at infinity as other fields normally would~\cite{Duff:1989ah, Duff:1995wd,Bousso:2000xa, Burgess:2015lda}. 
 
The field equations for $\cB_{\nu\lambda\rho}$ obtained by varying \pref{EF4D} are 
\be \label{4Dgaugeeq}
 \partial_\mu \Bigl[ \sqrt{-  g} \; Z \, \check G^{\mu\nu\lambda\rho}  \Bigr] =  \partial_\mu \Bigl[ \sqrt{-  g} \; \Bigl(Z \, G^{\mu\nu\lambda\rho} - X \,  \epsilon^{\,\mu\nu\lambda\rho}  \Bigr) \Bigr] = 0 \,.
\ee
Writing $G_{\mu\nu\lambda\rho} = f_4 \,  \epsilon_{\mu\nu\lambda\rho}$ shows that $Z f_4 - X$ is a constant (and so does not propagate). It is instead an integration constant, $K_4$, with
\be \label{f4express}
 f_4 = \frac{K_4 + X}{Z} \,.
\ee
The constant $K_4$ is fixed by matching to the full theory~\cite{Burgess:2015lda}, and it is through this comparison -- plus any quantisation of coefficients in $X(\chi)$ -- that one learns from the UV theory that $f_4$ takes quantised values.\footnote{Refs.~\cite{Bielleman:2015ina, Herraez:2018vae} note that the integration constants $K_4$ vanish in the examples they consider, but this is likely to be an artefact of the choices of fluxes investigated there (in particular because of the assumption of no IASD fluxes, so that 4D supersymmetry is broken only by fields in the 4D effective theory). $K_4 \neq 0$ in the example studied in~\cite{Burgess:2015lda}.}

Integrating out the 4-form field amounts to eliminating it\footnote{It is for this step that the surface term, $\cL_{st4}$, plays a crucial role~\cite{Duff:1989ah}.} using \pref{f4express}, leading to the first line of \pref{EF4D} with the replacement $V(\chi) \to U(\chi)$ with
\be\label{newU4D}
 U(\chi) := V(\chi) + \frac{1}{2 Z(\chi)} \Bigl[ K_4 + X(\chi) \Bigr]^2 \,.
\ee
This amounts to a simple change in the scalar potential, but in a way that knows (through $K_4$ and $X$) about extra-dimensional flux-quantisation. If flux-quantisation were not important and if scalar potentials were arbitrary then there would be no loss of generality in always integrating out 4-form fields and starting with a slightly different scalar potential. 

From the point of view of scale invariance, the 4D Lagrangian \pref{EF4D} scales as $\cL \to \lambda^2 \cL$ when the fields rescale as
\be
   g_{\mu\nu} \to \lambda^2 g_{\mu\nu} \,, \quad \hbox{and} \quad 
   G_{\mu\nu\lambda\rho} \to \lambda^{w_\ssG}  G_{\mu\nu\lambda\rho} \,,
\ee
provided that there exists a transformation of $\chi^a$ for which
\be \label{zetascaling}
  V(\chi) \to \lambda^{-2} V(\chi) \,, \quad Z(\chi) \to \lambda^{6-2w_\ssG}Z(\chi) \quad \hbox{and}\quad
  X(\chi) \to \lambda^{2-w_\ssG} X(\chi) \,,
\ee
and $\cG_{ab} \, \partial_\mu \chi^a \partial_\nu \chi^b$ is invariant. (These transformation properties are often inherited from the scaling properties of the higher-dimensional UV completion.) Eqs.~\pref{zetascaling} imply in particular that $(X^2/Z) \to \lambda^{-2}(X^2/Z)$, and so scales in the same way as does $V(\chi)$. Integrating out the 4-form field can therefore preserve scale-invariance, but only if the integration constants $K_4$ vanish. It is the presence of $K_4$ that passes the scale-breaking of flux quantisation down to the 4D EFT.

Notice also that the dependence of $U(\chi)$ on $X(\chi)$ is only quadratic, reflecting the fact that \pref{EF4D} depends quadratically on $G_{\mu\nu\lambda\rho}$ (and so also that the 10D UV completion depends only quadratically on the higher-dimensional form field-strength). This quadratic dependence of $U(\chi)$ on $X(\chi)$ has a clean interpretation in the supersymmetric versions of this argument, such as those that arise in 10D supergravity compactifications on Calabi-Yau spaces. In this supersymmetric context the 4-form induced term in \pref{newU4D} corresponds to a contribution to the auxiliary field part of the scalar potential~\cite{Bielleman:2015ina, Herraez:2018vae}. In this interpretation $X(\chi)$ corresponds to auxiliary fields that are functions of light 4D fields (and so can be described as contributions to the 4D superpotential that cause spontaneous symmetry breaking purely within the 4D effective supergravity), while $K_4$ represents higher-dimensional imaginary anti-self dual fluxes which breaks supersymmetry and do not depend directly on dynamical 4D fields.

The supermultiplets whose auxiliary fields appear in $X(\chi)$ are the ones whose axionic scalars also arise due to the same harmonic forms $\Omega$ and $\beta^{(a)}$ that appear in \pref{G7vsG4harm}. These are the fields that appear in the term linear in $G_4$ in the 4D action, and they do so because of the cross terms between the $G_4 \wedge \bar \beta$ piece and the $\delta C_2 \propto c^{(a)} \, \beta^{(a)}$ and $\delta B_2 \propto b^{(a)} \beta^{(a)}$ terms in the expansion $\delta G_3 = G_3 - \langle G_3 \rangle$ of $G_3$ about its background. 

For type IIB and heterotic vacua it is the complex-structure moduli that are associated with 3-cycles like $\beta^{(a)}$ (in type IIB also the dilaton is associated to $\Omega$), and this is the reason why these are the moduli that get stabilised by flux compactifications (and so appear in the low-energy 4D theory's superpotential, $W$). The same does {\it not} happen for the K\"ahler moduli because one needs an even-dimensional field-strength form to get a 4-form field to interfere with a scalar supermultiplet built using harmonic two-forms rather than three-forms. 

Type IIA vacua considered in the next section are different, however, because in this case both even- and odd-dimensional forms exist. As a consequence both complex-structure and K\"ahler moduli can get lifted by type IIA flux vacua (and so appear more generically in the low-energy superpotential). For more details see~\cite{Bielleman:2015ina, Herraez:2018vae}.

\subsection{Higher orders in $g_s$ and $\alpha'$}
\label{ssec:IIBHiOrd}

Let us now try to obtain the scaling with Im $\tau$ and $\cV_\ssE$ of contributions to the effective Lagrangian at higher orders in the $\alpha'$ and $g_s$ expansions. As we saw for the D-brane action, terms that arise at subdominant order in these expansions no longer transform as did the classical bulk action \pref{TypeIIBBulk}, but they do so in a predictable way because string loops involve specific higher powers of (Im $\tau)^{-1} = e^{\phi}$ and $\alpha'$ corrections involve higher derivatives (and so also higher powers of $g^{\ssM\ssN}$). The dependence on Im $\tau$ and $\cV_\ssE$ that this implies for the 4D action is obtained in the same way as above using the transformation under the two scalings of the corresponding term in the 10D action. 

\subsubsection{Corrections to bulk and brane actions}

To see how this works we repeat the exercise performed for heterotic vacua in \S\ref{ssec:HeteroHigherPowers} and consider the contribution of higher string loops and higher orders in $\alpha'$. In 10D string frame each additional closed-string loop costs a factor of $g_s^2 \propto e^{2\phi}$ (just like for heterotic loops) but each open-string loop costs only $g_s \propto e^\phi$. We consider corrections of this sort to both bulk and brane actions in turn.

\subsubsection*{Bulk action}

A bulk term suppressed (relative to tree-level) by $n$ powers of $e^\phi$, $m+1$ powers of curvature and $r$ powers of the Neveu-Schwarz 3-form field $H_3^2$ looks like
\be
 \cL_{nmr}^\ssB\propto \sqrt{-\hat g_{10}} \; e^{(n-2)\phi} \Bigl(  \hat g^{\circ \circ} {{\hat R}^\circ}_{\circ \circ\circ} \Bigr)^{m+1} \Bigl[  \hat g^{\circ \circ} \hat g^{\circ \circ} \hat g^{\circ \circ} H_{\circ\circ\circ} H_{\circ\circ\circ} \Bigr]^{r} \,,
\ee
where $\circ$ indicates appropriate index structure (whose details are not important for the scaling arguments made here). Transferring to the 10D EF using $\hat g_{\ssM\ssN} \propto e^{\phi/2} \tilde g_{\ssM\ssN}$ and noting that $H_3$ arises only within the combination $G_3 = F_3 + \tau \, H_3$, this becomes
\be
\label{IIBmnrCount}
 \cL_{nmr}^\ssB 
\propto \sqrt{-\tilde g_{10}} \; \left( \frac{1}{\hbox{Im}\, \tau} \right)^{(2n-m+r)/2} \Bigl( \tilde g^{\circ \circ} {{\tilde R}^\circ}_{\circ \circ\circ} \Bigr)^{m+1}  \Bigl[ \tilde g^{\circ \circ} \tilde g^{\circ \circ} \tilde g^{\circ \circ} G_{\circ\circ\circ} G_{\circ\circ\circ} \Bigr]^{r} \,.  
\ee
{\it Reality check:} specialising \pref{IIBmnrCount} to the case $(n,m,r) = (0,0,0)$ gives the 10D classical Einstein-Hilbert term, for which $\phi$ (or Im$\,\tau$) drops out completely in EF; and $(n,m,r) = (0,-1,1)$ is the classical $G_3$ kinetic term, whose proportionality to $e^{\phi} = (\hbox{Im}\,\tau)^{-1}$ agrees with \pref{TypeIIBBulk}.

Under the rescaling of \pref{IIBscalings} these transform as
\be \label{LBnmrscalingIIB}
\cL_{nmr}^\ssB \to
  \lambda^{4w_g-2n(w-w_g) +(m+r)(w-2w_g)}   \, \cL_{nmr}^\ssB \,,
\ee
which displays the classical bulk scaling result ($w_\ssL = 4w_g$) and the scaling factors associated with string loops and powers of $R$ or $G_3^2$. The implications of these scalings for corrections to the effective 4D theory are found by repeating the exercise leading to (\ref{TreeExpKSS}), using the compensator scaling property identified in \pref{BulkCompScale}.  Any contributions from $\cL_{nmr}^\ssB$ to $K$, $W$ or $\mathfrak{f}_{ab}$ therefore scale in the following way 
\bea
 \left( e^{-K/3} \right)_{nmr}^\ssB &\to&  \lambda^{2(2w_g+w)/3}  \left[ \lambda^{-2(w-w_g)} \right]^{n} \left( \lambda^{w-2w_g} \right)^{m+r} \left( e^{-K/3} \right)_{nmr}^\ssB \nn\\
 \hbox{and} \qquad  W_{nmr}^\ssB &\to&  \lambda^{w}  \left[ \lambda^{-2(w-w_g)} \right]^{n} \left( \lambda^{w-2w_g} \right)^{m+r} W_{nmr}^\ssB  \,,
 \label{TrProp}
\eea
if non-zero. 

Keeping in mind that shift symmetries imply $K$ depends (in perturbation theory) only on the imaginary part of $\tau$ and on the real parts of the moduli $T^i$, as well as the scaling relations $\cV_\ssE \to \lambda^{3w_g} \cV_\ssE$ and $\tau \to \lambda^{2(w-w_g)} \tau$, the transformation properties (\ref{TrProp}) imply
\be 
\setlength\fboxsep{0.25cm}
\setlength\fboxrule{0.8pt}
\boxed{
e^{-K/3} = (\hbox{Im}\,\tau)^{1/3} \cV_\ssE^{2/3}\sum_{nmr} \cA_{nmr} \left( \frac{1}{\hbox{Im}\,\tau} \right)^{n} \left[ \frac{(\hbox{Im}\,\tau)^{1/2}}{\cV_\ssE^{1/3}} \right]^{m+r}  \,,}
\label{KcorrIIB}
\ee
where the coefficients $\cA_{nmr}$ depend only on scale-invariant combinations of fields. 

\subsubsection*{Other fields and Spurions}

 At this point the dependence of the coefficients  $\cA_{nmr}$ on other fields can be arbitrary, provided only that they appear in $\cA_{nmr}$ (perhaps together with $\tau$ and $\cV_\ssE$) through scale-invariant combinations. The dependence of $\cA_{nmr}$ on these other fields can often be further constrained using additional information or symmetries, such as the generalized shift symmetries like \pref{betteryIIBshift} encountered above. For $W$ and $\mathfrak{f}_{ab}$ this extra information can be very constraining, leading in some situations to nonrenormalization theorems \cite{Witten:1985bz, Burgess:2005jx}. 
 
There can also be non-perturbative information, such as comes when building in what is believed to be the exact $SL(2,\mathbb{Z})$ symmetry of Type IIB vacua. Although this symmetry is not manifest order-by-order in string loops (it does, after all, include transformations like $\tau \to -1/\tau$), it is expected to constrain the form of $K$ and $W$ once non-perturbative corrections are included. See for instance~\cite{Sen:2013oza} for a discussion on how S-duality may provide further insights into string perturbation theory.


We briefly comment on the analogy to other EFT setups where spontaneously broken symmetries are present as organising principles of the EFT. The Goldstone bosons associated with these two scaling symmetries are the overall volume modulus and the dilaton which is in complete analogy to the pions in QCD. Spurions break these symmetries explicitly (in the sense as quark masses do in chiral perturbation theory) and make the would-be Goldstone bosons massive. The vacuum expectation values of the metric and the dilaton break these scaling symmetries spontaneously. This important class of spurion fields can appear in the $\cA_{nmr}$ coefficients. Tracking the possible spurion fields allows to analyse how and at which scale the symmetries are broken in the EFT.

An example of this type to consider is the non-Abelian spurion $f$, encountered for heterotic vacua in \pref{FieldStrHS}, and for IIB vacua in the discussion surrounding \pref{WD3stack} above. This spurion systematically appears in $K$ and $W$ only together with other non-Abelian fields, like $C^\ssI$ in the heterotic case or $Y^k$ for IIB vacua. 

Another important class of spurions come from supersymmetry breaking extra-dimensional fluxes, such as the expectation values for flux fields like $G_3.$ These often have nonzero background values, and when they generate nonzero contributions to the low-energy superpotential, they break both supersymmetry and some of the scale invariances (through expressions like \pref{WGVW}).

Having more fields at high energies that break the scaling symmetries means including more spurions in the low-energy theory, potentially undermining the conclusions drawn above about the form dictated by scaling for $K$ and $W$.\footnote{Additional spurion fields which can lead to no-scale breaking effect but that we did not analyse in this paper can for example originate from non-zero {\it v.e.v.}s of gauge fluxes on D7-branes (which induce moduli-dependent Fayet-Iliopoulos terms) or geometric and non-geometric fluxes.}

To see what is involved consider a spurion, $W_0$, representing an expectation for the field $G_3$ itself. When $W_0=0$ the dilaton remains a massless Goldstone boson (see \cite{Demirtas:2019sip} for an explicit example) while when the spurion $W_0$ is non-zero, one of the scaling symmetries is explicitly broken and the dilaton becomes massive. This spurion scales as $W_0 \to \lambda^w W_0$, as does $G_3$. Such a spurion can, and often does, appear in the low-energy superpotential, $W = W_0$, as found in explicit compactifications. The K\"ahler potential can also acquire a dependence on $W_0$, and it does so through having the coefficients $\cA_{nmr} = \cA_{nmr}(\cJ)$ of \pref{KcorrIIB} depend on the invariant combination
\be \label{cJdefSpurion}
  \cJ := \frac{ W_0}{(\hbox{Im } \tau)^{1/2} \cV_\ssE^{1/3}} = \frac{W_0}{\hbox{Im } \tau} \, \left[ \frac{(\hbox{Im }\tau)^{1/2}}{\cV_\ssE^{1/3}} \right]\,.
\ee
This shows that every factor of $W_0$ in $K$ comes with an automatic suppression by precisely one string-loop factor, $g_s$, and by one factor of $\alpha'$. Notice that this ensures that $\cJ$ is small even if the spurion $W_0$ itself is not.

In particular, a term in $e^{-K/3}$ involving $n$ loops, $m+1$ powers of curvature, and $2r$ powers of $G_3$ (of which $s \leq r$ of the $(G_3)^2$'s are replaced with their spurion {\it v.e.v.}~$W_0^2$, rather than as fluctuations) necessarily therefore also depends on $\tau$ and $\cV_\ssE$ as\footnote{It can happen that $W_0$ dependence actually enters into the 4D action in a way that cannot be captured by a contribution to $K$ or $W$. This occurs in particular when it contributes to terms that involve superspace derivatives in the invariants $\mathfrak{D}$ and $\mathfrak{F}$. In such cases the scaling conclusions drawn here ({\it e.g.}~that $W_0$ appears only through the invariant $\cJ$ of \pref{cJdefSpurion}) still apply, but should instead be interpreted as applying directly to $\mathfrak{D}$ or $\mathfrak{F}$, rather than to $K$ or $W$. }
\be 
\boxed{
e^{-K/3} = (\hbox{Im}\,\tau)^{1/3} \cV_\ssE^{2/3}\sum_{nmr} \cA_{nmr} \left( \frac{1}{\hbox{Im}\,\tau} \right)^{n} \left[ \frac{(\hbox{Im}\,\tau)^{1/2}}{\cV_\ssE^{1/3}} \right]^{m+r}\,\cJ^{2s}  \,, }
\label{KcorrIIBnew}
\ee
 In fact, in specific compactifications such terms can actually arise from effective 10D interactions involving powers of  $G_3$, with each factor of $G_3$ replaced by its {\it v.e.v.}, in which case a term involving $G_3^{2r}$ contributes as above, with $s=2r$.

\subsubsection*{Brane actions}

At face value an identical argument goes through for corrections to the brane actions, with the main difference relative to the bulk being the scaling weight of the initial brane action is $w_\ssL(\hbox{brane}) = (p-1)w_g - \frac12(p-3) w$ rather than $w_\ssL(\hbox{bulk}) = 4w_g$. Repeating the same exercise as for the bulk would then give a scaling behaviour 
\be
\cL_{nmr}^{{\rm D}p} \to
  \lambda^{(p-1)w_g- \frac12(p-3)w -2n(w-w_g) +(m+r)(w-2w_g)}   \, \cL_{nmr}^{{\rm D}p} \,,
\ee
from which conclusions like \pref{TrProp} can in principle be drawn. This kind of reasoning has the disadvantage that it assumes the brane and bulk contributions must enter additively in the 4D effective theory. They of course do so at leading order, where the 4D theory is simply the dimensional reduction of what ultimately is a local action in the higher dimensions. But extra-dimensional locality need not be preserved once Kaluza-Klein modes are integrated out, because these modes have wavelengths comparable to the size of the extra dimensions themselves. It is therefore preferable not to build in this assumption from the get-go in the 4D theory.

A better approach uses the identity 
\be \label{usefulwLident}
    (p-1)w_g - \frac12(p-3) w = 4w_g -2(w-w_g) + \frac12(p-7)(2w_g - w)
\ee
to recognize that the classical brane action scales in precisely the same way as does \pref{LBnmrscalingIIB} in the special case $n = 1$ and $m+r=\frac12(7-p)$ (which is an integer for $p$ odd). This happens because eq.~\pref{usefulwLident} simply states that the brane action differs from the scaling of the bulk because of its additional power of the 10D dilaton ($e^\phi$) and its different spacetime dimension (and the associated difference in scaling weight when the SF metric transforms). But since it is the scaling of these quantities that identify the $g_s$ and $\alpha'$ corrections, from the point of view of scaling behaviour brane contributions can be regarded as special cases of string-loop and $\alpha'$ corrections to the bulk.

From this point of view the dimensional reduction of a correction to the brane sector action suppressed by $\ell_o$ powers of $g_s \propto e^\phi$ and by $s$ powers of $\alpha'$ counts as the choice 
\be \label{gsalphapcopenstr}
   n = 1+\ell_o \qquad \hbox{and} \qquad 
   m+r = \frac12(7-p) + s \,,
\ee
in \pref{TrProp} ({\it i.e.} $m+r = 2+s$ for D3 branes or $m+r = s$ for D7s). Contributions to the gauge kinetic terms in particular scale as
\be
  \left( \mathfrak{f}_{ab} \right)_{nmr} \to  \lambda^{2w}  \left[ \lambda^{-2(w-w_g)} \right]^{n} \left( \lambda^{w-2w_g} \right)^{m+r}\left( \mathfrak{f}_{ab} \right)_{nmr} \,. 
\label{TrPropb}
\ee
{\it Reality check:} using \pref{gsalphapcopenstr} with the special case $\ell_o = s = 0$ --- which gives $n=1$ and $m+r = -\frac12(p-7)$ --- in \pref{TrPropb} then gives the scaling weight $w_{\mathfrak{f}} = (p-5)w_g -\frac12(p-7)w$, as found for classical gauge kinetic function  in \pref{gaugekinweight}.

Of course any corrections, particularly in the expansions of $W$ and $\mathfrak{f}_{ab}$ have zero coefficient unless they are also compatible with other symmetries, like axionic shifts; an expression of their non-renormalisation theorems. 

\subsubsection{Comparison with explicit calculations}

We next test these scaling forms against a variety of explicit calculations. Notice, when doing so, that because $n$ counts powers of $e^\phi$ in \pref{KcorrIIB} an $\ell_c$-loop contributions from closed-string loops have $n = 2\ell_c$ (because each closed-string loop carries a factor $g_s^2 \propto e^{2\phi}$), while for open-string loops we've seen $\ell_o$ loops corresponds to $n= \ell_o+1$. As usual the parameter $m+r$ sets the order in the $\alpha'$ expansion in the bulk, or is related to it by \pref{gsalphapcopenstr} for open-string contributions. There are several examples of corrections with which these scaling arguments can be compared: 

\subsubsection*{$\mathcal{O}$($\alpha'^3$) bulk corrections} 

In $\cN=2$ Calabi-Yau compactifications the first $\alpha'$ corrections are known to arise at $\mathcal{O}$($\alpha'^3$) which corresponds to $m+r=3$.  For arbitrary loop order $n$, (\ref{KcorrIIB}) predicts a correction to the K\"ahler potential of the form
\be
K = -\ln(\tau - \bar\tau) -2\ln\cV_\ssE - 3\ln\left[1+ \frac{\cA}{\cV_\ssE} \left(\frac{1}{\hbox{Im}\,\tau}\right)^{n-3/2}\right]\quad\hbox{so} \quad \delta K \sim \frac{g_s^{n-3/2}}{\cV_\ssE}\,,
\label{a3Pred}
\ee
which uses $g_s \propto e^\phi$. This matches several explicit computations of $\alpha'^3$ corrections like:
\begin{itemize}
\item Several $\cN=2$ string calculations using spherical string world sheet ({\it i.e.}~$n=0$) in the absence of background 3-form fluxes ({\it i.e.}~$r=0$) give results of the form $\delta K\sim g_s^{-3/2} /\cV_\ssE$~\cite{BBHL, Antoniadis:1997eg, BW}. These corrections to $K$ -- believed to originate from (curvature)${}^4$ terms in 10D -- change the kinetic terms for the low-energy fields. Once combined with the spurion $W_0 \propto J$ they also modify the 4D scalar potential of the 4D theory, with corrections to the potential having the form $V \sim g_s^{-1/2} |W_0|^2/\cV_\ssE^3$. Such terms originate from $R^3 G_3^2$ interactions in 10D. The existence of these kinds of corrections to the potential plays a crucial role in modulus stabilisation in LVS string vacua~\cite{LV, LVcorr, vonGersdorff:2005bf,LVLoops, LVLoops1}.

\item $\cN=2$ string calculations using toroidal string world sheet (closed string 1-loop level: $n=2\ell_c = 2$), again with vanishing 3-form fluxes ({\it i.e.}~$r=0$), give contributions of the form $\delta K\sim g_s^{1/2} /\cV_\ssE$~\cite{Antoniadis:1997eg};

\item $\cN=1$ string calculations with spherical world-sheet ({\it i.e.} closed-string tree level: $n=0$) and $G_3=0$ ({\it i.e.}~$r=0$) give contributions of the form $\delta K\sim g_s^{-3/2} /\cV_\ssE$~\cite{MPS};

\item $\cN=1$ contributions at open-string 1-loop level (i.e. $n=\ell_o+1=2$) for vanishing $G_3$ ({\it i.e.}~$r=0$) give contributions of the form $\delta K\sim g_s^{1/2} /\cV_\ssE$~\cite{HK}.

\item The determination of potential corrections which come from 10D terms proportional to powers of $G_3^2$ is a case where the spurion $W_0$ can play an important role, making \pref{KcorrIIBnew} more useful than (\ref{KcorrIIB}). For example, \pref{KcorrIIBnew} states that terms with $m+1$ power of curvature and $r$ powers of $G_3^2$ (of which $0 \leq s \leq r$ are replaced with their spurion {\it v.e.v.}) and loop order $n$ contribute to $K$ (or directly to $\mathfrak{D}$, if contributing through a term with extra superspace derivatives) with relative size
\be
\delta K =  \cA \left(\frac{1}{\hbox{Im}\,\tau}\right)^{n}\left[ \frac{(\hbox{Im}\,\tau)^{1/2}}{\cV_\ssE^{1/3}} \right]^{m+r} \left( \frac{W_0^2}{ \cV_\ssE^{2/3}\, \hbox{Im}\,\tau}  \right)^{s} \,,
\label{a3Pred}
\ee
and so in particular terms with $m+r=3$ give
\be
\delta K \sim \frac{g_s^{n-3/2}}{\cV_\ssE}\left( \frac{g_s\,W_0^2}{\cV_\ssE^{2/3}} \right)^s\,.
\label{deltaK}
\ee
This estimate applies in particular to tree-level corrections ({\it i.e.}~$n=0$) that generate 10D interactions like $R^3 G_3^2$ and $R^2 G_3^4$ and for both predicts for $s=1$ the result $\delta K \sim g_s^{-1/2}\,W_0^2/\cV_\ssE^{5/3}$. This agrees with what is found in~\cite{CLW, GMW}, both for the change to kinetic terms generated by the $R^3 G_3^2$ interactions, and for the changes to the scalar potential generated by the $R^2 G_3^4$ term. 

Notice also that these are examples of higher-derivative contributions to the low-energy 4D EFT since they involve higher powers of auxiliary fields, $F$. For instance, the correction to the scalar potential looks like $V \sim g_s^{1/2} |W_0|^4/\cV_\ssE^{11/3}$ and, as shown in \cite{CLW}, scales as $F^4$. In the limit where the superspace derivative expansion is under control, {\it i.e.} when the scale invariant combination $\cJ$ in \pref{cJdefSpurion} is small, as derived in \cite{Cicoli:2013swa}, this term is therefore subleading to the $n = r = 0$ and $m = 3$ correction computed in \cite{BBHL}, but could play a role in lifting leading-order flat directions \cite{Cicoli:2016chb}. Although it need not be true that these can be regarded as corrections to $K$ (see the discussion in~\cite{CLW,GMW}), this does not really matter for the power-counting argument above, which applies equally well when applied to $\mathfrak{D}$ as to $K$.

\item There are also reports of loop-generated corrections to $K$ that depend logarithmically on $\cV_\ssE$, rather than as a power law~\cite{Ibanez:1999pw, Antoniadis:2018hqy}. For instance~\cite{Antoniadis:2018hqy} find one-loop corrections to $K$ at $\mathcal{O}(\alpha'^3)$ that scale as $\delta K\sim g_s^{1/2} \ln \cV_\ssE/\cV_\ssE$. This kind of non-power dependence of the Lagrangian cannot be captured by \pref{KcorrIIB}, but neither should it be. Recall that the Wilsonian effective 4D Lagrangian is obtained by integrating out Kaluza-Klein energy scales and above, but does {\it not} include the effects of loops of 4D fields. While logarithmic dependence on expansion constants {\it can} arise in an action --- a familiar example of which is the famous $\alpha^5 \ln (1/\alpha)$ contribution to the Lamb shift --- such non-analytic dependence arises within a perturbative framework as a logarithm of length scales, $\ln(M/m)$, making them sensitive to the IR part of the theory. Ratios of mass scales get converted to logarithms of couplings once the relevant mass ratio is coupling dependent -- such as when $M \sim m_e$ and $m \sim \alpha\, m_e$ gives $\ln(M/m) \sim \ln(1/\alpha)$. We have not systematically assessed how our scaling arguments apply to these lower-energy contributions.

\end{itemize} 

\subsubsection*{$\mathcal{O}$($\alpha'^2$) corrections} 

A crucial ingredient to obtain $\cN=1$ 4D effective field theories in type IIB vacua is the presence of an orientifold projection, and this can allow $\mathcal{O}$($\alpha'^2$) corrections also to arise, corresponding to $m+r=2$. For these (\ref{KcorrIIBnew}) predicts an $\mathcal{O}$($\alpha'^2$) correction to the K\"ahler potential of the form 
\be
K = -\ln(\tau - \bar\tau) -2\ln\cV_\ssE - 3\ln\left[1+ \frac{\cA}{\cV_\ssE^{2/3}} \left(\frac{1}{\hbox{Im}\,\tau}\right)^{n-1}\right]\quad\hbox{so} \quad \delta K \sim \frac{g_s^{n-1}}{\cV_\ssE^{2/3}}\,,
\label{a2Pred}
\ee
As observed in~\cite{LVLoops1, Cicoli:2018kdo}, these types of corrections do not ruin the no-scale form of $K$. This is perhaps easiest to see when \pref{a2Pred} is written as in \pref{KcorrIIB}, which becomes
\be 
e^{-K/3} = (\hbox{Im}\,\tau)^{1/3} \left[ \cV_\ssE^{2/3} + \cA \left( \frac{1}{\hbox{Im}\,\tau} \right)^{n-1}  \right] \,,
\label{KcorrIIBx}
\ee
and shows that the correction to $e^{-K/3}$ is $\cV_\ssE$-independent (and so does not remove the zero eigenvector in the $\cV_\ssE$ direction of the matrix $\partial_i \partial_{\bar \jmath} e^{-K/3}$ that was ensured by the lowest order result). 

The result (\ref{a2Pred}) matches explicit computations of $\alpha'^2$ corrections in fluxless backgrounds (i.e. with $r=0$), such as:
\begin{itemize}
\item $\cN=2$ contributions at open string 1-loop level ({\it i.e.}  $n=\ell_o+1=2$) of the form $\delta K\sim g_s /\cV_\ssE^{2/3}$~\cite{BHK}. Because this loop correction is consistent with the no-scale result (\ref{a2Pred}) at first order, its contribution to the loop-corrected scalar potential vanishes. This cancellation of loop corrections to $K$ within the scalar potential has been called `extended no-scale structure' within the context of LVS models, where it was first seen~\cite{vonGersdorff:2005bf,LVLoops, LVLoops1}. This cancellation in the 4D scalar potential makes the $\mathcal{O}(g_s^2 \alpha'^2)$ terms in $\delta K$ negligible in $V$ relative to those arising from those arising at $\mathcal{O}(\alpha'^3)$ discussed above, which play the dominant role in stabilizing moduli. One-loop effects eventually lift leading-order flat directions through a scalar potential of order $V \sim g_s^3\,|W_0|^2/\cV_\ssE^{10/3}$~\cite{vonGersdorff:2005bf,LVLoops, LVLoops1}, with the corresponding moduli parametrically light even compared with other moduli (making them particularly attractive as inflaton candidates~\cite{,Cicoli:2008gp, Cicoli:2016chb}). 

While there is no general reason why even higher-loop corrections to the potential need also vanish, one speculates about circumstances for which $e^{-K/3}$ receives no corrections beyond one loop, in which case all higher-loop corrections to $K$ would turn out to be consistent with higher orders in the expansion of (\ref{a2Pred}). If so they would preserve the no-scale property to all string-loop orders. (This would remain interesting even if only true for a subclass of loop corrections to $K$. See for example~\cite{Halverson:2013qca} for an argument based on mirror symmetry to infer the absence of some $N=2$ corrections beyond $\alpha'^3$.)  
\item The largest possible contributions of the form \pref{KcorrIIBx} are the $\cN=1$ contributions that can arise at open string tree-level ({\it i.e.} $n=\ell_o+1=1$), and would be of order $\delta K \sim \cV_\ssE^{-2/3}$. Contributions to the effective Lagrangian of this size are known to exist, but in all known cases are understood to give corrections to the relationship between the supermultiplet variable $T_i$ and its scalar real and imaginary component moduli, $\chi_i$ and $a_i$, without also changing the dependence of $K$ on $\ov T_i+T_i$~\cite{GSW}.
\end{itemize}

\subsubsection*{$\mathcal{O}$($\alpha'^4$) corrections} 

These corrections correspond to $m+r=4$ and so at a generic string-loop order $n$ \pref{KcorrIIB} predicts a correction to the K\"ahler potential by an amount
\be
K = -\ln(\tau - \bar\tau) -2\ln\cV_\ssE - 3\ln\left[1+ \frac{\cA}{\cV_\ssE^{4/3}} \left(\frac{1}{\hbox{Im}\,\tau}\right)^{n-2}\right]\quad\hbox{so} \quad \delta K \sim \frac{g_s^{n-2}}{\cV_\ssE^{4/3}}\,.
\label{a4Pred}
\ee
This result matches the dilaton and volume scaling of the explicit computation of $\cN=2$ open string 1-loop ({\it i.e.} $n=\ell_o+1=2$) corrections to the K\"ahler potential in the absence of background 3-form fluxes ({\it i.e.} $r=0$) performed in~\cite{BHK}, which gives $\delta K \sim \cV_\ssE^{-4/3}$. The corresponding contributions to the 4D scalar potential scale like $V \sim g_s |W_0|^2/\cV_\ssE^{10/3}$.  

\subsubsection*{$\mathcal{O}$($\alpha'$) corrections?} 

To date no string corrections are known that arise at $\mathcal{O}$($\alpha'$), even for $\cN=1$ compactifications. However, if a 10D (curvature)${}^2$ term were present at any string loop order (corresponding to $m=1$ and $r=0$), then the general result (\ref{KcorrIIBnew}) would predict a correction to the K\"ahler potential of order
\be
K = -\ln(\tau - \bar\tau) -2\ln\cV_\ssE - 3\ln\left[1+ \frac{\cA}{\cV_\ssE^{1/3}} \left(\frac{1}{\hbox{Im}\,\tau}\right)^{n-1/2}\right]\quad\hbox{so} \quad \delta K \sim \frac{g_s^{n-1/2}}{\cV_\ssE^{1/3}}\,.
\label{a1Pred}
\ee
Such corrections, if present, could be important for LVS models~\cite{Cicoli:2018kdo}, since they contribute to the scalar potential at order $V \sim g_s^{n+1/2} |W_0|^2 /\cV_\ssE^{7/3}$, and so for large $\cV_\ssE$ could dominate the $\mathcal{O}(\alpha'^3)$ correction that is presently believed to dominate (and which scales in the potential like $V \sim g_s^{-1/2} |W_0|^2 /\cV_\ssE^3$). A central challenge to LVS constructions is the verification that this type of correction is absent, especially in the case where supersymmetry breaking background fluxes are present. 

\section{Descent from 10D type IIA supergravity}
\label{sec:IIA}

This section sketches how the scaling arguments of previous sections also go through for 10D type IIA vacua. The discussion here is relatively brief, partly because many of the issues already arise for the type IIB vacua discussed above. Unlike the IIB case, for the reasons discussed in \S\ref{sssec:SpaceFill} the K\"ahler moduli can, for type IIA vacua, also appear in the superpotential and so be stabilised much as are complex-structure moduli and the dilaton for type IIB.  We focus mostly on the parts of the argument that differ from the type IIB section, skipping over those parts that directly parallel our earlier description.

\subsection{Scaling in type IIA supergravity}
\label{ssec:IIAScaling}

The type IIA action also has the form of a sum of bulk and localised brane sources
\be
 S_{\rm IIA} =   S_{\rm bulk} +  S_{\rm loc}   \ ,
\ee
where $S_{\rm loc}$ describes the localised sources (D$p$-branes with $p$ even and O6-planes). We  examine the scaling properties of each of these in turn.

\subsubsection*{Scale invariances of the bulk}

The bosonic part of the type IIA bulk action~\cite{Grimm:2004uq} in 10D String frame is (schematically --- dropping numerical factors) contains a kinetic and Chern-Simons piece, $S_{\rm bulk} = S_{\rm kin} + S_{\CS}$, where
\be
\label{TypeIIAKin}
 S_\IIA = -\frac{1}{(\alpha')^4} \int \exd^{10}x \sqrt{-\hat g} \; \left\{e^{-2\phi} \Bigl[ \hat R - (\partial \phi)^2+ H_3^2  \Bigr] + \tilde F_2^2 + \tilde F_4^2 + \mathfrak{m}_0^2 \right\}   \ ,
\ee
where $F_{p+1} = \exd C_p$ and $H_3 = \exd B_2$ while 
\be
\tilde F_2 = F_2 + \mathfrak{m}_0 B_2 \quad\hbox{and} \quad
  \tilde F_4 = F_4 - C_1 \wedge H_3 - \frac{\mathfrak{m}_0}2 \, B_2 \wedge B_2 \,.
\ee
The CS term similarly has the scaling form
\be
  S_{\CS} =- \frac{1}{(\alpha')^4} \int \Bigl[ B_2 \wedge F_4 \wedge F_4 - \mathfrak{m}_0 B_2 \wedge B_2 \wedge B_2 \wedge \exd C_3  + \mathfrak{m}_0^2 B_2 \wedge B_2 \wedge B_2 \wedge B_2 \wedge B_2 \Bigr]  \,.
\ee 

In these expressions $\mathfrak{m}_0$ is the Romans mass parameter~\cite{Romans:1985tz}, which can also be regarded as the contribution to the action of a space-filling 10-form field strength. The 10D EF metric, $\tilde g_{\ssM\ssN}$, is then defined by $\hat g_{\ssM\ssN} = e^{\phi/2} \tilde g_{\ssM\ssN}$, in terms of which \pref{TypeIIAKin} takes the schematic\footnote{Numerical factors are not included here since our interest is in the action's scaling properties.} form
\be
\label{TypeIIAKinEF}
 S_{\IIA} = -\frac{1}{(\alpha')^4} \int \sqrt{-\tilde g} \;\left\{\tilde R + (\partial \phi)^2  + e^{-\phi} H_3^2 +e^{3\phi/2} \tilde F_2^2 + e^{\phi/2} \tilde F_4^2 + \mathfrak{m}_0^2\, e^{5\phi/2} \right\}   \,.
\ee

When $\mathfrak{m}_0 = 0$ the bulk action scales homogeneously under the rescalings
\be
\label{IIAscalings}
 \tilde g_{\ssM\ssN} \rightarrow \lambda^{w_g} \tilde g_{\ssM\ssN} \ , \qquad e^{-\phi} \rightarrow \lambda^{w_{\phi}} e^{-\phi} \ , \qquad B_2 \rightarrow \lambda^{w_{B_2}} B_2 \,, \qquad C_p \rightarrow \lambda^{w_{C_p}} C_p \,,
\ee
with the condition that $\tilde F_4$ scale homogeneously requiring
\be \label{F5B2relnIIA}
 w_{C_3} = w_{C_1} + w_{B_2}    \ .
\ee
Defining\footnote{Notice this is not quite the same definition as for IIB vacua.} $w_{B_2} = w$, the bulk action \pref{TypeIIAKinEF} scales as $S_{\IIA} \to \lambda^{4w_g} S_{\IIA}$ under \pref{IIAscalings}  provided that the various weights are related by the conditions
\be
    w_{C_1} = 2w_g - \frac{3w}2 \qquad \hbox{and} \qquad   w_\phi = 2(w_g - w) \ ,
\label{weights_F5_etcIIA}
\ee
leaving the two weights $w_g$ and $w$ arbitrary.  

Both scaling symmetries of the bulk action survive for non-zero $\mathfrak{m}_0$ only if $\mathfrak{m}_0$ is treated as a spurion, $\mathfrak{m}_0 \to \lambda^{w_m} \mathfrak{m}_0$, with scaling weight
\be
    w_m = 2 w_g - \frac{5w}2 \,.
\ee
Equivalently, only the specific transformation satisfying $w = \frac45 w_g$ survives as a symmetry when $\mathfrak{m}_0\neq 0$ does not transform. This is a 10D version of the breaking of the scaling symmetries by fluxes described in \S\ref{sssec:SpaceFill}. 

\subsubsection*{Scaling of localised sources}

The DBI part of the action for a D$p$-brane in 10D EF is the same as for type IIB supergravity, though with different dimension branes. The scaling transformations go through much as for the type IIB case, with an important difference. First, because $w_\phi$ here differs in sign from $w_\tau$ as was found in the IIB, the SF metric, $\hat g_{\ssM\ssN} = \tilde g_{\ssM\ssN} \, e^{\phi/2}$, in the IIA case scales as $\hat g_{\ssM\ssN} \to \lambda^{w_g - w_\phi/2} \hat g_{\ssM\ssN} = \lambda^w \hat g_{\ssM\ssN}$. Consistent scaling of the DBI action \pref{DBIaction} again requires $B_2$ must scale in the same way as does $\hat g_{\ssM\ssN}$, and this is an automatic consequence of the definition given above for $w$. With this in mind, both the DBI and WZ parts of the action then scale with weight
\be \label{wDBIdefIIA}
 w_\DBI(p)   
 =   2w_g + \frac12 (p-3)w   \,,
\ee
which is to be contrasted with the corresponding result \pref{wDBIdef} for the IIB theory. 

\subsubsection{Dimensional reduction}

The scaling properties under dimensional reduction are obtained for type IIA as for type IIB, with the metric given by \pref{GKP}, with scaling weights $\tilde g_{mn} \to \lambda^{w_g} \tilde g_{mn}$ and $\tilde g_{\mu \nu} \to \lambda^{w_g} \tilde g_{\mu \nu}$, and so the dimensionless EF extra-dimensional volume scales as  $\mathcal{V}_\ssE \to \lambda^{3w_g} \mathcal{V}_\ssE$, as in \pref{VEweight}. The 4D EF metric then scales as $g_{\mu\nu}  \to \lambda^{4w_g} g_{\mu\nu}$ while the leading Lagrangian scales as $\cL \to \lambda^{4w_g} \cL$. 

For IIA compactifications the moduli are captured by the complexified K\"ahler 2-form and holomorphic 3-form,
\begin{eqnarray}
\nonumber J_c&=& B_2+iJ\\
\Omega_c&=& C_3+i {\rm Re}(\mathfrak{C}\, \Omega)~,
\end{eqnarray}
where $\mathfrak{C} :=e^{-\phi} \cV^{1/2}_\ssS e^{K_{\rm cs}}$ and $K_{\rm cs} =-\log{\int\Omega \wedge\bar{\Omega}}$ with $\Omega$ denoting the usual Calabi-Yau holomorphic 3-form. Here $\cV_\ssS =\int J\wedge J\wedge J$ is the dimensionless extra-dimensional volume measured using the SF metric and $J$ is the Calabi-Yau K\"ahler form in string frame. For instance, the complex structure moduli and the dilaton are extracted from $\Omega_c$ through integrations over the bulk of the form
\begin{equation}
\int \Omega_c \wedge \beta_\ssL \,,
\end{equation}
where $\beta_\ssL$ are appropriate three-forms whose only property required here is that they do not scale (cf.~\cite{ibanezuranga}), and a similar statement relates $J_c$ to the K\"ahler moduli.

These definitions show that the 4D chiral superfield, $S$, that represents the dilaton and those, $U_a$, representing the complex-structure moduli both scale in the same way as does $\Omega_c$ (or $C_3$), while those superfields, $T_i$, representing the K\"ahler moduli scale as does $J_c$ (or $B_2$), implying 
\begin{equation} \label{eq:IIA4Dweights}
 w_\ssS = w_\ssU = w_{C_3} =  2w_g - \frac{w}2 \quad \hbox{and} \quad w_\ssT= w_{B_2} = w  \,.
 \end{equation}

For toroidal compactifications expressions for $S$, $T_i$ and $U_a$ can be made more explicit~\cite{Camara:2005dc} and are given in terms of ratios for the respective volumes of the tori, scaling as
\be
  S \sim e^{-\phi} \cV_\ssS^{1/2} + i a = e^{-\phi/4} \cV_\ssE^{1/2} +ia\quad \hbox{and} \quad
  T_i \sim \cV_\ssS^{1/3} + i b  =  e^{\phi/2}\cV_\ssE^{1/3} + i b\,,
\ee
where $b$ is the universal axion arising from $B_2$ while $a$ is the universal axion arising from $C_3$ and (as before) $\cV_\ssS$ denotes the SF extra-dimensional volume. Both real and imaginary parts therefore scale as indicated in \pref{eq:IIA4Dweights}. 

To determine the dependence of $K$ and $W$ on the moduli requires knowing how the compensator scales, which normally requires as input the scaling behaviour of one effective interaction from direct dimensional reduction. The superpotential produced by  $F_p$ and $H_3$ fluxes is known to be given by~\cite{Camara:2005dc, Derendinger:2004jn, Derendinger:2005ph, Villadoro:2005cu}
\begin{equation}
W=\int\Omega_c\wedge H_3+\sum_p\int e^{J_c}\wedge F_p\,,
\label{eq:IIAW}
\end{equation}
where $e^{J_c} \wedge F_p$ represents terms like $F_6$, $J_c \wedge F_4$, $J_c \wedge J_c \wedge F_2$ and so on. All of these terms scale in precisely the same way, leading to the conclusion that the superpotential scales as $W\to \lambda^{w_\ssW} W$ with weight
\be\label{WscaleIIA}
  w_\ssW = 2 w_g+\frac{w}{2} \,.
\ee

Given that the superpotential has weight $w_\ssW$, repeating the arguments given earlier for type IIB theories shows that the compensator $\Phi$ and $K$ must scale with weights
\be
  \Phi \to \lambda^{- \frac13(w_\ssW + 2w_g)} \Phi \quad \hbox{and} \quad
  e^{-K/3} \to \lambda^{\frac23(w_\ssW + 2w_g)} e^{-K/3} \,,
\ee
and so \pref{WscaleIIA} implies $e^{-K/3} \to \lambda^{w_\ssK} e^{-K/3}$ with weight
\be \label{KscaleIIA}
  w_\ssK = \frac23 \left(4 w_g + \frac{w}2 \right)  \,.
\ee
This is consistent with the behaviour found at lowest order in specific examples by dimensional reduction, which for toriodal examples gives expressions like~\cite{Camara:2005dc}
\be
  K_{\rm l.o.} = - \ln(S+\ol S) - \sum_{i=1}^3 \Bigl[ \ln (T_i+ \ol T_i) + \ln(U_i + \ol U_i) \Bigr]\,,
\ee
since this implies $e^{-K/3}$ scales with the weight $\frac43 w_\ssS + w_\ssT$, in agreement with \pref{KscaleIIA}.

\subsection{Higher orders in $g_s$ and $\alpha'$}

We can now repeat the scaling analysis to the effective Lagrangian at higher orders for the case of type IIA. The logic of the analysis proceeds much as for type IIB discussed in \S\ref{ssec:IIBHiOrd}, though with different weights for the scaling fields (and a greater variety of dimension of form-fields). Plus the possible presence of the spurion $\mathfrak{m}_0$. 

Consider the scaling weight of a higher-dimension 10D Lagrangian term of the form
\begin{eqnarray}
  {\cal L}_{nmrsq}  &\propto&  \sqrt{-\hat g_{10}} \; e^{(n-2)\phi} \Bigl(  \hat g^{\circ \circ} {{\hat R}^\circ}_{\circ \circ\circ} \Bigr)^{m+1} \Bigl[  \hat g^{\circ \circ} \hat g^{\circ \circ} \hat g^{\circ \circ} H_{\circ\circ\circ} H_{\circ\circ\circ} \Bigr]^{r} \nn\\
  &&\qquad\qquad\qquad\qquad\qquad \times \Bigl[  \hat g^{\circ \circ} \hat g^{\circ \circ} F_{\circ\circ} F_{\circ\circ} \Bigr]^{s}  \Bigl[  \hat g^{\circ \circ} \hat g^{\circ \circ} \hat g^{\circ \circ} \hat g^{\circ \circ} F_{\circ\circ\circ\circ} F_{\circ\circ\circ\circ} \Bigr]^{t} \mathfrak{m}_0^q \nn\\
  &\propto&  \sqrt{-\tilde g_{10}} \; e^{(2n-m-3r-2s-4t)\phi/2} \Bigl( \tilde g^{\circ \circ} {{\tilde R}^\circ}_{\circ \circ\circ} \Bigr)^{m+1}  \Bigl[ \tilde g^{\circ \circ} \tilde g^{\circ \circ} \tilde g^{\circ \circ} H_{\circ\circ\circ} H_{\circ\circ\circ} \Bigr]^{r} \\
  &&\qquad\qquad\qquad\qquad\qquad \times \Bigl[\tilde  g^{\circ \circ} \tilde g^{\circ \circ} F_{\circ\circ} F_{\circ\circ} \Bigr]^{s}\Bigl[  \tilde g^{\circ \circ} \tilde g^{\circ \circ} \tilde g^{\circ \circ} \tilde g^{\circ \circ} F_{\circ\circ\circ\circ} F_{\circ\circ\circ\circ} \Bigr]^{t} \mathfrak{m}_0^q\,.\nn
\end{eqnarray}
Taking the dilaton, 10D metric, form-field and spurion scalings from \S\ref{ssec:IIAScaling}, such a term scales as
\be 
  {\cal L}_{nmrsq} \to \lambda^{4w_g+2n(w-w_g)-(m+r)w+(2s+ 2t+q)(2w_g-\frac52w)} {\cal L}_{nmrsq}  \,.
\ee

Similar to the discussion for the IIB case, brane actions need not be considered separately but can instead be regarded as instances of higher-order corrections to the bulk action in powers of $e^\phi$ and $\alpha'$. For example, classical contributions coming from dimensionally reducing D$p$-brane actions scale with weight given in \pref{wDBIdefIIA}, and so scale differently than the classical bulk Einstein-Hilbert term (say) by one power of $e^\phi$ and $\frac12(7-p)$ fewer factors of the metric, and so correspond to the choices $n = 1$ and $m = \frac12(7-p)$. Notice in particular that $m$ takes the half-integer values $m=\frac52$ for leading-order D2 branes and $m = \frac12$ for leading contributions from D6 branes. Higher open-string loops and $\alpha'$ corrections correspond to shifting $n$ and $m$ by integers relative to these (and shifts of $r,s,t$ by integers away from zero).

Using the scaling of the compensator inferred earlier when deriving \pref{TrProp}, and repeating the steps gives the following scaling properties for the K\"ahler and superpotentials
\bea
 \left( e^{-K/3} \right)_{nmrsq} &\to&  \lambda^{\frac23 (4 w_g + \frac{w}2)}  \left[ \lambda^{2(w-w_g)} \right]^{n-2s-2t-q} \left( \lambda^{-w} \right)^{m+r+s+t+q/2} \left( e^{-K/3} \right)_{nmrsq} \nn\\
 \hbox{and} \qquad  W_{nmrsq} &\to&  \lambda^{2 w_g+\frac{w}{2}}  \left[ \lambda^{2(w-w_g)} \right]^{n-2s-2t-q} \left( \lambda^{-w} \right)^{m+r+s+t+q/2} W_{nmrsq}  \,.
 \label{TrPropIIA}
\eea

When comparing to the scaling behaviour \pref{eq:IIA4Dweights} for the 4D fields it is useful to consider the Lagrangian's dependence on a single scaling field of each weight, with all other fields written as a collection of scale-invariant ratios. Using $S_0 + \ol S_0$ and $T_0 + \ol T_0$ to denote the two basic scaling fields, with weights $w_\ssS = w_\ssU$ and $w_\ssT$  eq.~\pref{TrPropIIA} implies
\boxalign{\begin{align}\label{IIAKnmr}
e^{-K/3} &= (S_0 + \ol S_0)^{4/3}  (T_0+\ol T_0) \sum_{nmrsq} \cA_{nmrsq}  \,\mathfrak{m}_0^q \left[ \frac{(T_0+\ol T_0)^{3/2}}{S_0 + \ol S_0} \right]^{n} \nn \\ 
&\qquad\qquad\qquad\qquad\quad\times  \left( \frac1{T_0+\ol T_0} \right)^{m+r}  \left[ \frac{S_0 + \ol S_0}{(T_0 + \ol T_0)^2} \right]^{2(s+t)+q}, 
\end{align}}
where the coefficients $\cA_{nmrsq}$ are functions of the scale-invariant ratios of fields and the sums run over $n,r,s,q \geq 0$ and $m \geq -1$ (with $m$ taking half-integer values for D-brane contributions).

Superpotential forms consistent with the scaling behaviour similarly are
\be \label{IIAWnmr}
W  =  S_0 T_0 \sum_{nmrsq} \cA^\ssW_{nmrsq}  \,\mathfrak{m}_0^q \left( \frac{T_0^{3/2}}{S_0} \right)^{n}   \left( \frac1{T_0} \right)^{m+r}  \left( \frac{S_0}{T_0^2} \right)^{2(s+t)+q} \,.
\ee
For instance, a leading classical bulk contributions with $n=m+r=s=t=q=0$) leads to $W \propto S_0 T_0$, while the next order in $\alpha'$ corrections (from $n=s=t=q=0$ and $m+r = 1$) give $S_0$. Similarly, leading classical D6 contributions ($n =1$ and $m=\frac12$ with $r=s=t=q=0$) give $T_0^2$, and so on. These examples include the superpotentials found in explicit models~\cite{Camara:2005dc, Derendinger:2004jn, Derendinger:2005ph, Villadoro:2005cu}.
  
Of course the coefficients of all terms inconsistent with any axionic shift symmetries in \pref{IIAWnmr} must vanish, but these symmetries are also generically broken by extra-dimensional fluxes (with this breaking brought into the 4D EFT by space-filling fluxes, as described in \S\ref{sssec:SpaceFill}).

\section{Conclusions and outlook}
\label{sec:openquestions}

Determining the quantum corrections to EFTs having UV completions is one the main challenges to extracting robust implications from the UV theories in question. In particular, for string effective actions, the knowledge of perturbative corrections, especially to the K\"ahler potential, has long been one of the obstacles to extracting reliable predictions from these theories. The fact that the no-scale property is ubiquitous in these theories raises at least two important questions. The first asks for the origin of this no-scale property from the underlying UV theory; the second asks what are the dominant corrections to no-scale. Since no-scale implies vanishing tree-level scalar potentials, it makes the estimate of quantum (string loops and $\alpha'$) corrections crucial, since these can dominate the scalar potential and, therefore, most mechanisms for modulus stabilisation.

In this article we address both of these questions, which we regard to be pressing in order to assess if the current scenarios for moduli stabilisation are sufficiently robust, and/or if other scenarios could eventually emerge once further corrections are found. Overall we summarise our concrete findings as follows:

\begin{itemize}
\item We identify four distinct categories of no-scale property, summarised in Figure~\ref{fig:Summary}.
\item We examined the EFTs for a broad class of string vacua, and for all of them the no-scale property (if present) arises in the most restrictive, scaling, form that is traceable to the accidental scale invariances of the underlying UV completion.
\item Although most quantum corrections break scale invariance, some of the known corrections preserve the no-scale property at least to one higher order (this persistence of the no-scale property is known in the literature as `extended no-scale'). 
\item We use well-known scale invariances of the two-derivative higher-dimensional theory as the basis to organise quantum corrections. In 11D supergravity there is a single scaling symmetry (with roots in the $\alpha'$ expansion) that can (trivially) be used to organise the standard quantum corrections. The two scale invariances of 10D supergravity corresponding to the known perturbative string vacua correspond to the two generic expansions of weakly coupled vacua: the $\alpha'$ and string-loop expansions. We explicitly identify the scaling in 10D for both bulk and brane (DBI and WZ) actions.
\item We use these scalings to organise the quantum corrections to the 4D EFTs for heterotic, Type IIA and Type IIB string theories. We presented these expansions for the K\"ahler potentials in Eqs.~\pref{heteroticnmr2}, \pref{KcorrIIB} and \pref{IIAKnmr} for the heterotic, IIB and IIA respectively. They include both corrections from bulk and brane actions.
\item We test our general expressions by comparing them with various explicit calculations in the literature and found good agreement. We also identify potential quantum corrections that have not yet been computed, which could be relevant, in particular, to known calculations of moduli stabilisations (in which quantum corrections necessarily play a role).
\item Besides scaling symmetries and supersymmetry we also identify shift symmetries coming from Kalb-Ramond tensor fields. These include both standard axionic shift symmetries and generalised (approximate) shift symmetries. The generalised symmetries in particular ensure (for D3 branes in IIB string theory) that the K\"ahler potential appears as a function of the combination $T+\ov T- k(y, \ol y)$, with $T$ the volume modulus,  $y$ the D3 brane positions and $k(y,\ol y)$ the K\"ahler function of the Calabi-Yau manifold itself. Similar results hold for D7 branes and also for heterotic compactifications.
\item We also highlight a special role played by space-filling 4-forms in the 4D EFT, which bring to the 4D world extra-dimensional information about scale-invariance breaking and the presence of (quantised) fluxes.
\end{itemize}

Our main tool is the systematic exploitation of the importance of scaling and other symmetries in the 4D effective supergravity. These scaling symmetries underly the ubiquitous presence of no-scale supergravity in string effective actions at lowest order. Scale invariance turns out to be sufficient, and not necessary, for being no-scale. This allows some no-scale properties (like flat potentials) to persist beyond tree-level even after scaling symmetries are broken. In order to better clarify the relation between scale invariance and no-scale models, and thereby to better understand their origin and implications, we identify four nested categories of no-scale theories -- summarised in Figure~\ref{fig:Summary} -- of which scale-invariance ensures only the most restrictive.
 
Exact scale invariance leads to zero vacuum energy -- as seen from \pref{ScInvV2}). Unbroken supersymmetry similarly prefers non-positive vacuum energy. Anti-de Sitter vacua are obtained at leading order by badly breaking scale invariance in a supersymmetric way (such as by using multiple nonzero supersymmetric fluxes). Breaking supersymmetry more strongly than scale invariance at leading order leads to flat vacua. These observations help put into perspective the origins of tree-level no-go theorems for finding de Sitter space in supergravity and string theories (despite supersymmetric AdS solutions clearly existing).  For instance, the scaling symmetry derived in \S\ref{sec:IIA} shows that it is impossible to have metastable de-Sitter minima -- see also \eqref{ScInvV2} -- in compactifications of IIA string theory, in agreement with the findings of~\cite{Hertzberg:2007wc} (a result which has been interpreted as evidence for the so-called de-Sitter conjecture in~\cite{Obied:2018sgi}).

However, since these scaling symmetries are not exact it is also clear that the absence of metastable dS is very likely only a tree-level result. Not only do the accidental scaling symmetries imply that any metastable dS vacuum must arise at subdominant order in perturbation theory, they also ensure the leading-order vacuum is flat (and so within easy perturbative reach of de Sitter). Furthermore, the scale of any perturbatively achieved dS vacuum is necessarily suppressed relative to fundamental scales by the relevant small expansion parameters of the EFT.

Put differently, if tree-level contributions to the potential in a weakly coupled string theory are governed by the string scale, the dynamics around a potential de Sitter minimum in such a theory is almost inevitably governed by a scale hierarchically below the string scale (and therefore amenable to an EFT description). Therefore, although it is intellectually stimulating to study the 10D origin of these effects outside of an effective theory, it is strictly speaking unnecessary if there is a self-consistent understanding within the four dimensional effective field theory and its expansion parameters.\footnote{A similar comment applies to decribing 4D non-perturbative effects such as gaugino condensation, for which a 10D description is intriguing but largely beside the point.} Having an independent 10D understanding is not a precondition to understanding low-energy dS vacua, anymore than determining whether or not quarks are fundamental or composite at the Planck scale is a prerequisite to understanding the value of the QCD scale.

With this in mind, our discussion of positive scalar potentials in section \S\ref{sec:noscaleprops} may be relevant. The most general approach to understand no-scale (beyond scaling symmetries)  is due to the Barbieri, Cremmer, Ferrara~\cite{Barbieri:1985wq} theorem discussed in section \S\ref{sec:noscaleprops}, which states that supergravity theories give rise to semi-positive scalar potentials if  the matrix $-M$ defined in \pref{ExtendedNoScale} has one negative eigenvalue (with the rest positive, implying a negative determinant). This is actually the situation for the no-scale breaking of string EFTs through the presence of a cubic superpotential, such as appears in the heterotic and IIB compactifications. (In this case the K\"ahler potential is a particular case of~\eqref{sequester2} with $U$ corresponding to the complex structure moduli, $Z$ the D3 brane positions and $T$ K\"ahler moduli and similar for the heterotic case.) If the negative eigenvalue becomes zero (or the no-scale breaking fields are stabilised at $\langle Z\rangle =0$) the potential vanishes and there is a flat direction. This structure is a generalised no-scale model as discussed in \S\ref{sec:noscaleprops}. That section also identifies concrete supergravity models whose $F$-term potentials are semi-positive and give rise to vacua with flat directions with different levels of generality (including with de Sitter minima). Such models might be useful as waypoints towards obtaining de Sitter vacua from supersymmetric theories with a clean UV provenance. 

A main result of this article  is the use of scaling and other symmetries in string compactifications to constrain the form of explicit expressions for the $\alpha'$ and string-loop corrections for the low-energy effective action in both heterotic \pref{heteroticnmr2}, IIB \pref{KcorrIIB} and IIA \pref{IIAKnmr} string compactifications. In each case the combination of the dilaton  and (warped) volume $\cV_\ssE$ is identified that play the role of the string-loop and $\alpha'$ expansion parameters in the 4-dimensional EFT. The symmetries determine how the dilaton and $\cV_\ssE$ appear at each order in perturbation theory, and fix in particular their appearance in the tree-level effective action. Our constraints on their appearance in quantum corrections agrees with a great many general expressions for the combined $\alpha'$ and string loop expansions.

Besides providing a book-keeping measure that organises all the known corrections already computed, these general methods also highlight potentially important uncomputed corrections. An example of this type consists of the so-called $F^4$ corrections -- that are nicely captured by eq.~\eqref{KcorrIIBnew} -- obtained by keeping track of the $W_0$ dependence in higher-order corrections of the effective lagrangian. Our derivation fully agrees with the original calculations of~\cite{CLW}  but also shows how they scale to all orders, depending on the expansion parameter $\cJ = W_0\,(\hbox{Im}\,\tau)^{-1/2}\,\cV_\ssE^{-1/3}$. This parameter is always small within the string-loop and $\alpha'$ expansions, consistent with  the bound on  $W_0$ derived in~\cite{Cicoli:2013swa}.  We emphasise that some of the terms allowed by our scaling arguments might actually not be present, either because they arise with vanishing coefficient, or because they are allowed but redundant ({\it i.e.}~may be eliminated through appropriate field redefinitions). Moreover, further features likely appear -- such as the emergence of manifest $S$-duality -- once nonperturbative effects are included. Further study would be needed to identify the implications of any such nonperturbative information.

A number of other important questions are also left open. In particular, if evidence were to arise that corrections with $m+r=1$ in \pref{KcorrIIB} exist, then they would modify substantially the current approaches to modulus stabilisation for IIB vacua.  We hope to address more of these open questions, and to further explore the interplay between quantum corrections and accidental symmetries in future work.

\section*{Acknowledgements}

We thank Shanta de Alwis, Giuseppe Dibitetto, Anamar{\'{i}}a Font, Michael Haack, Jin U. Kang, Anshuman Maharana, Francesco Muia, Fabian Ruehle, Rafaelle Savelli, Andreas Schachner, Pramod Shukla, Roberto Valandro and Alexander Westphal  for helpful discussions. CB and SK thank the International Centre for Theoretical Physics (ICTP, Trieste) and CB thanks the Kavli Institute for Theoretical Physics (Santa Barbara) for providing such physically and intellectually engaging settings within which part of this research was completed. This work was partially supported by funds from the Natural Sciences and Engineering Research Council (NSERC) of Canada. Research at the Perimeter Institute is supported in part by the Government of Canada through NSERC and by the Province of Ontario through MRI. 

\appendix

\section{Axionic no-scale and linear multiplets}
\label{4D_no_scale_app}

This appendix sketches how to construct axionic no-scale models that do not fall into the standard no-scale category given in Definition~\ref{def2} of the main text. To do so we follow ref.~\cite{Ciupke:2015ora} and dualize the assumed axionic scalar field, $a$, to a Kalb-Ramond 2-form gauge potential, $B_{\mu\nu}$, and formulate the no-scale condition using these dual variables.\footnote{A brief reminder of how this duality works in 4D can be found in \S\ref{app:IIBshift}.} Since the natural supermultiplet for a Kalb-Ramond potential in 4D $\cN=1$ supergravity is the linear (not chiral) multiplet, the first step is to formulate the no-scale condition in terms of linear multiplets. 

As in the main text, we consider an $\mathcal{N}=1, D=4$ supergravity that involves three different types of chiral multiplets: $\{ T^i , G^{\hat{\imath}} , S^a \}$, where $i=1,\dots,n \ , \hat{\imath}=1,\dots,m$ and $a=1, \dots, p$. The three types of chiral multiplets are distinguished from one another by the following three conditions: 
\begin{enumerate}
\item The fields $\{T^i, G^{\hat{\imath}}\}$ are moduli for which axionic shift symmetries 
\be
T^i \rightarrow  T^i + i \lambda^i \ , \qquad G^{\hat{\imath}} \rightarrow G^{\hat{\imath}} + i \lambda^{\hat{\imath}}
\ee
restricts the K\"ahler potential to have the form $K = K(T+\bar T, G+\bar G)$. 
\item None of the fields $\{T^i, G^{\hat{\imath}}\}$ appear in the superpotential: $W = W(S^a)$. This second condition follows automatically from the first one if the corresponding shift symmetries are not anomalous. 
\item The K\"ahler potential for the fields $\{ T^i , G^{\hat{\imath}} , S^a \}$ depends only on a reduced number of $n+p$ variables. More precisely we ask
\be
 K(T + \ol{T}, G + \ol{G}, S,\ol{S}) = \cK\Bigl[ T^i + \bar{T}^i + \Sigma^i(G + \ol{G},S,\ol{S}) \Bigl] + \hat{\ssK}(S,\ol{S}) \,,
\ee
where $\Sigma^i$ are real-valued functions of $G^{\hat{\imath}} + \bar{G}^{\hat{\imath}},S^a$ and $\bar{S}^a$. We call a K\"ahler dependence with this reduced dependence a `coordinate degenerate' K\"ahler potential. 
\end{enumerate}

A special subclass of models with coordinate degeneracy are those where $m=0$ --- {\it i.e.}~no $G$ fields --- and the moduli space factorises into a product of manifolds $\mathfrak{M} = \mathfrak{M}_{T} \times \mathfrak{M}_{S}$. In general the moduli space does not factorise in this way and the fields $S^a$ and $T^i, G^{\hat{\imath}}$ have mixed kinetic terms. It is the complications to the no-scale description caused by this mixing that make it useful to discuss the no-scale property using the dual formulation wherein the chiral multiplets $T^i, G^{\hat{\imath}}$ are traded for linear multiplets. As we see below, this dualisation disentangles the chiral multiplets $S^a$ from the shift-symmetric directions $T^i, G^{\hat{\imath}}$, and so greatly simplifies the discussion of the scalar potential. In particular, dualisation allows the no-scale property to be stated in a much simpler form that permits the identification of all possible models of this type~\cite{Ciupke:2015ora}. 

\subsubsection*{Supergravity with linear multiplets}

We pause to briefly summarise how supergravity couples to linear and chiral multiplets. Linear multiplets are scalar superfields that satisfy the constraint equations\footnote{In this sense linear multiplets are what is left over in a scalar superfield once its left- and right-chiral parts are projected out.}~\cite{Cecotti:1987nw, Derendinger:1994gx, Binetruy:2000zx}
\be
(D^2 - 8\ol{R}) L_{\underline \ssI} = (\ol{D}^2 - 8R) L_{\underline \ssI} = 0 \ ,
\label{Def_linear_multi_app}
\ee
(where $R$ is the chiral superfield containing the Ricci scalar). Their bosonic degrees of freedom turn out to be $\left\{ L_{\underline \ssI}, B^{(2)}_{\underline \ssI} \right\}$, where $L_{\underline \ssI}$ is a real scalar and $B^{(2)}_{\underline \ssI}$ is a Kalb-Ramond two-form gauge potential.\footnote{We underline indices for linear multiplets to distinguish them from their dual counterparts, and capital letters collectively denote the axionic supermultiplets: $I = \{ i, \hat{\imath} \}$.}

To dualise consider a theory with chiral multiplets $S^a$ and linear multiplets $L_{\underline \ssI} = \{L^{(\ssT)}_{\underline i}, L^{(\ssG)}_{\underline{\hat{\imath}}}\}$ that are dual to the chiral multiplets $\Phi^\ssI = \{T^i, G^{\hat{\imath}}\}$. The Lagrangian of the theory is defined in terms of three different functions 
\be
\tilde{K}(L^{(\ssT)}_{\underline i}, L^{(\ssG)}_{\underline{\hat{\imath}}},S^a,\bar{S}^a) \ , \quad F(L^{(\ssT)}_{\underline i}, L^{(\ssG)}_{\underline{\hat{\imath}}},S^a,\bar{S}^a) \quad \hbox{and} \quad W(S^a) \ ,
\ee
where $\tilde{K}$ is a real-valued function --- called the Hesse potential --- which plays a role analogous to the K\"ahler potential in the purely chiral theory; $F$ is a real-valued function related to the choice of super-conformal frame; and $W$ is the holomorphic superpotential. EF normalisation for the Einstein-Hilbert action fixes $F$ to be
\be \label{FEFcond_app}
F = 1 - \frac13 \tilde{K}^{\underline i} L^{(\ssT)}_{\underline i} -\frac13 \tilde{K}^{\underline{\hat{\imath}}} L^{(\ssG)}_{\underline{\hat{\imath}}} \ ,
\ee
where (as before) $\tilde K^{\underline\ssI} := \partial \tilde K/\partial L_{\underline\ssI}$ and similarly for $F$ and $W$. 

It is also possible to define a function $K$ that coincides with the K\"ahler potential after dualisation and consequently takes the same name (even though in the formulation with linear multiplets there is no underlying K\"ahler geometry). It is defined as
\be
K = \tilde{K} + 3 F \ . 
\ee
In terms of $K$ the EF normalisation condition \pref{FEFcond_app} can be rewritten as
\be
 1 - \frac13 K^{\underline\ssI} L_{\underline\ssI} = F - F^{\underline\ssI} L_{\underline\ssI} \ .
\ee

The bosonic part of the 4D $\cN=1$ supergravity Lagrangian then is
\be
\begin{aligned}
\mathcal{L} & =  -\tfrac{1}{2}R*\mathbf{1} - \tilde K_{a {\bar{b}}}\, \exd S^a \wedge * \exd \bar{S}^{\bar{b}} + \tfrac{1}{4} \tilde K^{\underline\ssI \underline\ssJ}\,  \exd L_{\underline\ssI} \wedge * \exd L_{\underline\ssJ} - V * \mathbf{1} \\ 
& \quad + \tfrac{1}{4} \tilde K^{\underline\ssI \underline\ssJ}\, \exd B_{\underline\ssI}^{(2)} \wedge * \exd B_{\underline\ssJ}^{(2)}
- \tfrac{i}2\,  \exd B_{\underline\ssI}^{(2)} \wedge \big(\tilde {K^{\underline\ssI}}_a\, \exd S^a -\tilde {K^{\underline\ssI}}_{{\bar{b}}}\,\exd \bar{S}^{\bar{b}} \big)\ ,
\end{aligned}
\ee
with scalar potential 
\be
V = e^K \left[\tilde K^{a {\bar{b}}} D_a W D_{{\bar{b}}} \bar W - (3- L_{\underline\ssI} K^{\underline\ssI}) |W|^2  \right]\,.
\label{scalar_pot_app}
\ee
This scalar potential is positive semi-definite with flat directions at $V = 0$ if
\be
L_{\underline\ssI} K^{\underline\ssI} = 3 \ ,
\label{eq:no_scale_lm_app}
\ee
is satisfied identically for some of the fields. We use this condition to define the class of no-scale models of interest, and we show here that this is a broad enough definition to include some axionic generalised no-scale models that are not of the standard no-scale form once written using chiral multiplets.   

\subsubsection*{Dualisation of chiral and linear multiplets}

To understand how condition \pref{eq:no_scale_lm_app} relates to the no-scale definitions in the main text, we dualise the linear multiplets to get the dual description involving only chiral multiplets. The dualisation relates chiral and linear multiplets as follows:
\be
T^i + \bar{T}^i = \frac12\; \tilde{K}^{\underline i} \ , \qquad G^{\hat{\imath}} + \bar{G}^{\hat{\imath}} = \frac12 \;\tilde{K}^{\underline{\hat{\imath}}} \ ,
\label{dualisation_app}
\ee
where (as usual) $\tilde{K}^{\underline i} := \partial \tilde{K}/\partial L^{(\ssT)}_{\underline i}$ and $\tilde{K}^{\underline{\hat{\imath}}} := \partial \tilde{K}/\partial L^{(\ssG)}_{\underline{\hat{\imath}}}$. The K\"ahler potential of the chiral theory is then the Legendre transformation of $\tilde{K}$, and so
\be
K(T+\bar{T},G+\bar{G},S,\bar{S}) = \tilde{K}(L^{(\ssT)},L^{(\ssG)},S,\bar{S}) - 2(T^i + \bar{T}^i)L^{(\ssT)}_{\underline i} - 2(G^{\hat{\imath}} + \bar{G}^{\hat{\imath}}) L^{(\ssG)}_{\underline{\hat{\imath}}} \ ,
\ee
where $L^{(\ssT)}_{\underline i}$ and $L^{(\ssG)}_{\underline{\hat{\imath}}}$ are understood as functions of $T^i,G^{\hat{\imath}},S^a$, defined implicitly by eq.~\eqref{dualisation_app}. It is also useful to note the inverse relations
\be
K_i = - 2 L^{(\ssT)}_{\underline i} \ , \qquad K_{\hat{\imath}} = - 2 L^{(\ssG)}_{\underline{\hat{\imath}}} \ .
\ee

These dualisation relations allow the translation of eq.~\eqref{eq:no_scale_lm_app} into a condition for the purely chiral theory. After some algebra one finds that
\be  
\label{noscalechiralcomplicated_app}
L_{\underline\ssI} K^{\underline\ssI} = K^{\ssI \bar\ssJ} K_{\bar\ssJ} K_{\ssK} {\mathcal{M}^{\ssK}}_{\ssI} \,, 
\ee
where
\be  
{(\mathcal{M}^{-1})^{\ssI}}_{\ssJ} = \delta_{\ssJ}^{\ssI} - K^{\ssI {\bar{a}}} K_{\ssJ {\bar{a}}}- K^{a \bar\ssI} K_{a \bar\ssJ} \ .
\label{noscalechiralcomplicated2_app}
\ee
In these expressions the right-hand side involves the shift-symmetric chiral multiplets, $\Phi^\ssI = \{T^i,G^{\hat{\imath}}\}$, while the left-hand side involves the linear multiplets $L_{\underline\ssI} = \{ L^{(\ssT)}_{\underline i}, L^{(\ssG)}_{\underline{\hat{\imath}}}\}$. Using \pref{noscalechiralcomplicated_app} in \pref{eq:no_scale_lm_app} shows that the chiral version of the no-scale condition is
\be 
K^{\ssI \bar\ssJ} K_{\bar\ssJ} K_{\ssK} {\mathcal{M}^{\ssK}}_{\ssI} = 3\,,  
\label{noscalechiralcomplicated3_app}
\ee
where the matrix ${\mathcal{M}^\ssK}_\ssI$ is as in \pref{noscalechiralcomplicated2_app}. 

For the special case with $m=0$ (no $G$ multiplets) with a factorised space of moduli, $\mathfrak{M} = \mathfrak{M}_{T} \times \mathfrak{M}_{S}$, we have ${\cM^\ssI}_\ssJ = \delta^\ssI_\ssJ$ and so eq.~\pref{noscalechiralcomplicated3_app} reduces to the standard no-scale condition $K^{\ssI \bar\ssJ} K_{\ssI} K_{\bar\ssJ} = 3$. But in general eq.~\eqref{eq:no_scale_lm_app} --- or \pref{noscalechiralcomplicated3_app} --- gives a more general relation that ensures flat directions even when $K^{\ssI \bar\ssJ} K_{\ssI} K_{\bar\ssJ} \neq 3$, and so generalises the standard no-scale condition \pref{KKKnoscale}.

\subsection*{Scale invariance and dualised models}
\label{Sec23}

For the dual (linear multiplet) formulation, the no-scale relation of eq.~\eqref{eq:no_scale_lm_app} again involves homogeneous functions. In particular the no-scale condition turns out to require the functions $K$ and $F$ must have the form~\cite{Ciupke:2015ora}
\be
\label{eq:no_scale_lm_explicit_app}
e^K = H_3 \ , \qquad  F = H_1 \ ,
\ee
where $H_3$ is homogeneous of degree three and $H_1$ homogeneous of degree one in the fields $L_{\underline\ssI}$. The dependence on the chiral multiplets $S_a$ is arbitrary.  

We now identify scale transformations that are sufficient to ensure \pref{eq:no_scale_lm_explicit_app}. From the dualisation conditions eq.~\eqref{dualisation_app} we find that scale invariance in the linear multiplet formulation would require
\be
\label{eq:scale_app}
L^{(\ssT)}_{\underline i} \rightarrow \lambda^{-w_\ssT} L^{(\ssT)}_{\underline i} \ , \qquad L^{(\ssG)}_{\underline{\hat{\imath}}} \rightarrow \lambda^{-w_\ssG}  L^{(\ssG)}_{\underline{\hat{\imath}}} \, \qquad S^a \rightarrow S^a \quad \hbox{and} \quad  g_{\mu\nu} \rightarrow \lambda^2 g_{\mu \nu} \ , 
\ee
in which case $\mathcal{L} \rightarrow \lambda^2 \mathcal{L}$. Furthermore, the condition of coordinate degeneracy implies that $K$ is expressed entirely in terms of $L^{(\ssT)}_{\underline i}$ or
\be
K^{\underline{\hat{\imath}}} = \frac{\partial K}{\partial L^{(\ssG)}_{\underline{\hat{\imath}}}} = 0 \ .
\label{eq:coord_degen_app}
\ee
If the superpotential scales as $W\to \lambda^w\,W$, the form of the scalar potential in eq.~\eqref{scalar_pot_app} then allows the deduction
\be
e^K \rightarrow  \lambda^{-2(1+w)}\, e^K \ ,
\label{eq:scaling_eG_app}
\ee 
and so -- comparing with eq.~\eqref{eq:no_scale_lm_explicit_app} -- the scale invariant model is no-scale if
\be
w_\ssT = \frac23\,(1+w) \,.
\label{eq:weight_requirement_app}
\ee
The remaining weights for $L^{(\ssG)}_{\underline{\hat{\imath}}}$ (that is for the $G^{\hat{\imath}}$) are unfixed.

\section{Accidental low-energy shift symmetries}
\label{app:accidentalshift}
 
This appendix sketches the more detailed arguments for the existence of accidental shift symmetries relating moduli in heterotic and type IIB compactifications. In both cases the argument starts from the appearance of lower-dimension forms in the field strengths involving the relevant 4D axion field. Typically one has 
\be
   H = \exd B - \Xi(A) \,,
\ee
where $B$ is a $p$-form gauge potential and $\Xi$ is built from $q$-form potentials with $q<p$. These lower-dimensional forms -- denoted collectively by $A$ -- have gauge transformations $\delta A = \exd \Lambda_\ssA + \cdots$ (exterior derivative plus possible non-Abelian contributions). What is important is that $\Xi(A)$ is not itself gauge invariant but $\exd \Xi$ is: although $\delta \Xi$ is not zero, it satisfies $\exd \delta \Xi = 0$. As a result it is locally possible to find a quantity $\Sigma$ such that $\delta \Xi = \exd (\Lambda_\ssA \Sigma)$ and as a consequence it is always possible to ensure $H$ is invariant by assigning to $B_p$ the transformation rule $\delta B = \Lambda_\ssA \Sigma$ (in addition to its own gauge invariance $\delta B = \exd \Lambda_\ssB$).

These higher-dimensional gauge shifts induce a variety of shift symmetries in the 4D EFT whenever $B$ or $A$ contributes massless states to the low-energy theory. One way this can happen is if quantities like $\Lambda_\ssA \Sigma$ are non-zero when evaluated at the classical background configuration. Another relies on the existence of extra-dimensional harmonic forms, $\eta^\ssI$, that are typically associated with the existence of massless modes. Being harmonic means $\exd \eta^\ssI = 0$ but there does not exist globally defined quantities, $\varpi^\ssI$, such that $\eta^\ssI = \exd \varpi^\ssI$. Such forms allow transformations like $\delta B = \theta_\ssI \eta^\ssI$, which is a symmetry of $\exd B$ despite its {\it not} being a globally defined gauge transformation. We discuss examples for these mechanisms in the context of heterotic and Type~IIB respectively below.

\subsection{Heterotic compactifications}

Consider first the combination $H = \exd B - \Xi_\CS$ that arises in heterotic supergravity, where\footnote{All order unity numerical coefficients are for simplicity dropped from this discussion.} $B$ is a gauge 2-form and $\Xi_\CS$ is the Chern-Simons 3-form built from a non-Abelian gauge potential $A$. Under gauge transformations $\delta A = \exd \Lambda$ (for scalar gauge parameter $\Lambda$) the Chern-Simons form transforms as $\delta \Xi_\CS = \exd(\Lambda F)$ where $F$ is the 2-form gauge field-strength built from $A$. Invariance of $H$ requires $B$ to transform as $\delta B = \Lambda F$ under gauge transformations of $A$. 

The main text uses a symmetry transformation under which an axion mode in $B$ shifts by an amount linear in $A$ when $A$ shifts by a constant. To see how this might arise in 10D suppose $\eta$ is a harmonic 1-form\footnote{Harmonic 1-forms actually do not exist on the Calabi-Yau spaces of interest here, but a variation of the argument we now give nonetheless goes through with a few complications described below.} in the extra dimensions. Under the generalised (non-gauge) transformation\footnote{See for instance~\cite{Gaiotto:2014kfa}.} $\delta A = \eta$  the Chern-Simons form transforms as $\delta \Xi_\CS = \eta \wedge F$ (which the Bianchi identity $\exd F = 0$ ensures would agree with its gauge transformation rule --- with $\Lambda = -\varpi$ --- if there were to exist a $\varpi$ for which $\eta = \exd\varpi$). $H$ can be invariant under this transformation provided that $B$ transforms as $\delta B = \eta \wedge A$ so that $\exd \delta B = \eta \wedge F$ (using $\exd \eta = 0$). This transformation $\delta B$ is the shift linear in $A$ that we seek.

How does this apply for heterotic vacua, given that Calabi-Yau spaces do not have harmonic one-forms? In this case $A^{\beta}_\ssM$ transforms as an adjoint under the gauge group $E_8$, where $\beta$ is an $E_8$ index. For understanding the massless modes it is useful to decompose the 248-dimensional adjoint of $E_8$ according to its transformation properties under the subgroup $SU(3) \times E_6$, using $\mathbf{248} = (\mathbf{1}, \mathbf{ 78}) + (\mathbf{ 8} ,\mathbf{ 1}) + (\mathbf{ 3}, \mathbf{ 27}) + (\ol{\mathbf{ 3}}, \ol{\mathbf{ 27}})$.  This decomposition is useful because a Calabi-Yau space has $SU(3)$ holonomy, with simple compactifications identifying gauge and spin connections in the $SU(3)$ sector. This identification is not separately invariant under $SU(3)$ gauge and tangent-frame rotations, but is covariant under a diagonal $SU(3)$ subgroup that acts on both, allowing gauge and spacetime indices in the $SU(3)$ sector to be related by background quantities.

It then turns out that the low-energy matter fields arise as an expansion in terms of extra-dimensional harmonic forms of the form
\be
  B(x,z) =  b_\ssI(x) \, \omega^\ssI_{m\bar{n}}(z) \, \exd z^m  \exd \bar{z}^{\bar{n}} \quad \hbox{and} \quad
  A^{nr}(x,z) =   C^r_\ssI (x) (\omega^\ssI)_m^n(z)  \, \exd z^m
\ee
where $x^\mu$ and $z^m$ are respectively 4D and extra-dimensional coordinates, with $m,n = i , \bar\imath$ being $SU(3)$ triplet (or anti-triplet) indices and $r$ being a gauge index that runs over the $\mathbf{ 27}$ of $E_6$ (with the $E_8$ index given as the pair $\beta = (n,r)$). Finally, $\omega_{ml}$ is a harmonic (1,1)-form (a family of which does exist on Calabi-Yau spaces) whose index is raised using the Calabi-Yau metric. The index $I$ runs over a basis of independent harmonic (1,1)-forms. 

In terms of this decomposition the construction given above for the accidental shift symmetry goes through as before. The axionic shift symmetries $\delta b_\ssI = \theta_\ssI$ arise from the shift $\delta B_{mn} = \theta_\ssI \omega^\ssI_{mn}$ with constant $\theta_\ssI$. The accidental shift symmetry under coordinated shifts of $A$ and $B$ similarly has the form $\delta A = \eta$ and $\delta B_2 = \bar \eta \wedge A$, where
\be
   \delta A_j^{ir} = \eta^r \, \omega_j^i \quad \hbox{accompanied by} \quad
   \delta B_{i\bar\jmath} = \bar\eta_r \,C^r  \, \omega_{i\bar\jmath} \,,
\ee
for $\omega^\ssI$ the basis of harmonic two-forms and $\eta^r$ a collection of constant transformation parameters.

\subsection{Type IIB compactifications}

This section now presents a related (but different in detail) line of argument to establish a similar approximate shift symmetry for type IIB compactifications. Before giving this argument we first digress to remind the reader how axion/2-form duality works in 4 dimensions following the arguments of~\cite{Buscher:1987qj}. 

\subsubsection*{4D axion duality}
\label{app:IIBshift}

In 4D the duality between an axion $b$ and a 2-form gauge field $\cB _{\mu\nu}$ can be seen by starting from the action
\be\label{4Daxionduality}
  \frac{\cL_{\rm kin}}{\sqrt{-g}} =  - \frac{1}{2\cdot 3!} \,  \cH_{\mu \nu \lambda} \cH^{\mu \nu \lambda}   +  \frac{1}{3!} \, b\,\epsilon^{\mu\nu\lambda \rho}  \partial_\mu \cH_{\nu\lambda\rho}  \,,
\ee
where the functional integral is over an unconstrained 3-form field $\cH$ and a real scalar $b$. Integrating $b$ first imposes the Bianchi identity $\exd \cH = 0$ and so locally there exists a $\cB _{\mu\nu}$ such that $\cH = \exd \cB $, resulting in the usual 2-form formulation of the theory. 

The scalar dual is found by performing the functional integrals in the opposite order, starting with $\cH$. Dropping surface terms, the saddle point is at $\cH^{(c)}_{\mu\nu\lambda} = \epsilon_{\mu\nu\lambda\rho} \partial^\rho b$ and so integration leaves the Lagrangian
\be
  \frac{\cL_{\rm kin}}{\sqrt{-g}}   = \frac{1}{2\cdot 3!} \,   \epsilon_{\mu\nu\lambda\rho} \, \epsilon^{\mu\nu\lambda\sigma} \partial^\rho b\,  \partial_\sigma b = - \frac12 \, \partial^\mu b \, \partial_\mu b \,,
\ee
showing that $b$ as defined is canonically normalised. What is important is that this transformation works equally well for {\it any} Lagrangian built purely from $\cH$, although in general the integration over $\cH$ is then more difficult to evaluate. In the applications of interest terms cubic and higher in the fields are treated perturbatively, so the only terms in $\cL$ that affect the saddle point are those at linear and quadratic order.

Applications of this formalism to IIB compactifications run into complications due to the self-duality of the 5-form field strength. Because of its self-duality both fields $b$ and $\cB_{\mu\nu}$ appear simultaneously in the dimensional reduction. To keep track of such issues we adapt the framework of~\cite{Sen:2015nph, Sen:2019qit} to this dimensional reduction, in order to have a formulation that works when kinetic terms of the form $\cH_{\mu\nu\lambda} \cH^{\mu\nu\lambda}$ and $(\partial b)^2$ both appear simultaneously. 

To this end consider the following starting point as an alternative to \pref{4Daxionduality}:
\be \label{manifestdual}
 - \frac{\cL_{\rm kin}}{\sqrt{-g}} =   \frac{1}{4!} \,   \cH_{\mu \nu \lambda} \cH^{\mu \nu \lambda}  +  V_\mu \,V^\nu -  \frac{1}{3!} \,\epsilon^{\mu\nu\lambda \rho}  \cH_{\nu\lambda\rho} ( V_\mu - \partial_\mu b) \,,
\ee
with integrations over the three independent unconstrained fields $b$, $V_\mu$ and $\cH_{\mu\nu\lambda}$. Integrating out $V_\mu$ first leads to the saddle point condition
\be \label{Vsadpt}
   V_\mu^{(c)} = \frac{1}{2\cdot 3!} \,\epsilon_{\mu \nu\lambda\rho} \cH^{\nu\lambda\rho}
\ee
and (after integrating by parts) gives the Lagrangian \pref{4Daxionduality}. Integrating this over $b$ or $\cH$ then shows, as before, that \pref{manifestdual} is equivalent to both the Kalb-Ramond and the dual scalar formulation described above. This same conclusion can be drawn equally well by integrating first $b$ or $\cH_{\mu\nu\lambda}$ rather than $V_\mu$. 

The spirit of using $V_\mu$ and $\cH_{\mu\nu\lambda}$ is that interaction terms (such as obtained by dimensional reduction) give $\cL_{\rm int}(\cH, V)$ involving all possible powers of each type of field. All terms cubic or higher are to be treated perturbatively and so do not change the saddle point position and the rest of the arguments made above. What is of interest is when these interaction terms contain terms that are quadratic or linear in the fields, since these change the saddle point relations and so can change the leading-order part of the action.

\subsubsection*{D3 dimensional reduction}

We next sketch the dimensional reduction that produces these axion fields. Consider first the axion that appears with the volume modulus in $T$ that comes from $C_4$ in the compactified form
\be
  C_4(x,z) = b(x) \, \epsilon_{mnpqrs} \omega^{rs}(z)\, \exd z^m \wedge \exd z^n \wedge \exd z^p \wedge \exd z^q + \cB _{\mu\nu}(x) \; \omega_{mn}(z) \, \exd x^\mu \wedge \exd x^\nu \wedge \exd z^m \wedge \exd z^n 
\ee
and (as before) the $x^\mu$ denote the space-filling on-brane directions and $z^m$ are the 6 extra dimensions. The brane position is at $z^m = y^m(x)$ and we use the gauge where the 4 on-brane coordinates are equated to the 4 target-space space-filling coordinates $x^\mu$. The fields $\cB _{\mu\nu}$ and $b$ are not independent because they are dual to one another in the 4D sense, as is required because the field strength for $C_4$ is self-dual in 10D. 

The kinetic term for $\cB _{\mu\nu}$ comes from compactifying $\cL \propto \tilde F_{\mu\nu\lambda mn}^2$ and $\tilde F_5 = F_5 + $(lower-dimensional forms), where $F_5 = \exd C_4$. Dimensional reduction of $\tilde F_{\mu mnpq}^2$ also gives a kinetic term for $b$ proportional to $(\partial b)^2$. So keeping in mind these are not independent fields the axionic part of the total dimensional reduction of $\tilde F_5^2$ can be regarded as being
\be
  -\frac{\cL_{\rm kin}}{\sqrt{-g}} =   \frac{1}{4!} \,  \cH_{\mu \nu \lambda} \cH^{\mu \nu \lambda}   + V_\mu V^\mu    \,,
\ee
where $\cH ^2$ comes from $\tilde F_{\mu\nu\lambda mn}^2$ and $V^2$ comes from $\tilde F_{\mu mnpq}^2$ (with $\partial_\mu b$ replaced by $V_\mu$) once other fields are integrated out. 4D fields are assumed to be rescaled to be normalised as in the example above, which yields canonical normalisations once finally written in terms of either $\cH_{\mu\nu\lambda}$ or $\partial_\mu b$.

Consider now the coupling of $C_4$ to a space-filling D3 brane, which includes the term $\int C_4$ over the brane world-volume. This leads to the dimensionally reduced Lagrangian
\bea \label{LC4result}
 \frac{\cL_{C4}}{\sqrt{-g}} &=& \frac{1}{4!} \, \epsilon^{\mu\nu\lambda\rho} C_{\mu\nu\lambda\rho} \ni  - \frac{1}{3!}  \epsilon^{\mu\nu\lambda\rho} \Bigl(\cC_{mn} \cB _{\mu\nu} \partial_\lambda y^m \partial_\rho y^n + \cC_{mnpq} \, b \, \partial_\mu y^m \partial_\nu y^n \partial_\lambda y^p \partial_\rho y^q  \Bigr)  \nn\\
 &=&  \frac{1}{3!} \epsilon^{\mu\nu\lambda\rho} \Bigl(  \cC_{mn} \,\cH_{\mu\nu\lambda}  y^m \partial_\rho y^n + \tilde\cC_{mnpq} \, \partial_\mu b \,  y^m \partial_\nu y^n \partial_\lambda y^p \partial_\rho y^q  \Bigr)
\eea
where the last line integrates once by parts and the quantities $\cC_{mn}$ and $\tilde \cC_{mnpq}$ are coefficients involving $\omega_{mn}$ evaluated at the brane position. Replacing $\partial_\mu b \to V_\mu$ in \pref{LC4result} and adding the result to \pref{manifestdual} then gives
\be \label{manifestdual2}
 - \frac{\cL_{\rm kin}+\cL_{C_4}}{\sqrt{-g}} =  \frac{1}{4!} \,    \cH_{\mu\nu\lambda} \cH^{\mu\nu\lambda} +   V_\mu \,V^\mu +  \frac{1}{3!} \,\epsilon^{\mu\nu\lambda \rho}  \cH_{\nu\lambda\rho} \Bigl( \partial_\mu b - \cC_\mu  \Bigr)  -\frac{1}{3!} \epsilon^{\mu\nu\lambda\rho}\, V_\mu \Bigl( \cH_{\nu\lambda\rho} - \cC_{\nu\lambda\rho}  \Bigr) \,,
\ee
where
\be
  \cC_\mu := \cC_{mn}   \,y^m \partial_\mu y^n \quad\hbox{and} \quad
  \cC_{\nu\lambda\rho} := \cC_{mnpq} \,  y^m \partial_\nu y^n \partial_\lambda y^p \partial_\rho y^q \,.
\ee

Integrating out $V_\mu$ leads to the new saddle point  
\be \label{Vsadptv2}
   V^{(c)}_\mu  =  \frac{1}{2\cdot 3!} \,\epsilon_{\mu \nu\lambda\rho} \Bigl( \cH^{\nu\lambda\rho} - \cC^{\mu\nu\lambda}  \Bigr) 
\ee
leading to the Lagrangian
\be
 - \frac{\cL_{\rm kin}+\cL_{C_4}}{\sqrt{-g}} =  \frac{1}{2\cdot 3!} \,    \cH_{\mu\nu\lambda} \cH^{\mu\nu\lambda} - \frac{1}{2\cdot 3!}  \cH_{\nu\lambda\rho} \cC^{\nu\lambda\rho}  + \frac{1}{3!} \,\epsilon^{\mu\nu\lambda \rho}  \cH_{\nu\lambda\rho} \Bigl(  \partial_\mu b- \cC_{\mu}   \Bigr)  +\frac{1}{4!}   \cC^{\nu\lambda\rho}   \cC_{\nu\lambda\rho}    \,.
\ee
Finally, integrating out $\cH_{\mu\nu\lambda}$ using the saddle point
\be
   \cH^{(c)}_{\mu\nu\lambda} = \epsilon_{\mu\nu\lambda\rho} \Bigl(\partial^\rho b - \cC^\rho \Bigr) + \frac12 \, \cC_{\mu\nu\lambda}
\ee
gives
\be
  - \frac{\cL_{\rm kin}+\cL_{C_4}}{\sqrt{-g}} =   \frac{1}{2} \,  \Bigl(  \partial_\mu b- \cC_\mu  \Bigr)\Bigl(  \partial^\mu b- \cC^\mu \Bigr) + \frac{1}{2\cdot 3!} \,\epsilon^{\mu\nu\lambda \rho}  \cC_{\nu\lambda\rho} \Bigl(  \partial_\mu b- \cC_\mu \Bigr)   +\frac{1}{2\cdot 4!}   \cC^{\nu\lambda\rho}   \cC_{\nu\lambda\rho}   \,.
\ee

The terms involving $\cC_{\mu\nu\lambda}$ involve four or more derivatives and so can be dropped to the extent that 2-derivative terms are of interest. Notice also that $b$ only appears in the combination $\partial_\mu b - \cC_{mn} \, y^m \partial_\mu y^n$. Because of this $\cL$ has the symmetry
\be
  \delta y^m = \xi^m \quad \hbox{and} \quad \delta b = \cC_{mn}\, \xi^m y^n \,,
\ee
for constant $\xi^m$ (and we work in a basis where $y^m$ are real). Although quantities like $\cC_{mn}$ in principle depend on fields like $\omega_{mn}$ evaluated at $z = y(x)$, they are here treated as constants because we expand $y^m = y^m_{(c)} + \mfy^m$ and at leading order $\cC_{mn}$ depends only on the background $y_{(c)}$ while $y^m \partial_\nu y^n = \mfy^m \partial_\mu \mfy^n$ at lowest non-trivial order.  In complex coordinates ---  assuming $\cC_{i\bar\jmath} = -\cC_{\bar\jmath i} = \ov{\cC}_{\bar\imath j}$ are the only non-zero components --- $\cC_{mn} y^m \partial_\mu y^n$ becomes $\frac12 \,\cC_{i\bar\jmath} (y^i \partial_\mu \bar y^j - \bar y^j \partial_\mu y^i)$ and so $\partial_\mu b - \frac12\,\cC_{i\bar\jmath} (y^i \partial_\mu \bar y^j - \bar y^j \partial_\mu y^i)$ is invariant under
\be \label{app:NewShiftq}
 \delta y^i = \xi^i \,, \quad \delta \bar y^i = \bar \xi^i \quad \hbox{and} \quad
 \delta b =  \frac12\,\cC_{i\bar\jmath} (\xi^i  \bar y^j - \bar \xi^j   y^i) \,,
\ee
for constant $\xi^i$. This transformation of $b = \hbox{Im}\; T$ is consistent with a holomorphic transformation of $T$ if $\delta T = i\cC_{i\bar\jmath} \, \bar \xi^j  y^i$. 

Such a symmetry restricts how the fields $T$ and $y^m$ can appear in the 4D effective supergravity, implying in particular that any appearance of $T$ should come together with $\cC_{i\bar\jmath} y^i \bar y^{j}$ so as to remain invariant under \pref{app:NewShiftq}. Combined with the scaling information found in the main text this leads to a supergravity K\"ahler potential of the form
\be \label{app:KlogformD3}
   K = - \ln (\tau - \bar \tau) - 3 \ln(T+\ol T - i\cC_{i\bar\jmath} \, y^i \bar y^j ) \,,
\ee
in agreement with explicit dimensional reduction. 

\subsubsection*{Higher orders in $y$}

The previous section suggests that coupling to a D3 brane causes the axionic shift symmetry $\delta b =$ (constant) to be gauged by a $y$-dependent connection $\cC_\mu(y)$. Does this have a deeper symmetry origin in 10D? And, if so, what can be said about $\cC_\mu(y)$ at higher orders in the D3 fluctuation field $\mfy^n := y^n - y^n_{(c)}$? 

From the 10D point of view the axionic shift symmetry starts life as a shift of $C_4$ by a harmonic form in the extra dimensions: $\delta C_{mnpq} \propto \omega_{mnpq}$. This is a symmetry of the bulk action and also preserves the brane coupling $\int C_4$ in the special case of a static space-filling brane located at a specific spot, $z^m = y_{(c)}^m$, which does not depend on $C_{mnpq}$ because $\partial_\mu y^m_{(c)} = 0$. Invariance of the brane coupling to $C_4$ is more subtle once the brane is allowed to bend, however, since $y^m = y^m_{(c)} + \mfy^m(x)$ allows $\partial_\mu y^m$ to be non-zero and so allows the possibility that the integral over the pull-back of $\delta C_{mnpq}$ might be non-zero. But this apparent breaking is illusory because it can be cancelled by acting with an appropriate diffeomorphism $\delta y^i = \xi^i(y)$, which although a symmetry of the bulk action is itself spontaneously broken by the brane's presence at a fixed extra-dimensional position. 

From the 4D point of view the combined shifts of both $b$ and $y^m$ must appear as a non-linear realisation of the original shift symmetry~\cite{Coleman:1969sm}. Normally for coset-space sigma-models this involves the appearance of a composite gauge field like $\cC_\mu(y)$ whose appearance compensates for the position-dependence that enters into transformations due to the dependence on the local field $y^m(x)$ when non-linearly realised. What is at first sight puzzling is that a similar construction can also occur for the D3-position sigma-model, given that the target space explored by $y^m$ is not a coset space; it is instead a Calabi-Yau space having no isometries. A $y$-dependent non-linear realisation of the axion shift symmetry is nonetheless possible because the Calabi-Yau space is not really the supergravity target space; the Calabi-Yau space is instead what the supergravity target space becomes once scaling symmetries are projected out by fixing the compensator field $\Phi$. Before projection the $\cN=1$ supergravity target space always has at least one homothetic Killing vector (corresponding to scaling) and a Killing vector (associated with the $R$ symmetry that transforms all fields equally)~\cite{FvP}. The composite gauge field required to non-linearly realise the axionic shift symmetry is the one associated with this $R$-symmetry transformation. 

This connection with $R$ transformations can be seen at leading order in $\mfy^m$ in the previous section. We know that the leading coefficient of $\partial^\mu y^m \partial_\mu y^n$ coming from the DBI action is the Calabi-Yau metric, which in complex coordinates is proportional to\footnote{The proportionality constant here involves powers of the extra-dimensional warped volume.} $h_{i \bar\jmath} = \partial_i \partial_{\bar\jmath}k$, where $k(y,\bar y)$ is the CY K\"ahler potential. The previous section's discussion applies at leading order in $\mfy^i$ for which $k \simeq \delta_{i\bar\jmath} \, \mfy^i \bar \mfy^j+ \cdots$. In terms of $k$ the `composite' gauge field $\cC_\mu$ is consistent at leading order with
\be
  \cC_\mu(\mfy) = \frac12\,\cC_{i\bar\jmath} (\mfy^i \partial_\mu \bar \mfy^j - \bar \mfy^j \partial_\mu \mfy^i) \propto  \partial_{\bar\jmath} k \, \partial_\mu \bar y^j - \partial_i k \, \partial_\mu y^i \,,
\ee
which is the pull-back to the brane of the K\"ahler connection built from $k$, and is related to the $R$-symmetry Killing vector to this order.

This is connection is suggestive because the action of an $R$ symmetry on $k$ is to induce a K\"ahler transformation: $\delta k = \xi^i \partial_i k + \xi^{\bar\imath} \partial_{\bar\imath} k = r(y) + \ol{r(y)}$ for some holomorphic function $r(y)$~\cite{FvP}. This suggests that the appropriate non-linear generalisation of \pref{app:NewShiftq} should be 
\be
  \delta y^m = \xi^m(y) \quad \hbox{with} \quad \delta T = r(y) 
\ee
as suggested by~\cite{Kachru:2003sx, Baumann:2007ah}, generalising \pref{app:KlogformD3} to
\be
  K = -3 \ln\Bigl[ T+\ol T - k(y,\bar y) \Bigr]  \,. 
\ee
This does indeed correctly capture the leading kinetic and kinetic-mixing terms predicted for $T$ and $y^i$ by dimensional reduction. 

\subsubsection*{D7 couplings}

A very similar story also goes through for the couplings between D7 position moduli and the 10D axio-dilaton $\tau = C_0 + i e^{-\phi}$. In this section lattice indices $\mu,\nu$ denote space-filling 4D directions, middle-of-the-alphabet indices $m,n,p$ (or $i,j$ when complex) represent off-brane extra dimensions while $a,b,c$ represents on-brane extra dimensions.  

The direct coupling of a D7 brane to $C_0$ (as opposed to its field strength) comes from the Lagrangian term $\int \Omega_8$, where the 8-form $\Omega_8$ contains $C_0$ in two ways. It appears directly through the appearance in $\Omega_8$ of the dual $C_8$ --- where $\exd C_8 = \hat\star \exd C_0$ --- and it also appears through the lower-dimensional piece $C_0 \wedge B_2 \wedge B_2 \wedge B_2 \wedge B_2$ and its counterparts with $B_2$ replaced by the D7 gauge-boson field-strength $F_2$. The Lagrangian therefore contains both terms involving $C_0$ undifferentiated and others involve only derivatives of $C_0$ (through its dual).

In 4D the dual of $C_0$ is a two-form gauge potential $\cB_{\mu\nu}$ whose field strength $\cH = \exd \cB$ as usual is given by $\cH_{\mu\nu\lambda} = \epsilon_{\mu\nu\lambda\rho} \partial^\rho C_0$. The dimensional reduction of $C_8$ therefore contains $C_0$ through
\be 
  C_8(x,z) = \cB_{\mu\nu}(x) \omega_{mnpqrs}(z) \,\exd x^\mu \wedge \exd x^\nu \wedge \exd z^m \wedge \cdots \wedge \exd z^s
\ee
where $\omega_{mnpqrs} \propto \epsilon_{mnpqrs}$ is related to the volume-form of the extra dimensions. Pulled back to the brane world volume this becomes
\be
  C_{\mu\nu\lambda\rho \,abcd} = \cB_{\mu\nu}(x) \omega_{mnpqrs}(z)\, \partial_\lambda y^m \partial_\rho y^n \partial_a y^p \partial_b y^q \partial_c y^r \partial_d y^s \ni \cC^{\ssA\ssB} \cB_{\mu\nu} \partial_\lambda y_\ssA \partial_\rho y_\ssB \,,
\ee
where $y_\ssA$ are the D7 moduli defined by $\omega_{mn}(y) = y_\ssA \omega^\ssA_{mn}$ -- as in \pref{yAdef} -- where $\omega^\ssA_{mn}$ are harmonic 2-forms and $\omega_{mn}(y) = \Omega_{mnp} y^p$ where $\Omega$ is the Calabi-Yau's holomorphic 3-form, while the $\cC^{\ssA\ssB} = -\cC^{\ssB\ssA}$ involve an integral over the wrapped extra-dimensional 4-cycle weighted by the pull-backs of various harmonic forms.

The other contribution linear in the fluctuation $C_0$ arises in the form
\be
  C_0 \wedge B_2 \wedge B_2 \wedge B_2 \wedge B_2 \ni  \cC^{\ssA\ssB\ssC\ssD} C_0(x) \partial_\mu y_\ssA \partial_\nu y_\ssB \partial_\lambda y_\ssC \partial_\rho y_\ssD ~,
\ee
where the coefficient $\cC^{\ssA\ssB\ssC\ssD}$ contains integrations over the background configuration on the wrapped 4-cycle and the $y$-dependence comes from the pull back to the space-filling directions of the brane.

From here on the story goes much as it did for D3 branes, with a few minor differences. One difference is that the kinetic term for $C_0$ in the bulk does not appear in its dual form but directly as $(\partial C_0)^2$ coming from the $\partial \tau \partial \bar \tau$ term in the 10D Lagrangian, though this is not important when using the manifestly duality invariant framework of previous sections. Ignoring the dilaton factors (which can be re-instated later using the scaling properties of the main text), this gives
\be \label{manifestdual7}
 - \frac{\cL_{\rm kin}+\cL_{WZ}}{\sqrt{-g}} =  \frac{1}{4!} \,    \cH_{\mu\nu\lambda} \cH^{\mu\nu\lambda} +   V_\mu \,V^\mu +  \frac{1}{3!} \,\epsilon^{\mu\nu\lambda \rho}  \cH_{\nu\lambda\rho} \Bigl( \partial_\mu C_0 - \cC_\mu  \Bigr)  -\frac{1}{3!} \epsilon^{\mu\nu\lambda\rho}\, V_\mu \Bigl( \cH_{\nu\lambda\rho} - \cC_{\nu\lambda\rho}  \Bigr) \,,
\ee
where now
\be
  \cC_\mu :=  \cC^{\ssA\ssB} \,   y_\ssA \partial_\mu y_\ssB  \quad\hbox{and} \quad
  \cC_{\nu\lambda\rho} := \cC^{\ssA\ssB\ssC\ssD} \,  y_\ssA \partial_\nu y_\ssB \partial_\lambda y_\ssC \partial_\rho y_\ssD \,.
\ee
Integrating out $V_\mu$ and $\cH_{\mu\nu\lambda}$ as before then again gives
\bea
  - \frac{\cL_{\rm kin}+\cL_{WZ}}{\sqrt{-g}} &=&    \frac{1}{2} \,  \Bigl(  \partial_\mu C_0- \cC_\mu  \Bigr)\Bigl(  \partial^\mu C_0- \cC^\mu \Bigr) \\
  &&\qquad \qquad \qquad+ \frac{1}{2\cdot 3!} \,\epsilon^{\mu\nu\lambda \rho}  \cC_{\nu\lambda\rho} \Bigl(  \partial_\mu C_0- \cC_\mu \Bigr)   +\frac{1}{2\cdot 4!}   \cC^{\nu\lambda\rho}   \cC_{\nu\lambda\rho} \nn\,.
\eea

Notice that the 2-derivative terms depend only on the combination $\partial_\mu C_0 -\cC^{\ssA\ssB}   \,y_\ssA \partial_\mu y_\ssB$ and so have the symmetry $\delta y_\ssA = \xi_\ssA$ and $\delta C_0 = \cC^{\ssA\ssB}\, \xi_\ssA y_\ssB$, for constant $\xi_\ssA$ (in a basis for which the $y_\ssA$ are real). In a complex basis  --- assuming the only non-zero terms are $\cC^{\bar\ssA \ssB} = - \cC^{\ssB \bar \ssA} = \cC^{\ssA \bar\ssB *}$ --- the combination $\partial_\mu C_0 -\frac{1}2\, \cC^{\ssA\bar\ssB} (y_\ssA \partial_\mu \bar y_{\ssB} - \bar y_\ssB \partial_\mu y_\ssA)$ is invariant under
\be
   \delta y_\ssA = \xi_\ssA \,, \quad \delta \bar y_\ssA = \bar \xi_\ssA \quad
    \hbox{and} \quad \delta C_0 = \frac{1}2\, \cC^{\ssA \bar\ssB} (\bar \xi_\ssB y_\ssA - \xi_\ssA \bar y_\ssB) \,,
\ee
which can be promoted to a holomorphic version
\be
   \delta y_\ssA = \xi_\ssA \,, \quad \delta \bar y_\ssA = \bar \xi_\ssA \quad
    \hbox{and} \quad \delta \tau = \cC^{\ssA \bar\ssB} \bar \xi_\ssB y_\ssA \,,
\ee
where $C_0 = \hbox{Re}\;\tau$. Such a symmetry argues that $\tau$ and $y_\ssA$ must appear in the K\"ahler potential only in the combination $\tau - \bar \tau - \cC^{\ssA \bar\ssB} y_\ssA \bar y_\ssB$, and so argues that 
\be
 e^{-K/3} = (\tau - \bar \tau - \cC^{\ssA\bar\ssB} y_\ssA \bar y_\ssB)^{1/3} (T + \ol T - \cC_{i\bar\jmath} \,y^i \bar y^j).
\ee

\section{Scaling in the 10D IIB theory}
\label{app:10Dscaling}

This appendix goes through the scaling properties of various terms of the 10D type IIB action, for which details are not given in the main text. (We have also checked the fermionic terms, though do not reproduce this here since it is not needed elsewhere in the text.) The first part of this section shows that the WZ and the non-Abelian DBI action both scale in the same way as does the Abelian DBI action, for all D$p$-branes, without any further restrictions on the scaling weights. The second part identifies a pitfall associated with back-reaction that can arise when diagnosing the scaling behaviour of the couplings of branes to forms (and why this pitfall is not a problem in the present instance).

\subsection{Scaling and brane actions}
\label{app:WZscaling}

We consider first the WZ action, then turn to non-Abelian interactions amongst stacks of coincident branes.

\subsection*{WZ action}
\label{ssec:WZscalingcalc}

The Wess-Zumino part of the action is
\be
S_\WZ =   \mu_p \int_{S_p} \Omega_{p+1}\,e^{2\pi\alpha' F} \,,
\label{SCS}
\ee
where $\Omega_{p+1}$ represents the pull-back of a $(p+1)$-form given by
\be
\Omega_{p+1} = \sum C_n\, e^{B_2}\qquad n=0,2,4,6,8\,.
\label{Oop}
\ee 
After expanding the exponentials in (\ref{Oop}), the $(p+1)$-form $\Omega_{p+1}$ is given by a sum of terms which schematically look like
\be
\Omega_{p+1} = \sum_k C_n\,B_2^k \qquad \text{with}\qquad n+2k=p+1\,.
\label{Oopp1}
\ee

For example for $p = 3$ and $p=7$ the forms $\Omega_4$ and $\Omega_8$ are schematically\footnote{As in the main text all numerical factors are dropped since they do not affect scaling behaviour.}
\be
\Omega_4 = C_4  + C_2 B_2 + C_0 B_2^2\,, \qquad \Omega_8 = C_8 + C_6 B_2 + C_4 B_2^2 + C_2 B_2^3 + C_0 B_2^4 \nn \,,
\ee
where $C_8$ and $C_6$ are the 10D Hodge duals of $C_0$ and $C_2$, defined by
\be
    \exd C_8  = \hat*\exd C_0 \quad\hbox{and} \quad
    \exd C_6  = \hat*\exd C_2 \,,
\ee
where the hat is a reminder that the Hodge dual is defined using the SF metric. All the terms in (\ref{Oopp1}) scale in the same way because \pref{F5B2reln} and \pref{weights_F5_etc} imply 
\be
w_{C_n} = w_\tau+ \frac{n}{2}\, w_{B_2}\,.
\ee
Therefore, using the constraint in (\ref{Oopp1}), $S_\WZ$ for a $p$-brane action scales with weight
\be
w_\WZ (p) = w_\tau+ \frac12\left(p+1\right) w_{B_2} = w_\DBI(p) \,,
\ee
where the last equality uses \pref{weights_F5_etc} and \pref{wDBIdef}. Evidently the DBI and WZ contributions to the brane action scale the same way (for any $p$). Notice that this keeps being true also when expanding the exponential of the field strength $F$ in (\ref{SCS}) since, as can be seen from (\ref{openstringmaxwellwt}), $F$ scales with the same weight as the pull-back of $B_2$.

\subsection*{Non-Abelian brane action}
\label{ssec:NonAbDBIscalingcalc}

For a stack of D$p$-branes, the DBI action (\ref{DBIaction}) needs to be generalised to the non-Abelian case. This has been derived in~\cite{Myers:1999ps} in the static gauge where spacetime and worldvolume diffeomorphisms are used to set $y^a (\zeta^a) =  \zeta^a$, for $a = 0, 1,..., p$, while $y^i$, with $i = p + 1,..., 9$, are identified as fluctuations around $y^i=0$ in the directions transverse to the brane. Notice that the $y^i$'s are matrix valued and transform in the adjoint representation of the gauge group localised on the D$p$-brane stack which, depending on the type of orientifold projection, can be either $U(N)$ or $SO(N)$ or $USp(N)$.

Thus in static gauge the pull-back of the SF metric takes the form
\be
\hat g_{ab} = \hat g_{\ssM\ssN}\,\partial_a y^\ssM\,\partial_b y^\ssN = 
\hat g_{ab}+ \left(\hat g_{ia}\,\partial_b y^i +\hat g_{ib}\,\partial_a y^i\right)+ \hat g_{ij}\, \partial_a y^i\,\partial_b y^j\,,
\label{Pullback}
\ee
where all the terms in (\ref{Pullback}) scale homogeneously with the same weight as the metric.

The action for a stack of D$p$-branes in 10D string frame and in static gauge is $S_{\rm loc} = S_\DBI + S_\WZ$ where $S_\WZ$ denotes the Wess-Zumino contribution to which we return below, while the DBI term is given by
\be
S_\DBI = - \mu_p \int_{S_p} \mathrm{d}^{p+1} \zeta\; {\rm STr} e^{-\phi}\sqrt{-{\rm det} \left[P\left(E_{ab}+E_{ai}\left(Q^{-1}-1\right)^{ij}E_{jb}\right) +2\pi\alpha' F_{ab}\right]{\rm det} Q^i_j}  \,.
\label{SDBI}
\ee
Here $E_{ab}$ is the pull-back of $E_{\ssM\ssN}=\hat g_{\ssM\ssN} + B_{\ssM\ssN}$, the non-Abelian field strength is  $F =\exd A+f\,[A, A]$ and $Q$ is defined as
\be
Q^i_j := \delta^i_j + \frac{i\,f}{2\pi\alpha'}\left[y^i,y^k\right] E_{kj}\,.
\label{EMN}
\ee
Notice that, after conversion to EF, (\ref{SDBI}) reduces to (\ref{DBIaction}) in the Abelian case where $\left[y^i,y^k\right]=0$. Moreover, the appearance of the dimensionless parameter $f$ in front of the commutator between two $y$'s is justified by the fact that the action of all D$p$-branes with $p<9$ can be obtained by dimensional reduction from the D$9$-brane action in 10D with the identification $y^i=2\pi\alpha' A^i$~\cite{Witten:1995im, Bachas:1995kx, Tseytlin:1996it}.

The two contributions in $F_{ab}$ scale homogeneously only if the dimensionless parameter $f$ is considered as a spurion field which transforms as\footnote{The dimensionless non-Abelian spurion arises here for reasons similar to its presence in the heterotic discussion of \S\ref{ssec:hetroscale} and it can be regarded as being associated with the structure constants of the non-Abelian gauge group.}
\be 
f \rightarrow  \lambda^{w - 2 w_g}\, f \,.
\label{SpurionTransform}
\ee
Notice that with this transformation property the term proportional to $\left[y^i,y^k\right]$ in (\ref{EMN}) is automatically scale invariant. Therefore, thanks to the spurion field $f$, the Abelian and non-Abelian DBI actions scale with the same weight.

Returning now to the Wess-Zumino part of the action, (\ref{SCS}) is generalised to
\be
S_\WZ =   \mu_p \int_{S_p} {\rm Str} \left[ \Omega_{p+1}\,  e^{2\pi\alpha' F} \right] \,,
\label{SCSgen}
\ee
where $\Omega_{p+1}$ is now the pull-back of a $(p+1)$-form given by
\be
\Omega_{p+1} = e^{i f\, \iota_y \iota_y} \left(\sum C_n\, e^{B_2}\right)\,,\qquad n=0,2,4,6,8\,,
\label{Op}
\ee 
where $\iota_y$ represents the interior product with $y^i$. After expanding the exponentials in (\ref{Op}), the $(p+1)$-form $\Omega_{p+1}$ is given by a sum of terms which schematically look like
\be
\Omega_{p+1} = \sum_{k_1, k_2} C_n\,B_2^{k_1}\,\left(f\left[y,y\right]\right)^{k_2}\qquad \text{with}\qquad n+2\left(k_1-k_2\right)=p+1\,.
\label{Opp1}
\ee
For example for $p = 3$ the form $\Omega_4$ is schematically
\bea
\Omega_4 &=& C_4  + C_2  B_2 + C_0 B_2^2 + f\left[y,y\right] \left( C_6  + C_4 B_2 + C_2  B_2^2 + C_0 B_2^3 \right) \nonumber \\
&+& f^2\left[y,y\right]\left[y,y\right] \left( C_8  + C_6 B_2 + C_4 B_2^2 + C_2 B_2^3 + C_0 B_2^4 \right) + ... \nonumber
\eea
These terms all scale in the same way because \pref{weights_F5_etc} combined with (\ref{SpurionTransform}) imply 
\be
f\left[y,y\right]\to\lambda^{-w_{B_2}} \,f\left[y,y\right]\,.
\ee
Therefore, using the constraint in (\ref{Opp1}), also the non-Abelian WS action scales with the same weight as the Abelian DBI action for any $p$ since
\be
w_\WZ = w_\tau+\frac12 \left[n + 2\left(k_1-k_2\right)\right] w_{B_2}  = w_\DBI \,.
\ee

\subsection{Scaling and back-reaction: a dog that didn't bark}
\label{app:fluxscaling}

The case just made for the scaling of the WZ action takes for granted that the scaling properties of a localised source can be read off directly from the scaling properties of its action. We close this appendix with a description of a subtlety that could have invalidated this type of reasoning when assessing the scaling properties of Wess-Zumino-type couplings to localised sources (but which does not, in the above analysis). 

The subtlety in doing so arises once back-reaction of the source on the bulk fields becomes important (as if often true when using branes in compactifications). The problem arises when the back-reaction of the localised sources causes the solutions to the bulk field equations to vary rapidly enough near the source that the bulk action effectively acquires localised components. When this happens the bulk action also contributes to the complete energy density (and action) of the localised source and the complete scaling property of a back-reacting source requires tracking all of these contributions. This is particularly dangerous in cases where a brane directly couples to a field strength, $F = \exd C + \cdots$ (such as can occur, for example, in the coupling of vortex lines to electromagnetic fields). In the examples of~\cite{Burgess:2015gba, Burgess:2015kda} the sum of the bulk and the naive brane contributions turn out to cancel, making the naive scaling of the probe-brane action irrelevant to the source's real scaling properties.  

To see why consider a simple example coming from six-dimensional supergravity, though the details of the supergravity embedding are not important for the purposes of the argument. For details of the system here described see~\cite{Burgess:2015gba}. The stripped-down bulk action of interest is given by
\be \label{SB}
 S_{\rm bulk} = - \int \exd^{6}x \; \sqrt{-\hat g} \left[ \frac{1}{2\kappa^2} \; \hat g^{\ssM\ssN} \Bigl( \hat R_{\ssM \ssN} + \partial_\ssM \varphi \, \partial_\ssN \varphi \Bigr)  + \frac14  e^{-\varphi}  F_{\ssM \ssN} F^{\ssM \ssN}  \right] \,,
\ee
which is imagined to couple to a localised source with the following action
\be \label{eq:Sbranetoy}
 S_{\rm loc} = - \mu \int \d^4 \sigma \sqrt{-\gamma} +\zeta  \int F_4   \,,
\ee
where $\gamma_{ab}$ is the pull-back of the metric $\hat g_{\ssM\ssN}$ to the source, $\mu$ and $\zeta$ are constants (tension and magnetic coupling, respectively) and $F_4 = *F_2$ is the pull-back to the source of the Hodge dual of the bulk Maxwell field. 

The bulk field equations enjoy a classical scale invariance 
\be
 \hat g_{\ssM\ssN} \to \lambda^{w_g} \hat g_{\ssM\ssN} \,, \quad
 e^{\varphi} \to \lambda^{-w_g} e^\varphi \quad
 \hbox{and} \quad F_{\ssM\ssN} \to F_{\ssM \ssN} \,,
\ee
under which $S_{\rm bulk} \to \lambda^{2w_g} S_{\rm bulk}$. At face value this symmetry is also preserved by the tension term of the source since $\sqrt{-\gamma} \to \lambda^{2w_g}  \sqrt{-\gamma}$. But for $\zeta \ne 0$ the scale invariance appears to be broken by the flux coupling term. 

But the inclusion of back-reaction makes this conclusion of scale-breaking false: both bulk and source --- {\em including} the flux coupling term for generic $\zeta$ --- actually share the scale invariance of the bulk. 

The hard way to see why this is true is to explicitly solve all field equations. When this is done the extra-dimensional components of the Maxwell field acquire a localised component 
\be
  F_{mn} = F_{mn}^{\rm smooth} + k \, \epsilon_{mn} \, \delta^2(x) \,,
\ee
where $k \propto \zeta$ and the source is imagined to be placed at position $x^m = 0$ within the extra dimensions. But once this expression is used in the bulk action cross terms between $F_{mn}^{\rm smooth}$ and the delta-function term also contribute localised contributions to the system's action and energy density, with the crossed contribution of the Maxwell action competing with (and actually cancelling) the $\zeta$ term in $S_{\rm loc}$. 

This competition is subtle because of the need to regulate and understand the contribution of the delta function evaluated at the origin. This was done in great detail in~\cite{Burgess:2015gba}, essentially by computing a UV completion consisting of a Nielsen-Oleson vortex whose confined $U(1)$ gauge flux, $V_{\ssM\ssN}$, undergoes kinetic mixing with the external Maxwell field $F_{\ssM\ssN}$:  $\cL_{\rm mix} = - \frac{1}2 \,\xi \, V^{\ssM\ssN} F_{\ssM\ssN}$. Since this mixing is localised within the Nielsen-Oleson flux tube it is represented in the effective theory (for which the tube has vanishingly small size) by the effective flux coupling of \pref{eq:Sbranetoy}, with $\zeta \propto \xi$. Not surprisingly this off-diagonal mixing term in the source action gets cancelled by its counterpart in the bulk Maxwell action once the kinetic terms are diagonalised and written in terms of the eigenstate field 
\be \label{propeigen}
 \check F_{\ssM\ssN} = F_{\ssM\ssN} + c' V_{\ssM\ssN}
\ee
(with $c' \propto \xi$). This is the UV completion's version of the cancellations between the source action and localised terms in the bulk action.

\subsubsection*{A remedy}

When the above problem arises refs.~\cite{Burgess:2015gba, Burgess:2015kda} also provide a remedy. Scaling properties can be read simply from the naive source action if bulk form-fields are chosen for which Bianchi identities forbid the back-reaction from being localised near the brane. This can usually be arranged by appropriately defining the metric- and dilaton-dependence of the duality relations for bulk form fields. 

For instance, in the above example the complicated story is much simplified when the action is expressed in terms of the 4-form field found by dualising the Maxwell field in the bulk theory. In this case performing the dualisation --- such as along the lines of~\cite{Burgess:1993np} --- leads to the dual field
\be \label{eq:Fonshell}
 \check F_{\ssM\ssN\ssP\ssQ} = + \frac{1}{2} \, e^{-\varphi} \, \epsilon_{\ssM\ssN\ssP\ssQ\ssR\,\ssT} F^{\ssR\ssT} \,.
\ee
What is important about this field is that its Bianchi identity, $\exd \check F_4 = 0$, forbids it from having any $\delta^2(x)$ components. From the UV perspective the way this happens is that the dualisation gives $\check F_{\ssM\ssN\ssP\ssQ}$ in terms of the propagation eigenstate, $\check F_{\ssM\ssN}$ of \pref{propeigen}, rather than $F_{\ssM\ssN}$. 

The absence of $\delta^2(x)$ terms in $\check F_4$ forbids the appearance of any localised terms coming from evaluating the bulk Maxwell action using the back-reacted fields. As a result the symmetry properties of the source can be directly read off from $S_{\rm loc}$ once it is expressed in terms of the dual field $\check F_4$. 
In terms of $\check F_4$, the bulk and brane actions become
\be \label{SBdual}
 S_{\rm bulk} = - \int \exd^{6}x \; \sqrt{-\hat g} \left[ \frac{1}{2\kappa^2} \; \hat g^{\ssM\ssN} \Bigl( \hat R_{\ssM \ssN} + \partial_\ssM \varphi \, \partial_\ssN \varphi \Bigr)  + \frac1{2\cdot 4!}  \, e^{\varphi}  \check F_{\ssM \ssN\ssP\ssQ} \check F^{\ssM \ssN\ssP\ssQ}  \right] \,,
\ee
and
\be \label{eq:Sbranetoydual}
 S_{\rm loc} = - \check \mu \int  \d^4 \sigma \sqrt{-\gamma} +\zeta \int \check F_4  \,,
\ee
where $\check \mu$ differs from $\mu$ by a divergent renormalisation in the limit of zero source width.
 
Notice that the dilaton dependence makes $\check F_4$ not scale the same way as does $F_2$ or its Hodge dual $F_4 = *F_2$.  The bulk classical scale invariance in this case is given by
\be
 \hat g_{\ssM\ssN} \to \lambda^{w_g} \hat g_{\ssM\ssN} \,, \quad
 e^{\varphi} \to \lambda^{-w_g} e^\varphi \quad
 \hbox{and} \quad\check F_{\ssM\ssN\ssP\ssQ} \to \lambda^{2w_g} \check F_{\ssM \ssN \ssP \ssQ} \,,
\ee
under which both $S_{\rm bulk} \to \lambda^{2w_g} S_{\rm bulk}$ and $S_{\rm loc} \to \lambda^{2w_g} S_{\rm loc}$ now scale in the same way, showing that the coupling $\zeta$ does not break scale invariance even once back-reaction is included. 

In the end, the discussion given in \S\ref{ssec:WZscalingcalc} (and in the main text) evades these issues for several reasons. First, WZ couplings here are to form potentials rather than field strengths, and so the fields generated by the source-bulk coupling tend to be longer range and not as localised as for the examples considered in~\cite{Burgess:2015gba, Burgess:2015kda}. Performing the dualities in string frame similarly helps ensure that form-field Bianchi identities also work against effective brane-dilaton couplings sourcing localised configurations (even if brane couplings directly to bulk field strengths were entertained). 

\pagebreak


\begin{thebibliography}{99}

\bibitem{TheBook}
   {\it Introduction to Effective Field Theory (Thinking Effectively About Hierarchies of Scale)}, by C.P. Burgess, Cambridge University Press 2020. (https://www.physics.mcmaster.ca/$\sim$cburgess/cburgess/?page\_id=630)
   
\bibitem{PetrovBlechman}
 {\it Effective Field Theories}, by A.A. Petrov and A.E. Blechman, World Scientific 2014.
 
\bibitem{Weinberg:1978kz}
S.~Weinberg,
``Phenomenological Lagrangians,''
Physica A \textbf{96} (1979) no.1-2, 327-340

\bibitem{Donoghue:1994dn}
J.~F.~Donoghue,
``General relativity as an effective field theory: The leading quantum corrections,''
Phys. Rev. D \textbf{50} (1994), 3874-3888
[arXiv:gr-qc/9405057 [gr-qc]].

\bibitem{Burgess:2003jk}
C.~P.~Burgess,
``Quantum gravity in everyday life: General relativity as an effective field theory,''
Living Rev. Rel. \textbf{7} (2004), 5-56
[arXiv:gr-qc/0311082 [gr-qc]].

\bibitem{Donoghue:2017ovt}
J.~Donoghue,
``Quantum gravity as a low energy effective field theory,''
Scholarpedia \textbf{12} (2017) no.4, 32997
doi:10.4249/scholarpedia.32997

\bibitem{Vafa:2005ui}
C.~Vafa,
``The String landscape and the swampland,''
[arXiv:hep-th/0509212 [hep-th]].

\bibitem{Palti:2019pca}
E.~Palti,
``The Swampland: Introduction and Review,''
Fortsch. Phys. \textbf{67} (2019) no.6, 1900037
doi:10.1002/prop.201900037
[arXiv:1903.06239 [hep-th]].

\bibitem{Weinberg:1986cq}
S.~Weinberg,
``Superconductivity for Particular Theorists,''
Prog. Theor. Phys. Suppl. \textbf{86} (1986), 43

\bibitem{Polchinski:1992ed}
J.~Polchinski,
``Effective field theory and the Fermi surface,''
[arXiv:hep-th/9210046 [hep-th]].

\bibitem{Shankar:1993pf}
R.~Shankar,
``Renormalization group approach to interacting fermions,''
Rev. Mod. Phys. \textbf{66} (1994), 129-192
[arXiv:cond-mat/9307009 [cond-mat]].

    \bibitem{Witten:1985xb}
  E.~Witten,
  ``Dimensional Reduction of Superstring Models,''
  Phys.\ Lett.\  {\bf 155B} (1985) 151.

\bibitem{Burgess:1985zz}
  C.~P.~Burgess, A.~Font and F.~Quevedo,
  ``Low-Energy Effective Action for the Superstring,''
  Nucl.\ Phys.\ B {\bf 272} (1986) 661.
  
\bibitem{Nilles:1986cy}
  H.~P.~Nilles,
  ``The Role of Classical Symmetries in the Low-energy Limit of Superstring Theories,''
  Phys.\ Lett.\ B {\bf 180} (1986) 240.


  \bibitem{Cremmer:1983bf}
  E.~Cremmer, S.~Ferrara, C.~Kounnas and D.~V.~Nanopoulos,
  ``Naturally Vanishing Cosmological Constant in N=1 Supergravity,''
  Phys.\ Lett.\  {\bf 133B} (1983) 61.

  \bibitem{Barbieri:1982ac}
  R.~Barbieri, S.~Ferrara, D.~V.~Nanopoulos and K.~S.~Stelle,
  ``Supergravity, R Invariance and Spontaneous Supersymmetry Breaking,''
  Phys.\ Lett.\  {\bf 113B} (1982) 219.
  
  \bibitem{Chang:1983hk}
  N.~P.~Chang, S.~Ouvry and X.~z.~Wu,
  ``Nonminimal $N=1$ Supergravity: A Class of Models,''
  Phys.\ Rev.\ Lett.\  {\bf 51} (1983) 327.
  
  \bibitem{Ellis:1983sf}
  J.~R.~Ellis, A.~B.~Lahanas, D.~V.~Nanopoulos and K.~Tamvakis,
  ``No-Scale Supersymmetric Standard Model,''
  Phys.\ Lett.\  {\bf 134B} (1984) 429.
    
    
 \bibitem{Barbieri:1985wq}
  R.~Barbieri, E.~Cremmer and S.~Ferrara,
  ``Flat and Positive Potentials in $N=1$ Supergravity,''
  Phys.\ Lett.\  {\bf 163B} (1985) 143.
  

   \bibitem{Witten:1985bz}
  E.~Witten,
  ``New Issues in Manifolds of SU(3) Holonomy,''
  Nucl.\ Phys.\ B {\bf 268} (1986) 79.
  
\bibitem{Burgess:2005jx}
  C.~P.~Burgess, C.~Escoda and F.~Quevedo,
  ``Nonrenormalization of flux superpotentials in string theory,''
  JHEP {\bf 0606} (2006) 044
  [hep-th/0510213].
  
  
    \bibitem{Salam:1989fm}
  A.~Salam and E.~Sezgin,
  ``Supergravities In Diverse Dimensions. Vol. 1, 2,''
  Amsterdam, Netherlands: North-Holland (1989) 1499 p. Singapore, Singapore: World Scientific (1989) 1499 p
  
  \bibitem{Burgess:2011rv}
  Y.~Aghababaie, C.~P.~Burgess, J.~M.~Cline, H.~Firouzjahi, S.~L.~Parameswaran, F.~Quevedo, G.~Tasinato and I.~Zavala,
  ``Warped brane worlds in six-dimensional supergravity,''
  JHEP {\bf 0309} (2003) 037
  [hep-th/0308064];
 
 \bibitem{BMvNNQ}
  C.~P.~Burgess, A.~Maharana, L.~van Nierop, A.~A.~Nizami and F.~Quevedo,
  ``On Brane Back-Reaction and de Sitter Solutions in Higher-Dimensional Supergravity,''
  JHEP {\bf 1204} (2012) 018
  [arXiv:1109.0532 [hep-th]];
  
  \bibitem{GJZ}
  F.~F.~Gautason, D.~Junghans and M.~Zagermann,
  ``Cosmological Constant, Near Brane Behavior and Singularities,''
  JHEP {\bf 1309} (2013) 123
  [arXiv:1301.5647 [hep-th]].
  

  

\bibitem{EarlyScale}
F.~Cooper and G.~Venturi,
``Cosmology and Broken Scale Invariance,''
Phys.\ Rev.\ D \textbf{24} (1981), 3338

R.~Peccei, J.~Sola and C.~Wetterich,
``Adjusting the Cosmological Constant Dynamically: Cosmons and a New Force Weaker Than Gravity,''
Phys.\ Lett.\ B \textbf{195} (1987), 183-190

C.~Wetterich,
``Cosmology and the Fate of Dilatation Symmetry,''
Nucl.\ Phys.\ B \textbf{302} (1988), 668-696
[arXiv:1711.03844 [hep-th]].

C.~Wetterich,
``The Cosmon model for an asymptotically vanishing time dependent cosmological 'constant',''
Astron.\ Astrophys.\  \textbf{301} (1995), 321-328
[arXiv:hep-th/9408025 [hep-th]].

E.~Guendelman,
``Scale invariance and cosmology,''
AIP Conf.\ Proc.\  \textbf{493} (1999) no.1, 201
[arXiv:gr-qc/9901067 [gr-qc]].

\bibitem{Salvio:2014soa}
A.~Salvio and A.~Strumia,
``Agravity,''
JHEP \textbf{06} (2014), 080
doi:10.1007/JHEP06(2014)080
[arXiv:1403.4226 [hep-ph]].
  
  \bibitem{Weinberg:1988cp}
  S.~Weinberg,
  ``The Cosmological Constant Problem,''
  Rev.\ Mod.\ Phys.\  {\bf 61} (1989) 1.
    
  \bibitem{Burgess:2013ara}
  C.~P.~Burgess,
  ``The Cosmological Constant Problem: Why it's hard to get Dark Energy from Micro-physics,''
  arXiv:1309.4133 [hep-th].
  
  \bibitem{ScaleAnomaly}
  S. R. Coleman and R. Jackiw, 
  ``Why dilatation generators do not generate dilatations?,'' 
  Annals Phys. {\bf 67} (1971) 552;
  
  D. M. Capper and M. J. Duff, 
  ``Trace anomalies in dimensional regularization,'' 
  Nuovo Cim. {\bf A23} (1974) 173.
    
  \bibitem{SUSYNR}
  M.~T.~Grisaru, W.~Siegel and M.~Rocek,
  ``Improved Methods for Supergraphs,''
  Nucl.\ Phys.\ B {\bf 159} (1979) 429;
  
  
    N.~Seiberg,
  ``Naturalness versus supersymmetric nonrenormalization theorems,''
  Phys.\ Lett.\ B {\bf 318} (1993) 469
  [hep-ph/9309335].
  
    \bibitem{Groh:2012tf}
  K.~Groh, J.~Louis and J.~Sommerfeld,
  ``Duality and Couplings of 3-Form-Multiplets in N=1 Supersymmetry,''
  JHEP {\bf 1305} (2013) 001
  [arXiv:1212.4639 [hep-th]].
  
\bibitem{Ferrara:1994kg}
  S.~Ferrara, C.~Kounnas and F.~Zwirner,
  ``Mass formulae and natural hierarchy in string effective supergravities,''
  Nucl.\ Phys.\ B {\bf 429} (1994) 589
   Erratum: [Nucl.\ Phys.\ B {\bf 433} (1995) 255]
  [hep-th/9405188].
  
\bibitem{Covi:2008ea}
  L.~Covi, M.~Gomez-Reino, C.~Gross, J.~Louis, G.~A.~Palma and C.~A.~Scrucca,
  ``de Sitter vacua in no-scale supergravities and Calabi-Yau string models,''
  JHEP {\bf 0806} (2008) 057
  [arXiv:0804.1073 [hep-th]].
  
\bibitem{Burgess:2008ir}
  C.~P.~Burgess, J.~M.~Cline and M.~Postma,
  ``Axionic D3-D7 Inflation,''
  JHEP {\bf 0903} (2009) 058
  [arXiv:0811.1503 [hep-th]].

  \bibitem{LV}
  V. Balasubramanian, P. Berglund, J. P. Conlon and F. Quevedo, 
  ``Systematics of moduli stabilisation in Calabi-Yau flux compactifications,'' 
  JHEP 0503 (2005) 007 [hep-th/0502058].
  
	\bibitem{LVcorr}
  J. P. Conlon, F. Quevedo and K. Suruliz, 
  ``Large-volume flux compactifications: Moduli spectrum and D3/D7 soft supersymmetry breaking,'' 
  JHEP 0508 (2005) 007 [hep-th/0505076].
   
  \bibitem{LVLoops} 
    M. Berg, M. Haack and E. Pajer, 
   ``Jumping Through Loops: On Soft Terms from Large Volume Compactifications,'' 
   JHEP 0709 (2007) 031 
   [arXiv:0704.0737 [hep-th]];
  
  \bibitem{vonGersdorff:2005bf}
G.~von Gersdorff and A.~Hebecker,
Phys. Lett. B \textbf{624} (2005), 270-274
doi:10.1016/j.physletb.2005.08.024
[arXiv:hep-th/0507131 [hep-th]].
   
  \bibitem{LVLoops1} 
	M.~Cicoli, J.~P.~Conlon and F.~Quevedo,
``Systematics of String Loop Corrections in Type IIB Calabi-Yau Flux Compactifications,''
JHEP \textbf{01}, 052 (2008)
doi:10.1088/1126-6708/2008/01/052
[arXiv:0708.1873 [hep-th]];
M.~Cicoli, J.~P.~Conlon and F.~Quevedo,
``General Analysis of LARGE Volume Scenarios with String Loop Moduli Stabilisation,''
JHEP \textbf{10} (2008), 105
doi:10.1088/1126-6708/2008/10/105
[arXiv:0805.1029 [hep-th]].
	
  \bibitem{SLED}
  Y. Aghababaie, C. P. Burgess, S. L. Parameswaran and F. Quevedo, 
  ``Towards a naturally small cosmological constant from branes in 6-D supergravity,'' 
  Nucl.\ Phys.\ {\bf B680} (2004) 389 
  [arXiv:hep-th/0304256];
  
  C.P. Burgess, 
  ``Supersymmetric Large Extra Dimensions and the Cosmological Constant: An Update,'' 
  Ann. Phys. 313 (2004) 283-401 [arXiv:hep-th/0402200]; 
  ``Towards a natural theory of dark energy: Supersymmetric large extra dimensions,'' 
  in the proceedings of the Texas A\&M Workshop on String Cosmology, [arXiv:hep-th/0411140];

 C.~P.~Burgess and L.~van Nierop,
  ``Technically Natural Cosmological Constant From Supersymmetric 6D Brane Backreaction,''
  Phys.\ Dark Univ.\  {\bf 2} (2013) 1
  [arXiv:1108.0345 [hep-th]].
     
  
  \bibitem{SLEDLoops}
   D. Hoover and C. P. Burgess, 
   ``Ultraviolet sensitivity in higher dimensions,'' 
   JHEP 0601 (2006) 058 [hep-th/0507293].

  \bibitem{SLEDLoops1}
  C. P. Burgess and D. Hoover, 
  ``UV sensitivity in supersymmetric large extra dimensions: The Ricci-flat case,'' 
  Nucl.\ Phys.\ {\bf B772} (2007) 175 [hep-th/0504004].
  
  \bibitem{SLEDLoops2}
  C. P. Burgess, D. Hoover and G. Tasinato, 
  ``UV Caps and Modulus Stabilization for 6D Gauged Chiral Supergravity,'' 
  JHEP 0709 (2007) 124 [arXiv:0705.3212 [hep-th]];
  %
  ``Technical Naturalness on a Codimension-2 Brane,'' 
  JHEP 0906 (2009) 014. [arXiv:0903.0402 [hep-th]];
  
  \bibitem{SLEDLoops3}
   C.~P.~Burgess, L.~van Nierop, S.~Parameswaran, A.~Salvio and M.~Williams,
  ``Accidental SUSY: Enhanced Bulk Supersymmetry from Brane Back-reaction,''
  JHEP {\bf 1302} (2013) 120
  [arXiv:1210.5405 [hep-th].
  
\bibitem{Burgess:2015gba}
  C.~P.~Burgess, R.~Diener and M.~Williams,
  ``EFT for Vortices with Dilaton-dependent Localized Flux,''
  JHEP {\bf 1511} (2015) 054
  [arXiv:1508.00856 [hep-th]].

\bibitem{Burgess:2015kda}
  C.~P.~Burgess, R.~Diener and M.~Williams,
  ``A problem with $\delta$-functions: stress-energy constraints on bulk-brane matching (with comments on arXiv:1508.01124),''
  JHEP {\bf 1601} (2016) 017
  [arXiv:1509.04201 [hep-th]].

\bibitem{Burgess:2015lda}
  C.~P.~Burgess, R.~Diener and M.~Williams,
  ``Self-Tuning at Large (Distances): 4D Description of Runaway Dilaton Capture,''
  JHEP {\bf 1510} (2015) 177
  [arXiv:1509.04209 [hep-th]].
  
  
  \bibitem{Dine:1985he}
  M.~Dine and N.~Seiberg,
  ``Is the Superstring Weakly Coupled?,''
  Phys.\ Lett.\  {\bf 162B} (1985) 299.
  
  \bibitem{Goncharov:1985yu}
A.~Goncharov and A.~D.~Linde,
``Chaotic Inflation of the Universe in Supergravity,''
Sov.\ Phys.\ JETP \textbf{59} (1984), 930-933
    
\bibitem{Burgess:2016owb}
C.~Burgess, M.~Cicoli, S.~de Alwis and F.~Quevedo,
``Robust Inflation from Fibrous Strings,''
JCAP \textbf{05}, 032 (2016)
[arXiv:1603.06789 [hep-th]].
  
 
\bibitem{Burgess:2014tja}
C.~Burgess, M.~Cicoli, F.~Quevedo and M.~Williams,
``Inflating with Large Effective Fields,''
JCAP \textbf{11}, 045 (2014)
[arXiv:1404.6236 [hep-th]].
	
\bibitem{Freese:1990rb}
K.~Freese, J.~A.~Frieman and A.~V.~Olinto,
``Natural inflation with pseudo - Nambu-Goldstone bosons,''
Phys.\ Rev.\ Lett.\  \textbf{65} (1990), 3233-3236
  
  
\bibitem{Martin:2013tda}
  J.~Martin, C.~Ringeval and V.~Vennin,
  ``Encyclopaedia Inflationaris,''
  Phys.\ Dark Univ.\  {\bf 5-6} (2014) 75
  [arXiv:1303.3787 [astro-ph.CO]].
  
\bibitem{Martin:2013nzq}
  J.~Martin, C.~Ringeval, R.~Trotta and V.~Vennin,
  ``The Best Inflationary Models After Planck,''
  JCAP {\bf 1403} (2014) 039
  [arXiv:1312.3529 [astro-ph.CO]].
  
 \bibitem{Kallosh:2014rga}
  R.~Kallosh, A.~Linde and D.~Roest,
  ``Large field inflation and double $\alpha$-attractors,''
  JHEP {\bf 1408} (2014) 052
  [arXiv:1405.3646 [hep-th]].

  \bibitem{Burgess:2001vr}
  C.~P.~Burgess, P.~Martineau, F.~Quevedo, G.~Rajesh and R.~J.~Zhang,
  ``Brane - anti-brane inflation in orbifold and orientifold models,''
  JHEP {\bf 0203} (2002) 052
  [hep-th/0111025].

	\bibitem{Conlon:2005jm}
  J.~P.~Conlon and F.~Quevedo,
  ``Kahler moduli inflation,''
  JHEP {\bf 0601} (2006) 146
  [hep-th/0509012].

 
\bibitem{Cicoli:2008gp}
M.~Cicoli, C.~Burgess and F.~Quevedo,
``Fibre Inflation: Observable Gravity Waves from IIB String Compactifications,''
JCAP \textbf{03}, 013 (2009)
[arXiv:0808.0691 [hep-th]].


\bibitem{Cicoli:2011ct}
M.~Cicoli, F.~G.~Pedro and G.~Tasinato,
``Poly-instanton Inflation,''
JCAP \textbf{12}, 022 (2011)
[arXiv:1110.6182 [hep-th]].
	
 
\bibitem{Burgess:2013sla}
C.~Burgess, M.~Cicoli and F.~Quevedo,
``String Inflation After Planck 2013,''
JCAP \textbf{11}, 003 (2013)
[arXiv:1306.3512 [hep-th]].
	
\bibitem{Broy:2015zba}
  B.~J.~Broy, D.~Ciupke, F.~G.~Pedro and A.~Westphal,
  ``Starobinsky-Type Inflation from $\alpha'$-Corrections,''
  JCAP {\bf 1601} (2016) 001
  [arXiv:1509.00024 [hep-th]].
	
 
\bibitem{Cicoli:2016chb}
M.~Cicoli, D.~Ciupke, S.~de Alwis and F.~Muia,
``$\alpha'$ Inflation: moduli stabilisation and observable tensors from higher derivatives,''
JHEP \textbf{09}, 026 (2016)
[arXiv:1607.01395 [hep-th]].
	
	  \bibitem{Ellis:2020xmk}
J.~Ellis, D.~V.~Nanopoulos, K.~A.~Olive and S.~Verner,
``Phenomenology and Cosmology of No-Scale Attractor Models of Inflation,''
[arXiv:2004.00643 [hep-ph]].

	
 \bibitem{Kachru:2003sx}
  S.~Kachru, R.~Kallosh, A.~D.~Linde, J.~M.~Maldacena, L.~P.~McAllister and S.~P.~Trivedi,
  ``Towards inflation in string theory,''
  JCAP {\bf 0310} (2003) 013
  [hep-th/0308055].

\bibitem{Giddings:2005ff}
S.~B.~Giddings and A.~Maharana,
Phys. Rev. D \textbf{73} (2006), 126003
doi:10.1103/PhysRevD.73.126003
[arXiv:hep-th/0507158 [hep-th]].

\bibitem{Denef:2004ze}
F.~Denef and M.~R.~Douglas,
``Distributions of flux vacua,''
JHEP \textbf{05} (2004), 072
[arXiv:hep-th/0404116 [hep-th]].

\bibitem{MartinezPedrera:2012rs}
D.~Martinez-Pedrera, D.~Mehta, M.~Rummel and A.~Westphal,
``Finding all flux vacua in an explicit example,''
JHEP \textbf{06} (2013), 110
[arXiv:1212.4530 [hep-th]].

\bibitem{Cicoli:2013cha}
M.~Cicoli, D.~Klevers, S.~Krippendorf, C.~Mayrhofer, F.~Quevedo and R.~Valandro,
``Explicit de Sitter Flux Vacua for Global String Models with Chiral Matter,''
JHEP \textbf{05}, 001 (2014)
[arXiv:1312.0014 [hep-th]].

\bibitem{Cicoli:2012sz}
M.~Cicoli, M.~Goodsell and A.~Ringwald,
``The type IIB string axiverse and its low-energy phenomenology,''
JHEP \textbf{10}, 146 (2012)
[arXiv:1206.0819 [hep-th]].

\bibitem{Cicoli:2018tcq}
M.~Cicoli, D.~Ciupke, C.~Mayrhofer and P.~Shukla,
``A Geometrical Upper Bound on the Inflaton Range,''
JHEP \textbf{05}, 001 (2018)
[arXiv:1801.05434 [hep-th]].

\bibitem{BHK} 
M.~Berg, M.~Haack and B.~Kors,
  ``String loop corrections to Kahler potentials in orientifolds,''
  JHEP {\bf 0511} (2005) 030
  [hep-th/0508043];
	
  \bibitem{BBHL}
  K.~Becker, M.~Becker, M.~Haack and J.~Louis,
  ``Supersymmetry breaking and alpha-prime corrections to flux induced potentials,''
  JHEP {\bf 0206}, 060 (2002)
  [hep-th/0204254].

	  \bibitem{CLW}
  D.~Ciupke, J.~Louis and A.~Westphal,
  ``Higher-Derivative Supergravity and Moduli Stabilization,''
  JHEP {\bf 1510} (2015) 094
  [arXiv:1505.03092 [hep-th]].

  \bibitem{Burgess:2010sy}
  C.~P.~Burgess, A.~Maharana and F.~Quevedo,
  ``Uber-naturalness: unexpectedly light scalars from supersymmetric extra dimensions,''
  JHEP {\bf 1105} (2011) 010
  [arXiv:1005.1199 [hep-th]].

\bibitem{Berg:2005yu}
M.~Berg, M.~Haack and B.~Kors,
``On volume stabilization by quantum corrections,''
Phys. Rev. Lett. \textbf{96} (2006), 021601
[arXiv:hep-th/0508171 [hep-th]].


\bibitem{Cicoli:2014sva}
M.~Cicoli, K.~Dutta and A.~Maharana,
``N-flation with Hierarchically Light Axions in String Compactifications,''
JCAP \textbf{08}, 012 (2014)
[arXiv:1401.2579 [hep-th]].

\bibitem{Cicoli:2011it}
M.~Cicoli, M.~Kreuzer and C.~Mayrhofer,
``Toric K3-Fibred Calabi-Yau Manifolds with del Pezzo Divisors for String Compactifications,''
JHEP \textbf{02}, 002 (2012)
[arXiv:1107.0383 [hep-th]].


\bibitem{Coughlan:1983ci}
  G.~D.~Coughlan, W.~Fischler, E.~W.~Kolb, S.~Raby and G.~G.~Ross,
  ``Cosmological Problems for the Polonyi Potential,''
  Phys.\ Lett.\  {\bf 131B} (1983) 59.
	
\bibitem{Banks:1993en}
  T.~Banks, D.~B.~Kaplan and A.~E.~Nelson,
  ``Cosmological implications of dynamical supersymmetry breaking,''
  Phys.\ Rev.\ D {\bf 49} (1994) 779
  [hep-ph/9308292].
	
\bibitem{deCarlos:1993wie}
  B.~de Carlos, J.~A.~Casas, F.~Quevedo and E.~Roulet,
  ``Model independent properties and cosmological implications of the dilaton and moduli sectors of 4-d strings,''
  Phys.\ Lett.\ B {\bf 318} (1993) 447
  [hep-ph/9308325].

\bibitem{Endo:2006zj}
  M.~Endo, K.~Hamaguchi and F.~Takahashi,
  ``Moduli-induced gravitino problem,''
  Phys.\ Rev.\ Lett.\  {\bf 96} (2006) 211301
  [hep-ph/0602061].

\bibitem{Nakamura:2006uc}
  S.~Nakamura and M.~Yamaguchi,
  ``Gravitino production from heavy moduli decay and cosmological moduli problem revived,''
  Phys.\ Lett.\ B {\bf 638} (2006) 389
  [hep-ph/0602081].


  \bibitem{BCKMQ}
  R.~Blumenhagen, J.~P.~Conlon, S.~Krippendorf, S.~Moster and F.~Quevedo,
  ``SUSY Breaking in Local String/F-Theory Models,''
  JHEP {\bf 0909} (2009) 007
  [arXiv:0906.3297 [hep-th]].
  
\bibitem{Aparicio:2014wxa}
L.~Aparicio, M.~Cicoli, S.~Krippendorf, A.~Maharana, F.~Muia and F.~Quevedo,
``Sequestered de Sitter String Scenarios: Soft-terms,''
JHEP \textbf{11}, 071 (2014)
[arXiv:1409.1931 [hep-th]].
  
	
\bibitem{Reece:2015qbf}
M.~Reece and W.~Xue,
 ``SUSY's Ladder: reframing sequestering at Large Volume,''
 JHEP {\bf 04} (2016), 045
[arXiv:1512.04941 [hep-ph]].
  
 \bibitem{Randall:1998uk}
  L.~Randall and R.~Sundrum,
  Nucl.\ Phys.\ B {\bf 557} (1999) 79
  [hep-th/9810155].
  
 
\bibitem{Anisimov:2001zz}
  A.~Anisimov, M.~Dine, M.~Graesser and S.~D.~Thomas,
  ``Brane world SUSY breaking,''
  Phys.\ Rev.\ D {\bf 65} (2002) 105011
  [hep-th/0111235].
	

\bibitem{Anisimov:2002az}
  A.~Anisimov, M.~Dine, M.~Graesser and S.~D.~Thomas,
  ``Brane world SUSY breaking from string / M theory,''
  JHEP {\bf 0203} (2002) 036
  [hep-th/0201256].
 

\bibitem{Jockers:2004yj}
  H.~Jockers and J.~Louis,
  ``The Effective action of D7-branes in N = 1 Calabi-Yau orientifolds,''
  Nucl.\ Phys.\ B {\bf 705} (2005) 167
  [hep-th/0409098].
  
\bibitem{Jockers:2005zy}
  H.~Jockers and J.~Louis,
  ``D-terms and F-terms from D7-brane fluxes,''
  Nucl.\ Phys.\ B {\bf 718} (2005) 203
  [hep-th/0502059].

\bibitem{Grimm:2004uq}
  T.~W.~Grimm and J.~Louis,
  ``The Effective action of N = 1 Calabi-Yau orientifolds,''
  Nucl.\ Phys.\ B {\bf 699} (2004) 387
  [hep-th/0403067].
  
\bibitem{Grimm:2005fa}
  T.~W.~Grimm,
  ``The Effective action of type II Calabi-Yau orientifolds,''
  Fortsch.\ Phys.\  {\bf 53} (2005) 1179
  [hep-th/0507153].

\bibitem{DallAgata:2013jtw}
G.~Dall'Agata and F.~Zwirner,
``New Class of $N=1$ No-Scale Supergravity Models,''
Phys. Rev. Lett. \textbf{111} (2013) no.25, 251601
doi:10.1103/PhysRevLett.111.251601
[arXiv:1308.5685 [hep-th]].

\bibitem{Ciupke:2015ora}
  D.~Ciupke and L.~Zarate,
  ``Classification of Shift-Symmetric No-Scale Supergravities,''
  JHEP {\bf 1511} (2015) 179
  [arXiv:1509.00855 [hep-th]].
	
\bibitem{Blaback:2015zra}
  J.~Bl\r{a}b\"ack, U.~H.~Danielsson, G.~Dibitetto and S.~C.~Vargas, 
  ``Universal dS vacua in STU-models,''
  JHEP {\bf 1510} (2015) 069
  [arXiv:1505.04283 [hep-th]].
 

\bibitem{LinearMult}
  S. Ferrara, J. Wess and B. Zumino, Phys. Lett. B51 (1974) 239;
  
  W. Siegel, Phys. Lett. B85 (1979) 333.
  
\bibitem{NewMinimal}
  M.F.~Sohnius and P.C.~West, 
``An alternative minimal off-shell version of N=1 supergravity,''
  Phys.\ Lett.\ {\bf 105B} (1981) 353.

\bibitem{Cecotti:1987nw}
  S.~Cecotti, S.~Ferrara and M.~Villasante,
  ``Linear Multiplets and Super Chern-Simons Forms in 4D Supergravity,''
  Int.\ J.\ Mod.\ Phys.\ A {\bf 2} (1987) 1839.
  
  \bibitem{Ovrut:1988fd}
  B.~A.~Ovrut and C.~Schwiebert,
  ``Linear Multiplets Coupled To New Minimal Supergravity,''
  Nucl.\ Phys.\ B {\bf 321} (1989) 163.
	
\bibitem{Burgess:1995kp}
  C.~P.~Burgess, J.-P.~Derendinger, F.~Quevedo and M.~Quiros,
  ``Gaugino condensates and chiral linear duality: An Effective Lagrangian analysis,''
  Phys.\ Lett.\ B {\bf 348} (1995) 428
  [hep-th/9501065].


\bibitem{Freedman:1976xh}
D.~Z.~Freedman, P.~van Nieuwenhuizen and S.~Ferrara,
``Progress Toward a Theory of Supergravity,''
Phys. Rev. D \textbf{13} (1976), 3214-3218

\bibitem{Deser:1976eh}
S.~Deser and B.~Zumino,
``Consistent Supergravity,''
Phys. Lett. B \textbf{62} (1976), 335

\bibitem{Stelle:1978wj}
K.~Stelle and P.~C.~West,
``Relation Between Vector and Scalar Multiplets and Gauge Invariance in Supergravity,''
Nucl. Phys. B \textbf{145} (1978), 175-188

\bibitem{Ferrara:1978jt}
S.~Ferrara and P.~van Nieuwenhuizen,
``Tensor Calculus for Supergravity,''
Phys. Lett. B \textbf{76} (1978), 404

\bibitem{Cremmer:1978hn}
E.~Cremmer, B.~Julia, J.~Scherk, S.~Ferrara, L.~Girardello and P.~van Nieuwenhuizen,
``Spontaneous Symmetry Breaking and Higgs Effect in Supergravity Without Cosmological Constant,''
Nucl. Phys. B \textbf{147} (1979), 105

\bibitem{Salam:1974jj}
A.~Salam and J.~Strathdee,
``On Superfields and Fermi-Bose Symmetry,''
Phys. Rev. D \textbf{11} (1975), 1521-1535

 \bibitem{Cheung:2011jp}
  C.~Cheung, F.~D'Eramo and J.~Thaler,
  ``Supergravity Computations without Gravity Complications,''
  Phys.\ Rev.\ D {\bf 84} (2011) 085012
  [arXiv:1104.2598 [hep-ph]].
  

\bibitem{Kugo:1982mr}
  T.~Kugo and S.~Uehara,
  ``Improved Superconformal Gauge Conditions in the $N=1$ Supergravity {Yang-Mills} Matter System,''
  Nucl.\ Phys.\ B {\bf 222} (1983) 125.


  
  \bibitem{Gross:1984dd}
  D.~J.~Gross, J.~A.~Harvey, E.~J.~Martinec and R.~Rohm,
  ``The Heterotic String,''
  Phys.\ Rev.\ Lett.\  {\bf 54} (1985) 502.
  
  \bibitem{Gross:1985fr}
  D.~J.~Gross, J.~A.~Harvey, E.~J.~Martinec and R.~Rohm,
  ``Heterotic String Theory. 1. The Free Heterotic String,''
  Nucl.\ Phys.\ B {\bf 256} (1985) 253.
  
  \bibitem{Gross:1985rr}
  D.~J.~Gross, J.~A.~Harvey, E.~J.~Martinec and R.~Rohm,
  ``Heterotic String Theory. 2. The Interacting Heterotic String,''
  Nucl.\ Phys.\ B {\bf 267} (1986) 75.
  
\bibitem{Green:2016tfs}
  M.~B.~Green and A.~Rudra,
  ``Type I/heterotic duality and M-theory amplitudes,''
  JHEP {\bf 1612} (2016) 060
  [arXiv:1604.00324 [hep-th]].
  
 \bibitem{Chamseddine:1980cp}
  A.~H.~Chamseddine,
  ``N=4 Supergravity Coupled to N=4 Matter,''
  Nucl.\ Phys.\ B {\bf 185} (1981) 403.
  
  \bibitem{Bergshoeff:1981um}
  E.~Bergshoeff, M.~de Roo, B.~de Wit and P.~van Nieuwenhuizen,
  ``Ten-Dimensional Maxwell-Einstein Supergravity, Its Currents, and the Issue of Its Auxiliary Fields,''
  Nucl.\ Phys.\ B {\bf 195} (1982) 97.
  
 \bibitem{Chapline:1982ww}
  G.~F.~Chapline and N.~S.~Manton,
  ``Unification of Yang-Mills Theory and Supergravity in Ten-Dimensions,''
  Phys.\ Lett.\  {\bf 120B} (1983) 105.
  
\bibitem{Cai:1986sa}
  Y.~Cai and C.~A.~Nunez,
  ``Heterotic String Covariant Amplitudes and Low-energy Effective Action,''
  Nucl.\ Phys.\ B {\bf 287} (1987) 279.
    
  \bibitem{Gross:1986mw}
  D.~J.~Gross and J.~H.~Sloan,
  ``The Quartic Effective Action for the Heterotic String,''
  Nucl.\ Phys.\ B {\bf 291} (1987) 41.

   \bibitem{Russo:1997mk}
  J.~G.~Russo and A.~A.~Tseytlin,
  ``One loop four graviton amplitude in eleven-dimensional supergravity,''
  Nucl.\ Phys.\ B {\bf 508} (1997) 245
  [hep-th/9707134].

\bibitem{Bergshoeff:1989de}
   E.~A.~Bergshoeff and M.~de Roo,
  ``The Quartic Effective Action of the Heterotic String and Supersymmetry,''
  Nucl.\ Phys.\ B {\bf 328} (1989) 439.
  
 \bibitem{Chemissany:2007he}
  W.~A.~Chemissany, M.~de Roo and S.~Panda,
  ``alpha'-Corrections to Heterotic Superstring Effective Action Revisited,''
  JHEP {\bf 0708} (2007) 037
  [arXiv:0706.3636 [hep-th]].
  
   \bibitem{Ibanez:1986xy}
  L.~E.~Ibanez and H.~P.~Nilles,
  ``Low-Energy Remnants of Superstring Anomaly Cancellation Terms,''
  Phys.\ Lett.\  {\bf 169B} (1986) 354.
  
 \bibitem{Green:1984sg}
  M.~B.~Green and J.~H.~Schwarz,
  ``Anomaly Cancellation in Supersymmetric D=10 Gauge Theory and Superstring Theory,''
  Phys.\ Lett.\  {\bf 149B} (1984) 117.
  
 \bibitem{Chang:1986ac}
  D.~Chang and H.~Nishino,
  ``Heterotic String $O(\alpha')$ Corrections to $D = 10$, $N=1$ Supergravity,''
  Phys.\ Lett.\ B {\bf 179} (1986) 257.

\bibitem{Giddings:2001yu}
  S.~B.~Giddings, S.~Kachru and J.~Polchinski,
  ``Hierarchies from fluxes in string compactifications,''
  Phys.\ Rev.\ D {\bf 66} (2002) 106006
  [hep-th/0105097].

\bibitem{Dasgupta:1999ss}
  K.~Dasgupta, G.~Rajesh and S.~Sethi,
  ``M theory, orientifolds and G - flux,''
  JHEP {\bf 9908} (1999) 023
  [hep-th/9908088].
  

\bibitem{Berg:2004ek}
M.~Berg, M.~Haack and B.~Kors,
``Loop corrections to volume moduli and inflation in string theory,''
Phys. Rev. D \textbf{71} (2005), 026005
[arXiv:hep-th/0404087 [hep-th]].

 \bibitem{Baumann:2006th}
D.~Baumann, A.~Dymarsky, I.~R.~Klebanov, J.~M.~Maldacena, L.~P.~McAllister and A.~Murugan,
``On D3-brane Potentials in Compactifications with Fluxes and Wrapped D-branes,''
JHEP \textbf{11} (2006), 031
[arXiv:hep-th/0607050 [hep-th]].


\bibitem{Gukov:1999ya}
  S.~Gukov, C.~Vafa and E.~Witten,
  ``CFT's from Calabi-Yau four folds,''
  Nucl.\ Phys.\ B {\bf 584} (2000) 69
   Erratum: [Nucl.\ Phys.\ B {\bf 608} (2001) 477]
  [hep-th/9906070].
  
  \bibitem{Candelas:1985en}
  P.~Candelas, G.~T.~Horowitz, A.~Strominger and E.~Witten,
  ``Vacuum Configurations for Superstrings,''
  Nucl.\ Phys.\ B {\bf 258} (1985) 46.

  
\bibitem{Cicoli:2012vw}
M.~Cicoli, S.~Krippendorf, C.~Mayrhofer, F.~Quevedo and R.~Valandro,
``D-Branes at del Pezzo Singularities: Global Embedding and Moduli Stabilisation,''
JHEP \textbf{09}, 019 (2012)
[arXiv:1206.5237 [hep-th]].

\bibitem{Grana:2003ek}
  M.~Grana, T.~W.~Grimm, H.~Jockers and J.~Louis,
  ``Soft supersymmetry breaking in Calabi-Yau orientifolds with D-branes and fluxes,''
  Nucl.\ Phys.\ B {\bf 690} (2004) 21
  [hep-th/0312232].

\bibitem{FvP}
  {\it Supergravity}, by D.Z. Freedman and A. van Proeyen, Cambridge University Press 2012.
  
\bibitem{Baumann:2007ah}
D.~Baumann, A.~Dymarsky, I.~R.~Klebanov and L.~McAllister,
``Towards an Explicit Model of D-brane Inflation,''
JCAP \textbf{01} (2008), 024
[arXiv:0706.0360 [hep-th]].

\bibitem{ibanezuranga}
  L.~E.~Ibanez and A.~M.~Uranga,
  ``String theory and particle physics: An introduction to string phenomenology,''
  Cambridge University Press (2012)
  
 \bibitem{Jockers:2005pn}
H.~Jockers,
``The Effective Action of D-branes in Calabi-Yau Orientifold Compactifications,''
Fortsch. Phys. \textbf{53} (2005), 1087-1175
[arXiv:hep-th/0507042 [hep-th]].

   \bibitem{Kaloper:2008fb}
  N.~Kaloper and L.~Sorbo,
  ``A Natural Framework for Chaotic Inflation,''
  Phys.\ Rev.\ Lett.\  {\bf 102} (2009) 121301
  [arXiv:0811.1989 [hep-th]];
 
   N.~Kaloper, A.~Lawrence and L.~Sorbo,
  ``An Ignoble Approach to Large Field Inflation,''
  JCAP {\bf 1103} (2011) 023
  [arXiv:1101.0026 [hep-th]].

\bibitem{Bielleman:2015ina}
  S.~Bielleman, L.~E.~Ibanez and I.~Valenzuela,
  ``Minkowski 3-forms, Flux String Vacua, Axion Stability and Naturalness,''
  JHEP {\bf 1512} (2015) 119
  [arXiv:1507.06793 [hep-th]].

 
 
  \bibitem{Herraez:2018vae}
  A.~Herraez, L.~E.~Ibanez, F.~Marchesano and G.~Zoccarato,
  ``The Type IIA Flux Potential, 4-forms and Freed-Witten anomalies,''
  JHEP {\bf 1809} (2018) 018
  [arXiv:1802.05771 [hep-th]].

    \bibitem{Bousso:2000xa}
    R.~Bousso and J.~Polchinski,
  ``Quantization of four form fluxes and dynamical neutralization of the cosmological constant,''
  JHEP {\bf 0006}, 006 (2000)
  [hep-th/0004134].

  
  \bibitem{Duff:1989ah}
M.~J.~Duff,
``The Cosmological Constant Is Possibly Zero, but the Proof Is Probably Wrong,''
Phys.\ Lett.\ B \textbf{226} (1989), 36
  
	
\bibitem{Antoniadis:1997eg}
  I.~Antoniadis, S.~Ferrara, R.~Minasian and K.~S.~Narain,
  ``R**4 couplings in M and type II theories on Calabi-Yau spaces,''
  Nucl.\ Phys.\ B {\bf 507} (1997) 571
  [hep-th/9707013].
  
  \bibitem{Sen:2013oza}
A.~Sen,
``S-duality Improved Superstring Perturbation Theory,''
JHEP \textbf{11} (2013), 029
doi:10.1007/JHEP11(2013)029
[arXiv:1304.0458 [hep-th]].
 
\bibitem{BW}
F.~Bonetti and M.~Weissenbacher,
  ``The Euler characteristic correction to the K\"ahler potential revisited,''
  JHEP {\bf 1701} (2017) 003
  [arXiv:1608.01300 [hep-th]].
  
  \bibitem{MPS}
  R.~Minasian, T.~G.~Pugh and R.~Savelli,
  ``F-theory at order $\alpha'^3$,''
  JHEP {\bf 1510} (2015) 050
  [arXiv:1506.06756 [hep-th]].

   \bibitem{HK}
  M.~Berg, M.~Haack, J.~U.~Kang and S.~Sjörs,
  ``Towards the one-loop K\"ahler metric of Calabi-Yau orientifolds,''
  JHEP {\bf 1412} (2014) 077
  [arXiv:1407.0027 [hep-th]];
	
	M.~Haack and J.~U.~Kang,
  ``One-loop Einstein-Hilbert term in minimally supersymmetric type IIB orientifolds,''
  JHEP {\bf 1602} (2016) 160
  [arXiv:1511.03957 [hep-th]];
	
	M.~Haack and J.~U.~Kang,
  ``Field redefinitions and K\"ahler potential in string theory at 1-loop,''
  JHEP {\bf 1808} (2018) 019
  [arXiv:1805.00817 [hep-th]].

   
  \bibitem{GMW}
  T.~W.~Grimm, K.~Mayer and M.~Weissenbacher,
  ``One-modulus Calabi-Yau fourfold reductions with higher-derivative terms,''
  JHEP {\bf 1804} (2018) 021
  [arXiv:1712.07074 [hep-th]].

  T.~W.~Grimm, K.~Mayer and M.~Weissenbacher,
  ``Higher derivatives in Type II and M-theory on Calabi-Yau threefolds,''
  JHEP {\bf 1802} (2018) 127
  [arXiv:1702.08404 [hep-th]].

\bibitem{Cicoli:2013swa}
M.~Cicoli, J.~P.~Conlon, A.~Maharana and F.~Quevedo,
``A Note on the Magnitude of the Flux Superpotential,''
JHEP \textbf{01} (2014), 027
[arXiv:1310.6694 [hep-th]].

  
  \bibitem{Ibanez:1999pw}
L.~E.~Ibanez, R.~Rabadan and A.~Uranga,
``Sigma model anomalies in compact D = 4, N=1 type IIB orientifolds and Fayet-Iliopoulos terms,''
Nucl. Phys. B \textbf{576} (2000), 285-312
[arXiv:hep-th/9905098 [hep-th]].
  
    \bibitem{Antoniadis:2018hqy}
  I.~Antoniadis, Y.~Chen and G.~K.~Leontaris,
  ``Perturbative moduli stabilisation in type IIB/F-theory framework,''
  Eur.\ Phys.\ J.\ C {\bf 78} (2018) no.9,  766
  [arXiv:1803.08941 [hep-th]];
   I.~Antoniadis, Y.~Chen and G.~K.~Leontaris,
  ``Logarithmic loop corrections, moduli stabilisation and de Sitter vacua in string theory,''
  arXiv:1909.10525 [hep-th];
  %
``Logarithmic loop corrections, moduli stabilisation and de Sitter vacua in string theory,''
JHEP \textbf{01} (2020), 149
[arXiv:1909.10525 [hep-th]].


	\bibitem{Cicoli:2018kdo}
M.~Cicoli, S.~De Alwis, A.~Maharana, F.~Muia and F.~Quevedo,
``De Sitter vs Quintessence in String Theory,''
Fortsch. Phys. \textbf{67}, no.1-2, 1800079 (2019)
[arXiv:1808.08967 [hep-th]].
	
	\bibitem{GSW}
T.~W.~Grimm, R.~Savelli and M.~Weissenbacher,
  ``On $\alpha'$ corrections in N=1 F-theory compactifications,''
  Phys.\ Lett.\ B {\bf 725} (2013) 431
  [arXiv:1303.3317 [hep-th]];

T.~W.~Grimm, J.~Keitel, R.~Savelli and M.~Weissenbacher,
  ``From M-theory higher curvature terms to $\alpha'$ corrections in F-theory,''
  Nucl.\ Phys.\ B {\bf 903} (2016) 325
  [arXiv:1312.1376 [hep-th]]. 

	
  
\bibitem{Romans:1985tz}
  L.~J.~Romans,
  ``Massive N=2a Supergravity in Ten-Dimensions,''
  Phys.\ Lett.\  {\bf 169B} (1986) 374.
  


  \bibitem{Camara:2005dc}
  P.~G.~Camara, A.~Font and L.~E.~Ibanez,
  ``Fluxes, moduli fixing and MSSM-like vacua in a simple IIA orientifold,''
  JHEP {\bf 0509} (2005) 013
  [hep-th/0506066].

\bibitem{Derendinger:2004jn}
J.~Derendinger, C.~Kounnas, P.~Petropoulos and F.~Zwirner,
``Superpotentials in IIA compactifications with general fluxes,''
Nucl. Phys. B \textbf{715} (2005), 211-233
[arXiv:hep-th/0411276 [hep-th]].

\bibitem{Derendinger:2005ph}
J.~Derendinger, C.~Kounnas, P.~Petropoulos and F.~Th,
``Fluxes and gaugings: N=1 effective superpotentials,''
Fortsch. Phys. \textbf{53} (2005), 926-935
[arXiv:hep-th/0503229 [hep-th]].

\bibitem{Villadoro:2005cu}
G.~Villadoro and F.~Zwirner,
``N=1 effective potential from dual type-IIA D6/O6 orientifolds with general fluxes,''
JHEP \textbf{06} (2005), 047
[arXiv:hep-th/0503169 [hep-th]].

    
\bibitem{Hertzberg:2007wc}
  M.~P.~Hertzberg, S.~Kachru, W.~Taylor and M.~Tegmark,
  ``Inflationary Constraints on Type IIA String Theory,''
  JHEP {\bf 0712} (2007) 095
  [arXiv:0711.2512 [hep-th]].  
  
 \bibitem{Obied:2018sgi}
  G.~Obied, H.~Ooguri, L.~Spodyneiko and C.~Vafa,
  ``De Sitter Space and the Swampland,''
  arXiv:1806.08362 [hep-th]. 

 
	\bibitem{Derendinger:1994gx}
J.~P.~Derendinger, F.~Quevedo and M.~Quiros,
``The Linear multiplet and quantum four-dimensional string effective actions,''
Nucl. Phys. B \textbf{428} (1994), 282-330
[arXiv:hep-th/9402007 [hep-th]].

\bibitem{Binetruy:2000zx}
  P.~Binetruy, G.~Girardi and R.~Grimm,
  ``Supergravity couplings: A Geometric formulation,''
  Phys.\ Rept.\  {\bf 343} (2001) 255
  [hep-th/0005225].


 \bibitem{Gaiotto:2014kfa}
D.~Gaiotto, A.~Kapustin, N.~Seiberg and B.~Willett,
``Generalized Global Symmetries,''
JHEP \textbf{02} (2015), 172
doi:10.1007/JHEP02(2015)172
[arXiv:1412.5148 [hep-th]].
 
 \bibitem{Buscher:1987qj}
T.~Buscher,
``Path Integral Derivation of Quantum Duality in Nonlinear Sigma Models,''
Phys. Lett. B \textbf{201} (1988), 466-472

 \bibitem{Sen:2015nph}
A.~Sen,
``Covariant Action for Type IIB Supergravity,''
JHEP \textbf{07} (2016), 017
[arXiv:1511.08220 [hep-th]].

\bibitem{Sen:2019qit}
A.~Sen,
``Self-dual forms: Action, Hamiltonian and Compactification,''
J. Phys. A \textbf{53} (2020) no.8, 084002
[arXiv:1903.12196 [hep-th]].

\bibitem{Coleman:1969sm}
S.~R.~Coleman, J.~Wess and B.~Zumino,
``Structure of phenomenological Lagrangians. 1.,''
Phys. Rev. \textbf{177} (1969), 2239-2247
%
C.~G.~Callan, Jr., S.~R.~Coleman, J.~Wess and B.~Zumino,
``Structure of phenomenological Lagrangians. 2.,''
Phys. Rev. \textbf{177} (1969), 2247-2250

 \bibitem{DallAgata:1998ahf}
  G.~Dall'Agata, K.~Lechner and M.~Tonin,
  ``D = 10, N = IIB supergravity: Lorentz invariant actions and duality,''
  JHEP {\bf 9807} (1998) 017
  [hep-th/9806140].
  


\bibitem{Myers:1999ps}
  R.~C.~Myers,
  ``Dielectric branes,''
  JHEP {\bf 9912} (1999) 022
  [hep-th/9910053].
	
\bibitem{Witten:1995im}
  E.~Witten,
  ``Bound states of strings and p-branes,''
  Nucl.\ Phys.\ B {\bf 460} (1996) 335
  [hep-th/9510135].
	
\bibitem{Bachas:1995kx}
  C.~Bachas,
  ``D-brane dynamics,''
  Phys.\ Lett.\ B {\bf 374} (1996) 37
  [hep-th/9511043].
	
\bibitem{Tseytlin:1996it}
  A.~A.~Tseytlin,
  ``Selfduality of Born-Infeld action and Dirichlet three-brane of type IIB superstring theory,''
  Nucl.\ Phys.\ B {\bf 469} (1996) 51
  [hep-th/9602064].

\bibitem{Burgess:1993np}
  C.~P.~Burgess and F.~Quevedo,
  ``Bosonization as duality,''
  Nucl.\ Phys.\ B {\bf 421} (1994) 373
  [hep-th/9401105].

\bibitem{AbdusSalam:2020ywo}
S.~AbdusSalam, S.~Abel, M.~Cicoli, F.~Quevedo and P.~Shukla,
[arXiv:2005.11329 [hep-th]].

\bibitem{Duff:1995wd}
M.~J.~Duff, J.~T.~Liu and R.~Minasian,
Nucl. Phys. B \textbf{452} (1995), 261-282
doi:10.1016/0550-3213(95)00368-3
[arXiv:hep-th/9506126 [hep-th]].
	

\bibitem{Halverson:2013qca}
J.~Halverson, H.~Jockers, J.~M.~Lapan and D.~R.~Morrison,
Commun. Math. Phys. \textbf{333} (2015) no.3, 1563-1584
doi:10.1007/s00220-014-2157-z
[arXiv:1308.2157 [hep-th]]. 

\bibitem{Demirtas:2019sip}
M.~Demirtas, M.~Kim, L.~Mcallister and J.~Moritz,
Phys. Rev. Lett. \textbf{124} (2020) no.21, 211603
doi:10.1103/PhysRevLett.124.211603
[arXiv:1912.10047 [hep-th]].

\end{thebibliography}
\end{document}